\documentclass[epj]{svjour}
\usepackage{graphicx}
\usepackage{color}

\def\mb#1{\setbox0=\hbox{$#1$}\kern-.025em\copy0\kern-\wd0
\kern-0.05em\copy0\kern-\wd0\kern-.025em\raise.0233em\box0}

\begin{document}
   \title{Thermodynamics of the HMF model with a magnetic field}

 \author{P.H. Chavanis}

\institute{Laboratoire de Physique Th\'eorique (IRSAMC), CNRS and UPS, Universit\'e de Toulouse, F-31062 Toulouse, France\\
\email{chavanis@irsamc.ups-tlse.fr}
}

\titlerunning{The HMF model with a magnetic field}

   \date{To be included later }

   \abstract{We study the thermodynamics of the Hamiltonian Mean Field
   (HMF) model with an external potential playing the role of a
   ``magnetic field''. If we consider only fully stable states, this
   system does not present any phase transition. However, if we take
   into account metastable states (for a restricted class of
   perturbations), we find a very rich phenomenology. In particular,
   the system displays a region of negative specific heats in the
   microcanonical ensemble in which the temperature decreases as the
   energy increases. This leads to ensembles inequivalence and to
   zeroth order phase transitions similar to the ``gravothermal
   catastrophe'' and to the ``isothermal collapse'' of
   self-gravitating systems. In the present case, they correspond to
   the reorganization of the system from an ``anti-aligned'' phase
   (magnetization pointing in the direction opposite to the magnetic
   field) to an ``aligned'' phase (magnetization pointing in the same
   direction as the magnetic field). We also find that the magnetic
   susceptibility can be negative in the microcanonical ensemble so
   that the magnetization decreases as the magnetic field
   increases. The magnetic curves can take various shapes depending on
   the values of energy or temperature. We describe hysteretic cycles
   involving positive or negative susceptibilities. We also show that
   this model exhibits gaps in the magnetization at fixed energy,
   resulting in ergodicity breaking.   }

   \maketitle
%

\section{Introduction}
\label{sec_model}

The statistical mechanics of systems with long-range interactions has
recently been the object of an intense activity
\cite{houches,assisebook,oxford,cdr}. In particular, the Hamiltonian
Mean Field (HMF) model \cite{ms,kk,ik,inagaki,pichon,ar,cvb}, which is
a prototype for systems with long-range interactions, has been
particularly studied by statistical mechanicians in order to
illustrate the beauty and the richness of these systems. The HMF model
can be viewed as a $XY$ spin system with infinite range
interactions or as a one dimensional model of
particles moving on a ring and interacting via a long-range potential
truncated to one Fourier mode (cosine potential). In that second
interpretation, it shares many analogies with self-gravitating systems
\cite{ik,inagaki,pichon,ar,cvb} but is much simpler to study since it
avoids difficulties linked with the singular nature of the
gravitational potential at the origin and the absence of a natural
confinement \cite{paddy,katzrev,ijmpb}.

The HMF model has been shown to possess many interesting features both
at equilibrium and out-of-equilibrium. The dynamical evolution of the
HMF model in the microcanonical ensemble displays non Boltzmannian
quasistationary states (QSS) \cite{ar,lrtbis,lrt} that are stable
steady states of the Vlasov equation
\cite{ik,cvb,choi,yamaguchi,incomplete,jain,cd,campastab}, out-of-equilibrium phase transitions explained by Lynden-Bell's theory of violent relaxation
\cite{lb,epjb,precommun,antobis,anto,proc,staniscia1} or by dynamical
processes \cite{bachelard,firpo,leoncini,bachelardbis}, re-entrant
phases \cite{epjb,staniscia1}, negative kinetic specific heats
\cite{prlneg,cc}, incomplete violent relaxation \cite{incomplete,cc},
vanishing of the Lyapunov exponent
\cite{latoralyap,firpolyap},  aging \cite{aging}, glassy dynamics \cite{glassy}, collective oscillations \cite{mk}, non-ergodicity \cite{pluch,figue}, dynamical phase transitions \cite{campadyn}, algebraic velocity
correlation functions and anomalous diffusion
\cite{lrr,yamaseul,plr,latora,news,bd,ybd}, front structure of the velocity distribution tails
\cite{cl} and slow convergence towards the Boltzmann distribution
\cite{lrtbis,lrt,yamaguchi,yamaseul,cgm} explained by kinetic theory
\cite{cdr,bgm,chavkin}. The influence of an external noise and the
coupling with a thermal bath (canonical ensemble) have also been
considered \cite{cvb,bo1,bo2,bco,cbo}, as well as the effect of
collisions between particles \cite{gm}. At equilibrium, the caloric
curve $\beta(E)$ displays a second order phase transition between a
homogeneous phase at high energies/temperatures and a clustered phase
at low energies/temperatures
\cite{ms,kk,ik,inagaki,pichon,ar,cvb,largedev,isostab}. This is similar to the
collapse (or Jeans instability) of self-gravitating systems
\cite{ik,pichon,cvb,cd}. However, contrary to self-gravitating systems \cite{paddy,katzrev,ijmpb}, the caloric curve does not exhibit negative specific heats and the ensembles are
equivalent \cite{cdr}.

New features arise when the system is submitted to an external
potential playing the role of a magnetic field. The magnetic field
imposes a specific direction to the magnetization and breaks the
rotational symmetry of the original HMF model. In this paper, we study
in detail the thermodynamics of this system. If we consider only fully
stable states, there is no phase transition. However, if we take into
account metastable states (for a restricted class of perturbations),
we find a very rich phenomenology. In particular, the system displays
a region of negative specific heat in the microcanonical ensemble in
which the temperature decreases as the energy increases. This leads to
ensembles inequivalence and to zeroth order phase transitions similar
to the ``gravothermal catastrophe'' and ``isothermal collapse'' of
self-gravitating systems. In the present case, they correspond to the
reorganization of the system from an ``anti-aligned'' phase
(magnetization pointing in the direction opposite to the magnetic
field) to an ``aligned'' phase (magnetization pointing in the same
direction as the magnetic field). We also find that the magnetic
susceptibility can be negative in the microcanonical ensemble so that
the magnetization decreases as the magnetic field increases. The
magnetic curves can take various shapes depending on the values of
energy or temperature. We describe hysteretic cycles involving both
positive and negative magnetic susceptibilities. We also show that
this model exhibits gaps in the magnetization at fixed energy,
resulting in ergodicity breaking.

A preliminary study of this model has been performed by Velazquez \&
Guzman \cite{vg} who derived several interesting results. We believe,
however, that their study carries out some mistakes\footnote{We find,
indeed, different conditions of stability. We think that the reason
comes from the fact that Velazquez \& Guzman \cite{vg} work in terms
of the internal energy, while we work in terms of the total energy
taking into account the contribution of the magnetic field.} and that
complements are needed. This is the object of the present paper. Our
paper closely follows the presentation and the methodology exposed in
\cite{cvb,isostab} for the usual HMF model without magnetic field.

\section{Thermodynamical approach}
\label{sec_ta}

\subsection{The HMF model with a magnetic field}
\label{sec_m}

The HMF model is a system of $N$ particles of unit mass $m=1$ moving on a circle and interacting via a cosine potential. When an external potential (named ``magnetic field'') $H$ is imposed, the dynamics of these particles is governed by the Hamilton equations
\begin{eqnarray}
\frac{d\theta_i}{dt}=\frac{\partial {\cal H}}{\partial v_i}, \qquad \frac{d v_i}{dt}=-\frac{\partial {\cal H}}{\partial \theta_i},\qquad\qquad\nonumber\\
{\cal H}=\frac{1}{2}\sum_{i=1}^{N} v_i^2-\frac{k}{4\pi}\sum_{i\neq j}\cos(\theta_i-\theta_j)-H\sum_{i=1}^{N} \cos\theta_i,
\label{m1}
\end{eqnarray}
where $\theta_i\in [-\pi,\pi]$ and $-\infty<v_i<+\infty$ denote  the position (angle) and the velocity of particle $i$ and $k$ is the coupling constant (we assume here that $k>0$). We have assumed, without loss of generality, that the field ${\bf H}$ is directed along the $x$-axis. Like for the usual HMF model, the proper thermodynamic limit corresponds to $N\rightarrow +\infty$ in such a way that the rescaled energy $\epsilon=8\pi E/kM^2$ and  the rescaled inverse temperature $\eta=\beta kM/4\pi$ remains of order unity \cite{cvb}. We can take $k\sim 1/N$ \cite{ms} which is the Kac prescription. In that case, the energy is extensive ($E/N\sim 1$) and the temperature intensive ($\beta\sim 1$) but the system remains fundamentally non-additive \cite{cdr}. For $N\rightarrow +\infty$, the mean field approximation is exact \cite{ms,bh} and the $N$-body distribution function  is a product of $N$ one-body distributions: $P_{N}(\theta_1,v_1,...,\theta_N,v_N,t)=P_{1}(\theta_1,v_1,t)...P_{1}(\theta_N,v_N,t)$.

Let us introduce the distribution function $f(\theta,v,t)=NP_{1}(\theta,v,t)$.  For a fixed interval of time and $N\rightarrow +\infty$, the evolution of the distribution function $f(\theta,v,t)$ is governed by the Vlasov equation \cite{bh}:
\begin{eqnarray}
\label{m2}
\frac{\partial f}{\partial t}+v\frac{\partial f}{\partial\theta}-\frac{\partial\Phi}{\partial\theta}\frac{\partial f}{\partial v}-\frac{\partial\Phi_{ext}}{\partial\theta}\frac{\partial f}{\partial v}=0,
\end{eqnarray}
where
\begin{equation}
\Phi(\theta,t)=-\frac{k}{2\pi}\int_{0}^{2\pi}\cos(\theta-\theta')\rho(\theta',t)\, d\theta',\label{m3}
\end{equation}
is the self-consistent potential generated by the density of particles $\rho(\theta,t)=\int f(\theta,v,t)\, dv$ and
\begin{eqnarray}
\label{m4}
\Phi_{ext}(\theta)=-H\cos\theta,
\end{eqnarray}
is the external potential. The mean force acting on a particle located in $\theta$ is $\langle F\rangle(\theta,t)=-\partial\Phi/\partial\theta(\theta,t)-\Phi_{ext}'(\theta)$. Expanding the cosine function in equation (\ref{m3}), we obtain
\begin{eqnarray}
\label{m5}
\Phi(\theta,t)=-B_x\cos\theta-B_y \sin\theta,
\end{eqnarray}
where
\begin{eqnarray}
\label{m6}
B_x=\frac{k}{2\pi}\int_0^{2\pi} \rho(\theta,t)\cos\theta\, d\theta,
\end{eqnarray}
\begin{eqnarray}
\label{m7}
B_y=\frac{k}{2\pi}\int_0^{2\pi}\rho(\theta,t)\sin\theta\, d\theta,
\end{eqnarray}
are proportional to the magnetization $b_x=\frac{1}{N}\int\rho\cos\theta\, d\theta$ and $b_y=\frac{1}{N}\int\rho\sin\theta\, d\theta$. The magnetization can be viewed as the order parameter of the HMF model.

Let us introduce the mass
\begin{eqnarray}
\label{m8}
M[\rho]=\int \rho \, d\theta,
\end{eqnarray}
and the mean field energy
\begin{eqnarray}
\label{m9}
E[f]=\frac{1}{2}\int f v^2\, d\theta dv+\frac{1}{2}\int \rho\Phi\, d\theta+\int \rho\Phi_{ext}\, d\theta,
\end{eqnarray}
where the first term is the kinetic energy, the second term the potential energy of interaction and the third term the potential energy due to the external field. Using equations (\ref{m5})-(\ref{m7}), the potential energy can be expressed in terms of the magnetization as
\begin{eqnarray}
\label{m10}
W=-\frac{\pi B^2}{k}-\frac{2\pi}{k}B_x H.
\end{eqnarray}
We also introduce the Boltzmann entropy
\begin{eqnarray}
\label{m11}
S[f]=-\int f\ln \left (\frac{f}{N}\right )\, d\theta dv,
\end{eqnarray}
and the Boltzmann free energy
\begin{eqnarray}
\label{m12}
F[f]=E[f]-T S[f],
\end{eqnarray}
where $T=1/\beta>0$ is the temperature. In the microcanonical ensemble, the statistical equilibrium state is determined by the maximization problem
\begin{eqnarray}
\label{m13}
\max_f\left\lbrace S\lbrack f\rbrack\, |\, E\lbrack f\rbrack=E,\, M\lbrack f\rbrack=M\right\rbrace.
\end{eqnarray}
In the canonical ensemble,   the statistical equilibrium state is determined by the minimization problem
\begin{eqnarray}
\label{m14}
\min_f\left\lbrace F\lbrack f\rbrack\, |\, M\lbrack f\rbrack=M\right\rbrace.
\end{eqnarray}
The Boltzmann entropy functional (\ref{m11}) and the maximum entropy principle (\ref{m13}) can be justified by a standard combinatorial analysis (see, e.g., \cite{cvb}). The distribution function $f(\theta,v)$ that is solution of (\ref{m13}) is the {\it most probable macroscopic state}, i.e. the macrostate that is the most represented at the microscopic level, assuming that the accessible  microstates (with the proper values of mass and energy) are equiprobable.

\subsection{Critical points}
\label{sec_c}

We shall first determine the {\it critical points} of these variational problems.  The critical points of the maximization problem (\ref{m13}) are determined by the variational principle
\begin{eqnarray}
\label{c1}
\delta S-\beta\delta E-\alpha\delta M=0,
\end{eqnarray}
where $\beta=1/T$ and $\alpha$ are Lagrange multipliers associated with the conservation of energy and mass. The critical points of the minimization problem (\ref{m14}) are determined by the variational principle
\begin{eqnarray}
\label{c2}
\delta F+\alpha T\delta M=0,
\end{eqnarray}
where $\alpha$ is a Lagrange multiplier associated with the conservation of mass. Since $T$ is fixed in the canonical ensemble, it is clear that equation (\ref{c2}) is equivalent to equation (\ref{c1}). Therefore, the optimization problems (\ref{m13}) and (\ref{m14}) have the {\it same} critical points. Performing the variations in  equations  (\ref{c1}) and (\ref{c2}), we find that the critical points are given by the mean field Maxwell-Boltzmann distribution
\begin{eqnarray}
\label{c3}
f(\theta,v)=A'\, e^{-\beta\left \lbrack\frac{v^2}{2}+\Phi_{tot}(\theta)\right \rbrack},
\end{eqnarray}
where $A'=M e^{-1-\alpha}$ is a constant and $\Phi_{tot}(\theta)=\Phi(\theta)+\Phi_{ext}(\theta)$ is the total potential in $\theta$. Integrating over the velocities,
we get the mean field Boltzmann distribution
\begin{eqnarray}
\label{c4}
\rho(\theta)=A\, e^{-\beta\Phi_{tot}(\theta)},
\end{eqnarray}
where $A=(2\pi/\beta)^{1/2}A'$. Using the expressions (\ref{m4}) and (\ref{m5}) of the potential, the distribution function (\ref{c3}) can be rewritten
\begin{eqnarray}
\label{c5}
f(\theta,v)=A'\, e^{-\beta\left \lbrack \frac{v^2}{2}-(B_x+H)\cos\theta-B_y\sin\theta\right \rbrack}.
\end{eqnarray}
In the following, we consider critical points whose magnetization
${\bf B}$ is parallel to the magnetic field ${\bf H}$ so that $B_y=0$
and $B_x=B$ (we show in Appendix \ref{sec_abs} that there is no
critical point with $B_y\neq 0$).  In that case, the foregoing
expression takes the form
\begin{eqnarray}
\label{c6}
f(\theta,v)=A'\, e^{-\beta\left \lbrack\frac{v^2}{2}-(B+H)\cos\theta\right \rbrack}.
\end{eqnarray}
The corresponding density profile is
\begin{eqnarray}
\label{c7}
\rho(\theta)=A\, e^{\beta (B+H)\cos\theta}.
\end{eqnarray}
The amplitude $A$ and the magnetization $B$ are determined by substituting equation (\ref{c7}) in equations (\ref{m8}) and (\ref{m6}). This yields
\begin{equation}
A=\frac{M}{2\pi I_{0}(\beta (B+H))},
\label{c8}
\end{equation}
and
\begin{equation}
\frac{2\pi B}{kM}=\frac{I_{1}(\beta (B+H))}{I_{0}(\beta (B+H))},
\label{c9}
\end{equation}
where $I_n(x)$ is the modified Bessel function of order $n$. Equation (\ref{c9}) determines the magnetization $B$ as a function of the temperature $T$. Then, $A$ is given by equation (\ref{c8}). Finally, the distribution function and the density profile can be written
\begin{eqnarray}
\label{c10}
f(\theta,v)=\left (\frac{\beta}{2\pi}\right )^{1/2} \, \rho({\theta})\, e^{-\beta\frac{v^2}{2}},
\end{eqnarray}
\begin{eqnarray}
\rho(\theta)=\frac{M}{2\pi I_0(\beta (B+H))} e^{\beta (B+H)\cos\theta},
\label{c11}
\end{eqnarray}
where $B$ is determined in terms of $T$ by equation (\ref{c9}). The study of the self-consistency relation (\ref{c9}) will be performed graphically in Section \ref{sec_cs}.

\subsection{Thermodynamical parameters}
\label{sec_t}

Let us now determine the expressions of the energy, entropy and free energy. For the Maxwell-Boltzmann distribution (\ref{c10}), the kinetic energy is
\begin{eqnarray}
\label{t1}
K=\frac{1}{2}MT.
\end{eqnarray}
Combining this relation with equation (\ref{m10}), we find that the total energy $E=K+W$ is given by
\begin{eqnarray}
\label{t2}
E=\frac{1}{2}MT-\frac{\pi B^2}{k}-\frac{2\pi}{k}BH.
\end{eqnarray}
The series of equilibria giving $T$ as a function of $E$ is determined by equations (\ref{c9}) and (\ref{t2}) by eliminating $B$. The relation that gives the magnetization $B$ as a function of the energy $E$ is determined by equations (\ref{c9}) and (\ref{t2}) by eliminating $T$. Finally, using equations (\ref{m11}) and (\ref{c10}), the entropy is given by
\begin{eqnarray}
\label{t3}
S=\frac{1}{2}M\ln T-\int\rho\ln\rho\, d\theta,
\end{eqnarray}
up to a term $\frac{1}{2}M+\frac{1}{2}M \ln (2\pi)+M\ln M$. Using equation (\ref{c11}), it can be rewritten
\begin{eqnarray}
\label{t4}
S=\frac{1}{2}M\ln T+M\ln I_0(\beta (B+H))-\frac{2\pi}{kT}B(B+H),\nonumber\\
\end{eqnarray}
up to a term $\frac{1}{2}M+\frac{3}{2}M \ln (2\pi)$. The relation between the entropy $S$ and  the energy $E$ is determined by equations (\ref{t4}), (\ref{t2}) and (\ref{c9}) by eliminating $T$ and $B$. Using equations (\ref{t4}) and (\ref{t2}), the free energy (\ref{m12}) is given by
\begin{eqnarray}
\label{t5}
F=\frac{1}{2}M T-\frac{1}{2}MT\ln T-MT\ln I_0(\beta (B+H))+\frac{\pi B^2}{k},\nonumber\\
\end{eqnarray}
up to a term $-\frac{1}{2}MT-\frac{3}{2}MT \ln (2\pi)$. The relation between the free energy $F$ and the temperature $T$ is determined by equations (\ref{t5}) and (\ref{c9}) by eliminating $B$.

\begin{figure}
\begin{center}
\includegraphics[clip,scale=0.3]{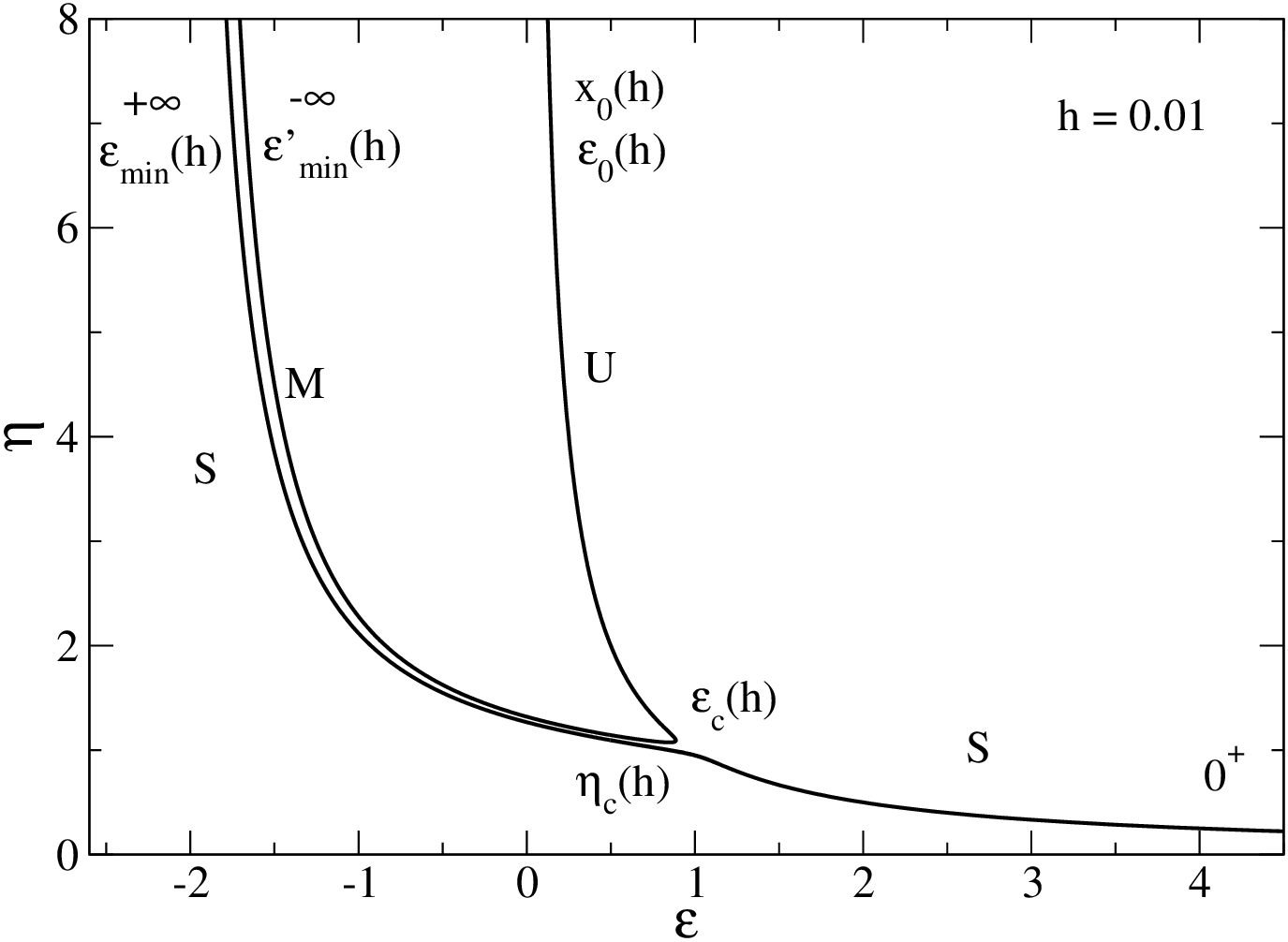}
\caption{Series of equilibria (caloric curve) giving the inverse temperature $\eta$ as a function of the energy $\epsilon$ for $h=0.01$. In this figure, and in the following figures, we have indicated the values of $x$ that parameterizes these curves.}
\label{epsilon-etaNEW}
\end{center}
\end{figure}

\begin{figure}
\begin{center}
\includegraphics[clip,scale=0.3]{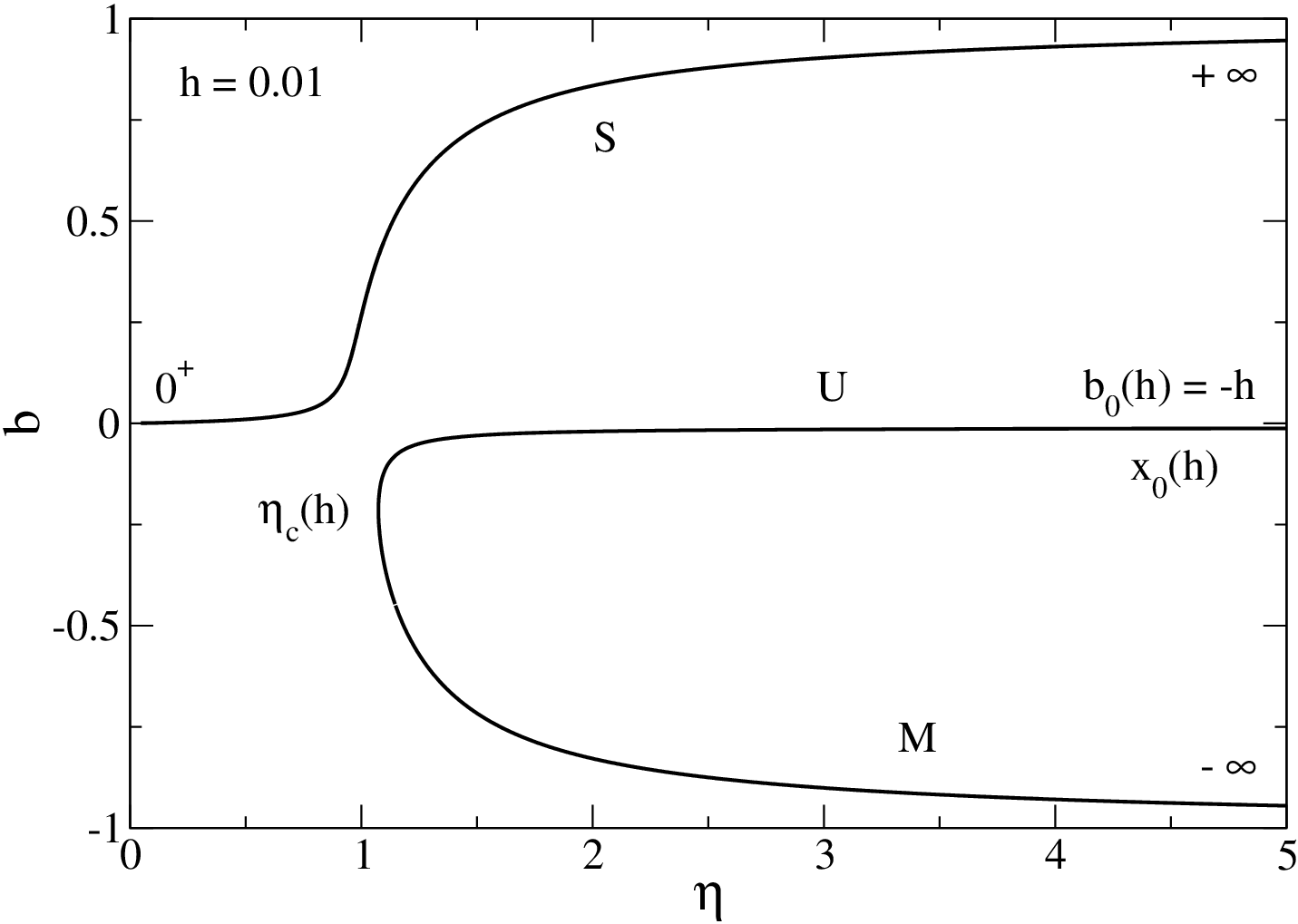}
\caption{Magnetization $b$ (order parameter) as a function of the inverse temperature $\eta$ for $h=0.01$. It exhibits a turning point of temperature at $\eta_c(h)$.}
\label{eta-bNEW}
\end{center}
\end{figure}

\begin{figure}
\begin{center}
\includegraphics[clip,scale=0.3]{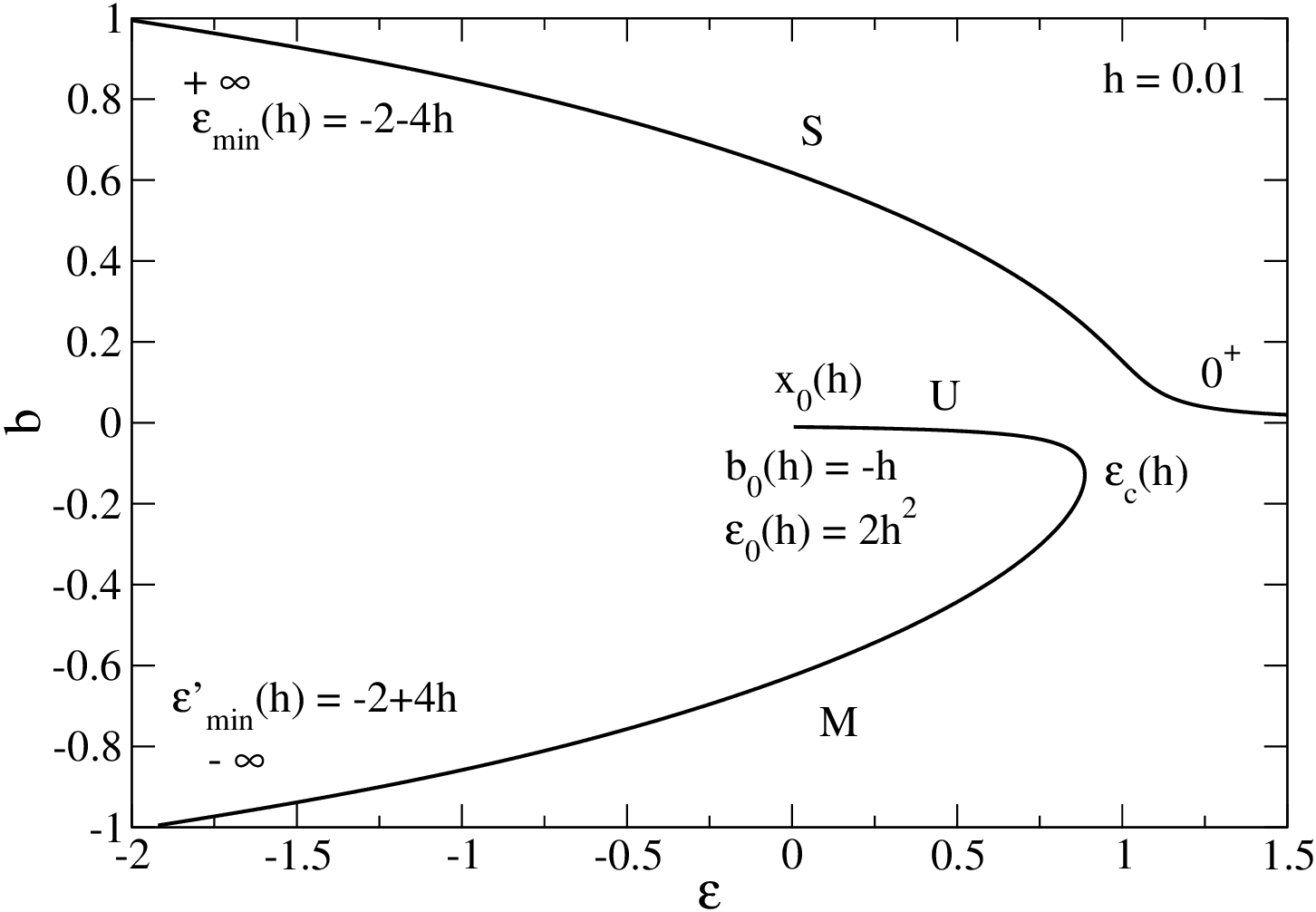}
\caption{Magnetization $b$ (order parameter) as a function of the energy $\epsilon$ for $h=0.01$. It exhibits a turning point of energy at $\epsilon_c(h)$.}
\label{epsilon-bNEW}
\end{center}
\end{figure}

\begin{figure}
\begin{center}
\includegraphics[clip,scale=0.3]{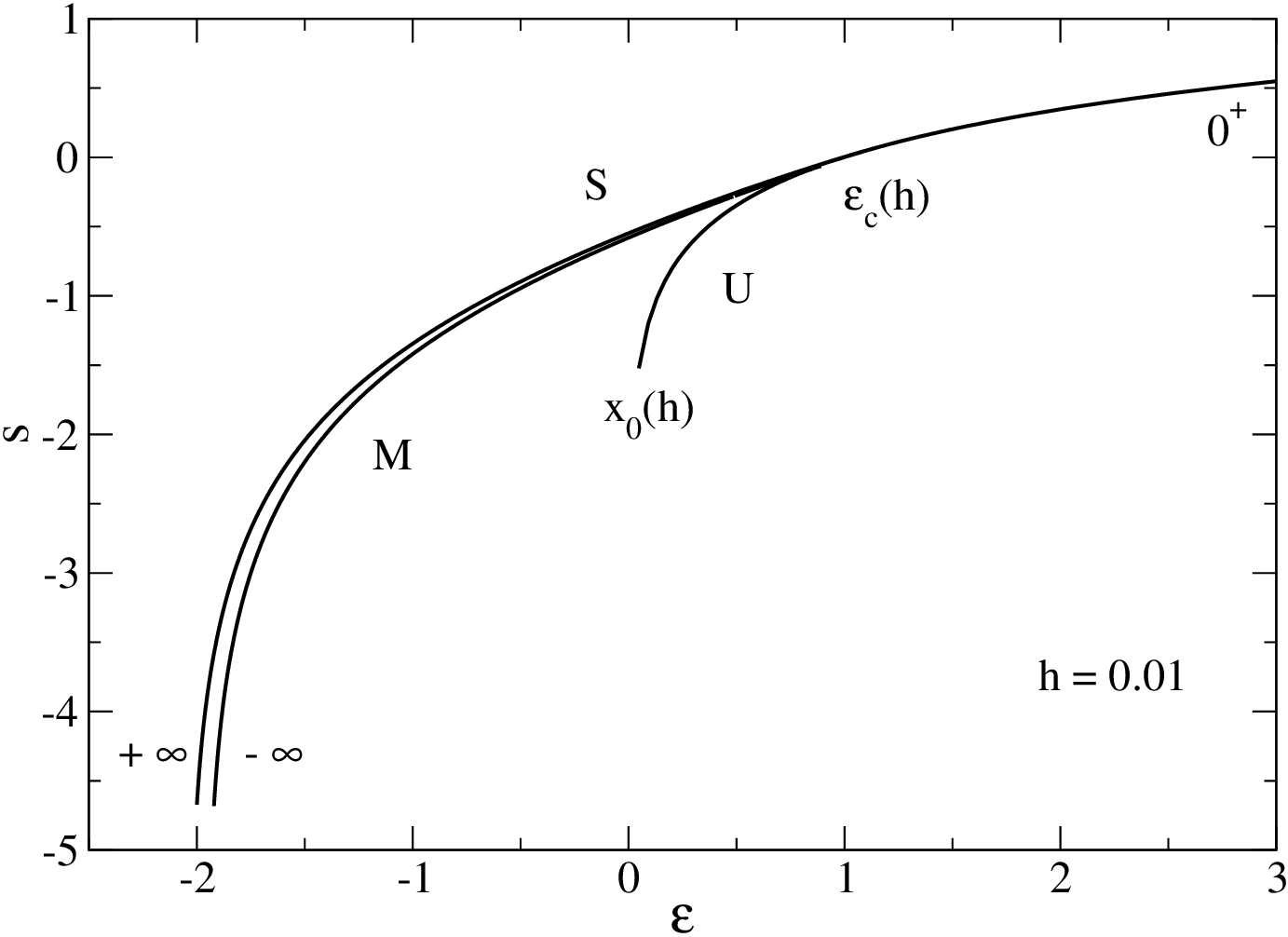}
\caption{Entropy $s$ as a function of the energy $\epsilon$ for $h=0.01$.}
\label{entropie}
\end{center}
\end{figure}

\begin{figure}
\begin{center}
\includegraphics[clip,scale=0.3]{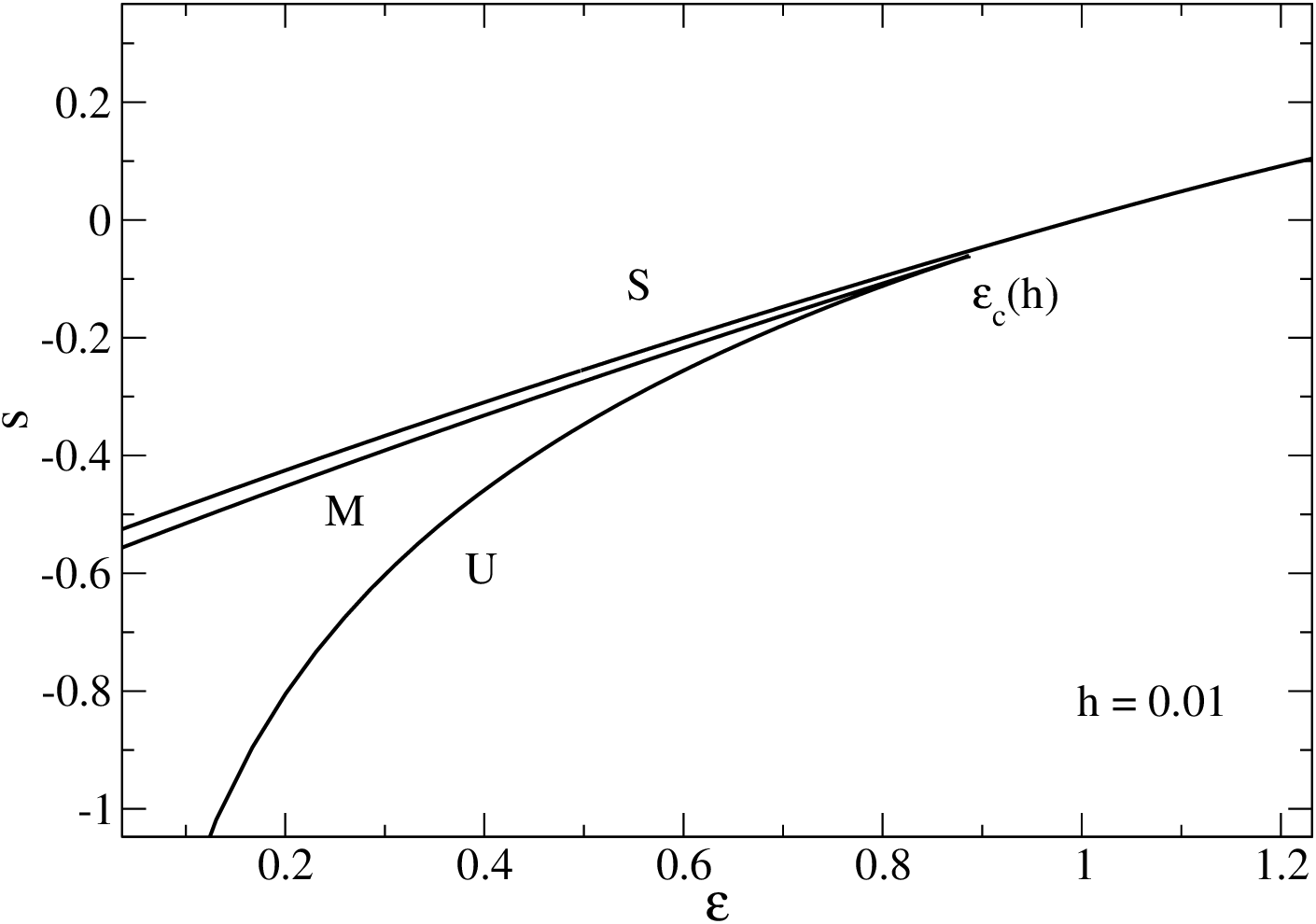}
\caption{Same as Figure \ref{entropie}  (zoom).}
\label{entropieZOOM}
\end{center}
\end{figure}

\begin{figure}
\begin{center}
\includegraphics[clip,scale=0.3]{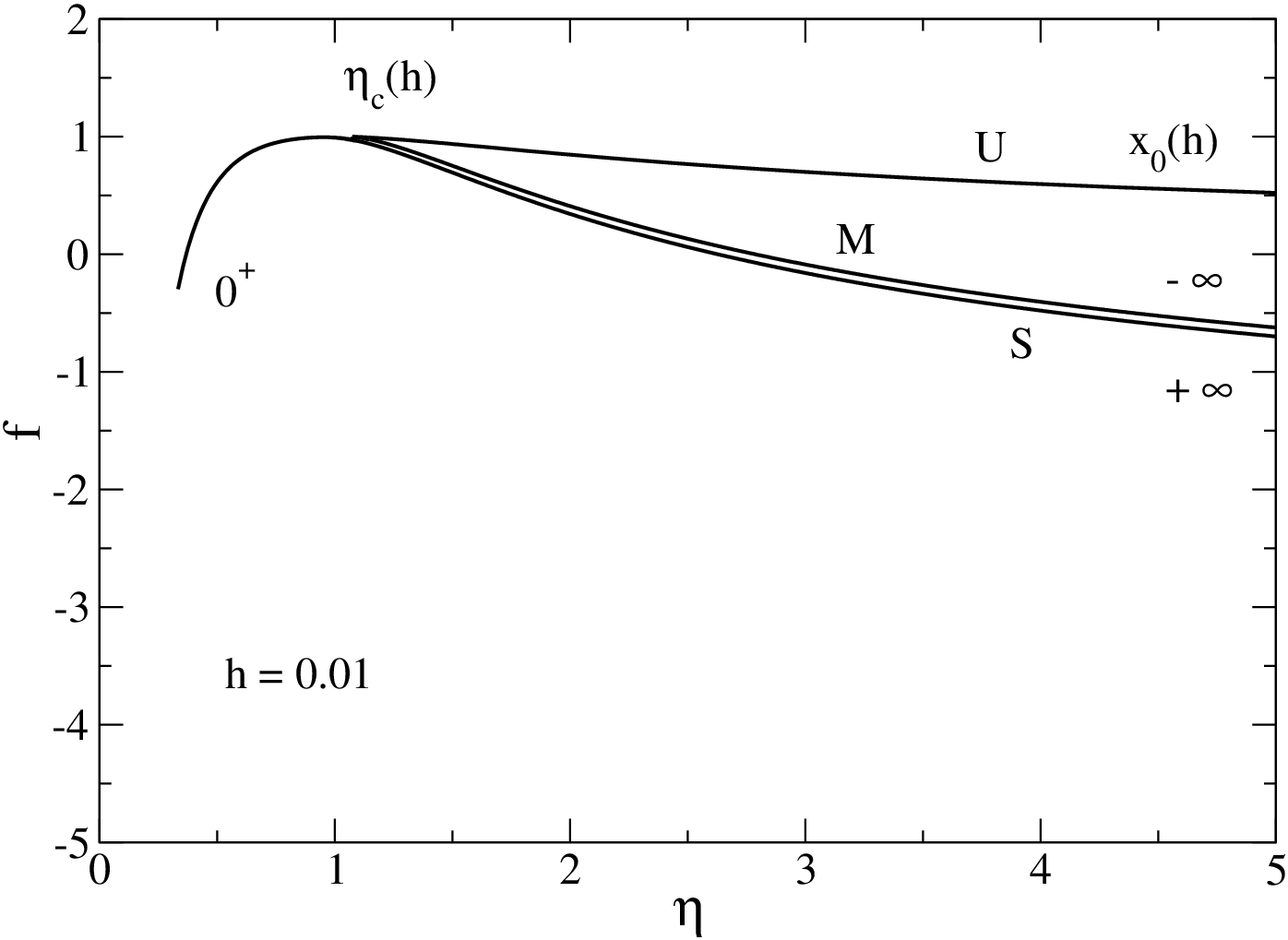}
\caption{Free energy $f$ as a function of the inverse temperature $\eta$ for $h=0.01$.}
\label{f}
\end{center}
\end{figure}

\begin{figure}
\begin{center}
\includegraphics[clip,scale=0.3]{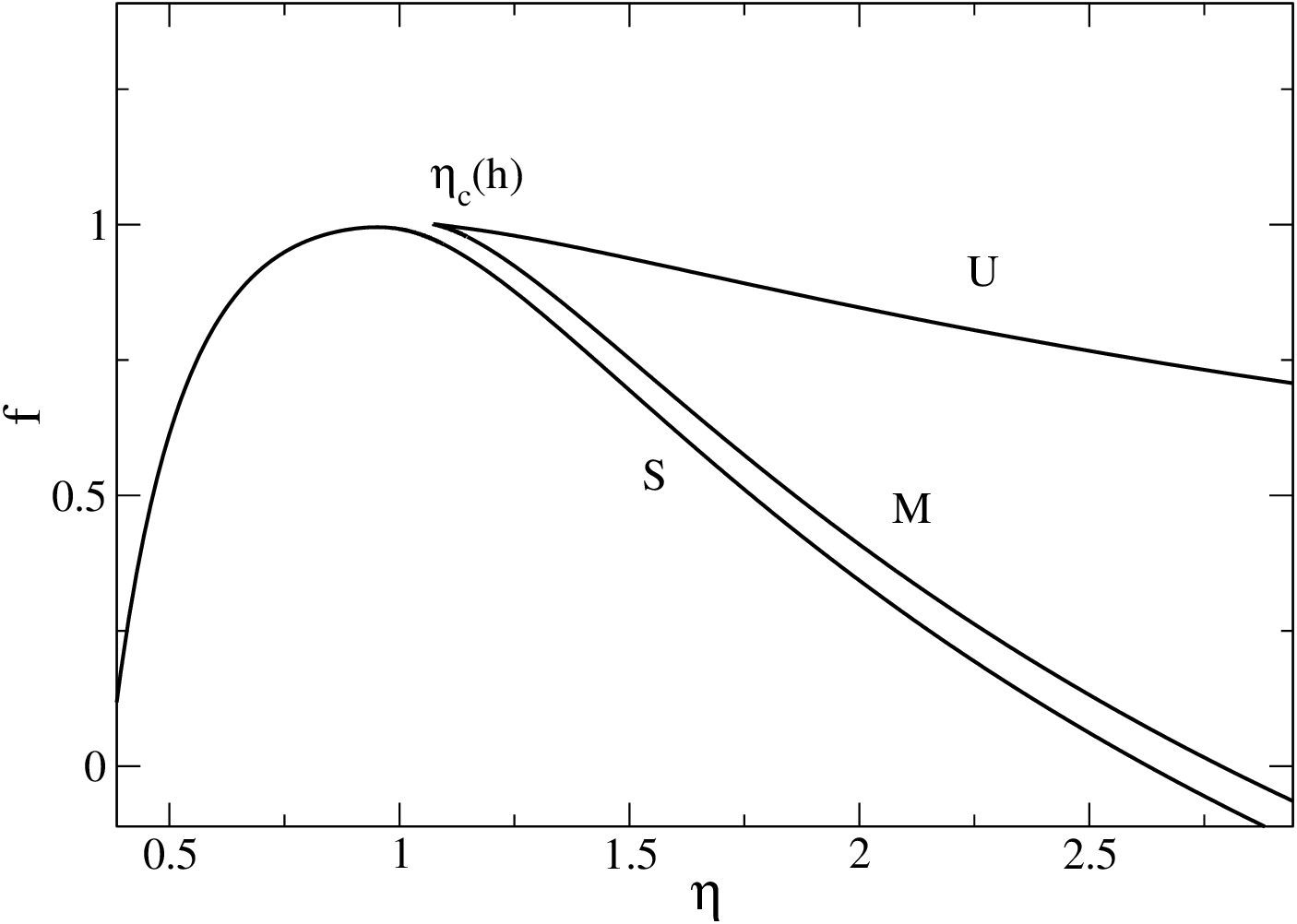}
\caption{Same as Figure \ref{f} (zoom).}
\label{fZOOM}
\end{center}
\end{figure}

It is convenient to write these equations in parametric form by introducing the parameter $x\equiv \beta (B+H)$. Then, defining $h\equiv 2\pi H/(kM)$, we obtain
\begin{eqnarray}
\label{t6}
b\equiv \frac{2\pi B}{k M}=\frac{I_1(x)}{I_0(x)},
\end{eqnarray}
\begin{eqnarray}
\label{t7}
\eta\equiv \frac{\beta k M}{4\pi}=\frac{x}{2(b(x)+h)},
\end{eqnarray}
\begin{eqnarray}
\label{t8}
\epsilon\equiv \frac{8\pi E}{kM^2}=\frac{1}{\eta(x)}-2b(x)^2-4hb(x),
\end{eqnarray}
\begin{eqnarray}
\label{t9}
s\equiv \frac{S}{M}=-\frac{1}{2}\ln\eta(x)+\ln I_0(x)-x b(x),
\end{eqnarray}
\begin{eqnarray}
\label{t10}
f\equiv \frac{8\pi F}{kM^2}=\epsilon(x)-\frac{2}{\eta(x)}s(x),
\end{eqnarray}
where unimportant constants have been omitted in the expression of the
entropy.  From these relations, we can obtain the curves $T(E)$,
$B(T)$, $B(E)$, $S(E)$ and $F(T)$ in parametric form. These curves are
plotted in Figures \ref{epsilon-etaNEW}-\ref{fZOOM} for $h=0.01$. We
have chosen a small value of $h$ in order to show the connection with
the results of the ordinary HMF model ($h=0$)
\cite{cvb,isostab}. Other curves are represented in Section
\ref{sec_poincare}.  The relations (\ref{t6})-(\ref{t10}) characterize
{\it all the critical points} of (\ref{m13}) and (\ref{m14}). The
structure of the series of equilibria will be studied in Section
\ref{sec_se}. Then, we must select among the critical points those
that are (local) {\it maxima} of $S$ at fixed $E$ and $M$
(microcanonical ensemble) and those that are (local) {\it minima} of
$F$ at fixed $M$ (canonical ensemble). This will be the object of
Section \ref{sec_poincare}.

\subsection{Series of equilibria}
\label{sec_se}

By definition, the series of equilibria is formed by all the critical points of entropy at fixed mass and energy or, equivalently, by all the critical points of free energy at fixed mass (we have already indicated that the variational problems (\ref{m13}) and (\ref{m14}) have the same critical points). The stability of these critical points, in each ensemble, will be investigated in Section \ref{sec_poincare} by using the Poincar\'e theorem and in Sections \ref{sec_cano} and \ref{sec_micro} by studying the sign of the second order variations of entropy or free energy. In this section, we simply describe the structure of the series of equilibria.

Let us assume that $h>0$ to fix the ideas (the case $h<0$ can be
treated symmetrically). To understand the following discussion, it is required to look in parallel at
Figures \ref{epsilon-etaNEW}-\ref{fZOOM} and at the asymptotic expansions of Appendix \ref{sec_asy}. The
graphical construction  of Figure  \ref{blambda} can also be useful.
We need to distinguish two curves: (i) the
series of equilibria with positive magnetization (same sign as the
imposed field) corresponds to $0\le x<+\infty$. It will be called the
``aligned'' phase. For $x\rightarrow 0$, we find that $b\rightarrow
0$, $\eta\rightarrow 0$ and $\epsilon\rightarrow +\infty$. More
precisely, $\eta\sim 1/\epsilon$, $b\sim h\eta$ and $b\sim
h/\epsilon$. For $x\rightarrow +\infty$, we find that $b\rightarrow 1$,
$\eta\rightarrow +\infty$ and $\epsilon\rightarrow
\epsilon_{min}(h)\equiv -2-4h$. More precisely, $\eta\sim
2/(\epsilon-\epsilon_{min}(h))$, $1-b\sim 1/(4(1+h)\eta)$ and $1-b\sim
(\epsilon-\epsilon_{min}(h))/(8(1+h))$; (ii) the series of equilibria
with negative magnetization (sign opposite to the imposed field)
corresponds to $-\infty<x\le x_0(h)$. It will be called the
``anti-aligned'' phase. For $x\rightarrow -\infty$, we find that
$b\rightarrow -1$, $\eta\rightarrow +\infty$ and $\epsilon\rightarrow
\epsilon'_{min}(h)\equiv -2+4h$. More precisely, $\eta\sim
2/(\epsilon-\epsilon'_{min}(h))$, $-1-b\sim -1/(4(1-h)\eta)$ and
$-1-b\sim -(\epsilon-\epsilon'_{min}(h))/(8(1-h))$. For $x\rightarrow
x_0(h)$, we find that $b\rightarrow b_0(h)\equiv -h$, $\eta\rightarrow +\infty$
and $\epsilon\rightarrow \epsilon_0(h)\equiv 2h^2$. More precisely,
$\eta\sim 1/(\epsilon-\epsilon_{0}(h))$, $b_0(h)-b\sim
-x_0(h)/(2\eta)$ and $b_0(h)-b\sim
-(x_0(h)/2)(\epsilon-\epsilon_{0}(h))$.  The series of equilibria with
negative magnetization presents a turning point of temperature at
$\eta=\eta_c(h)$ corresponding to $x=x_{c}^{cano}(h)$ and a turning
point of energy at $\epsilon=\epsilon_{c}(h)$ corresponding to
$x=x_{c}^{micro}(h)$. Note that $x_{c}^{cano}(h)<x_{c}^{micro}(h)$
so that the turning points ($\epsilon(\eta_c(h)),\eta_c(h)$) and
($\epsilon_{c}(h),\eta(\epsilon_c(h))$) differ. The ``anti-aligned''
phase exists only for $h<1$. For $h\rightarrow 1^{-}$, we find that
$\eta_c(h)\rightarrow +\infty$, $\epsilon_c(h)\rightarrow 2$, $\epsilon_0(h)\rightarrow 2$, $\epsilon'_{min}(h)\rightarrow 2$ and $b_0(h)\rightarrow -1$.

The asymptotic results described previously can be understood easily.

At very high temperatures $T\rightarrow +\infty$
(i.e. $\eta\rightarrow 0$) in the canonical ensemble or at very high
energies $\epsilon\rightarrow +\infty$ in the microcanonical ensemble,
the long-range interaction is negligible with respect to thermal
motions and the system behaves essentially like a noninteracting
perfect gas. The equilibrium state is a spatially homogeneous
configuration ($b=0$) with Maxwellian velocity distribution.  This is
the unique (global) entropy maximum at fixed mass and energy or the
unique (global) free energy minimum at fixed mass.  In that high
energy/temperature limit, $\epsilon\sim 1/\eta$.

Let us now consider the zero temperature state $T=0$ (i.e. $\eta\rightarrow +\infty$) in the canonical ensemble or the
minimum energy state in the microcanonical ensemble. To that purpose, we have to determine the
minimum energy at fixed mass. The kinetic energy is minimized by assigning the velocity $v=0$ to each particle. Then, we have to minimize the potential energy $w=-2(b^2+2bh)$.  It is easy to show that the global energy minimum is
$f(\theta,v)=M\delta(v)\delta(\theta)$ corresponding to a
magnetization $b=+1$ and an energy $\epsilon_{min}(h)=-2-4h$. There also
exists a local energy minimum
$f(\theta,v)=M\delta(v)\delta(\theta-\pi)$ corresponding to a
magnetization $b=-1$ and an energy $\epsilon'_{min}(h)=-2+4h$  and a local
energy maximum $f(\theta,v)=\frac{1}{2}M\delta(v)\lbrack
\delta(\theta-\theta_0)+\delta(\theta+\theta_0)\rbrack$ with $\cos\theta_0=-h$
corresponding to a magnetization $b=-h$ and an energy
$\epsilon_0(h)=2h^2$.

For $h\rightarrow 0$, we recover the results of \cite{cvb,isostab}
valid for $h=0$. In particular, $\epsilon_{min}=\epsilon_{min}'=-2$,
$\epsilon_{0}=0$ and $\epsilon_c=\eta_c=1$. In that case, the points
($\epsilon(\eta_c),\eta_c$) and ($\epsilon_{c},\eta(\epsilon_c)$)
coincide and the curve $\eta(\epsilon)$ forms a ``spike'' at
$(\epsilon_c,\eta_c)=(1,1)$ leading to a second order phase transition
\cite{cdr}. On the other hand, the branch of inhomogeneous solutions
becomes degenerate due to the invariance by rotation of the phase of
the magnetization.

\subsection{Specific heat}
\label{sec_sh}

\begin{figure}
\begin{center}
\includegraphics[clip,scale=0.3]{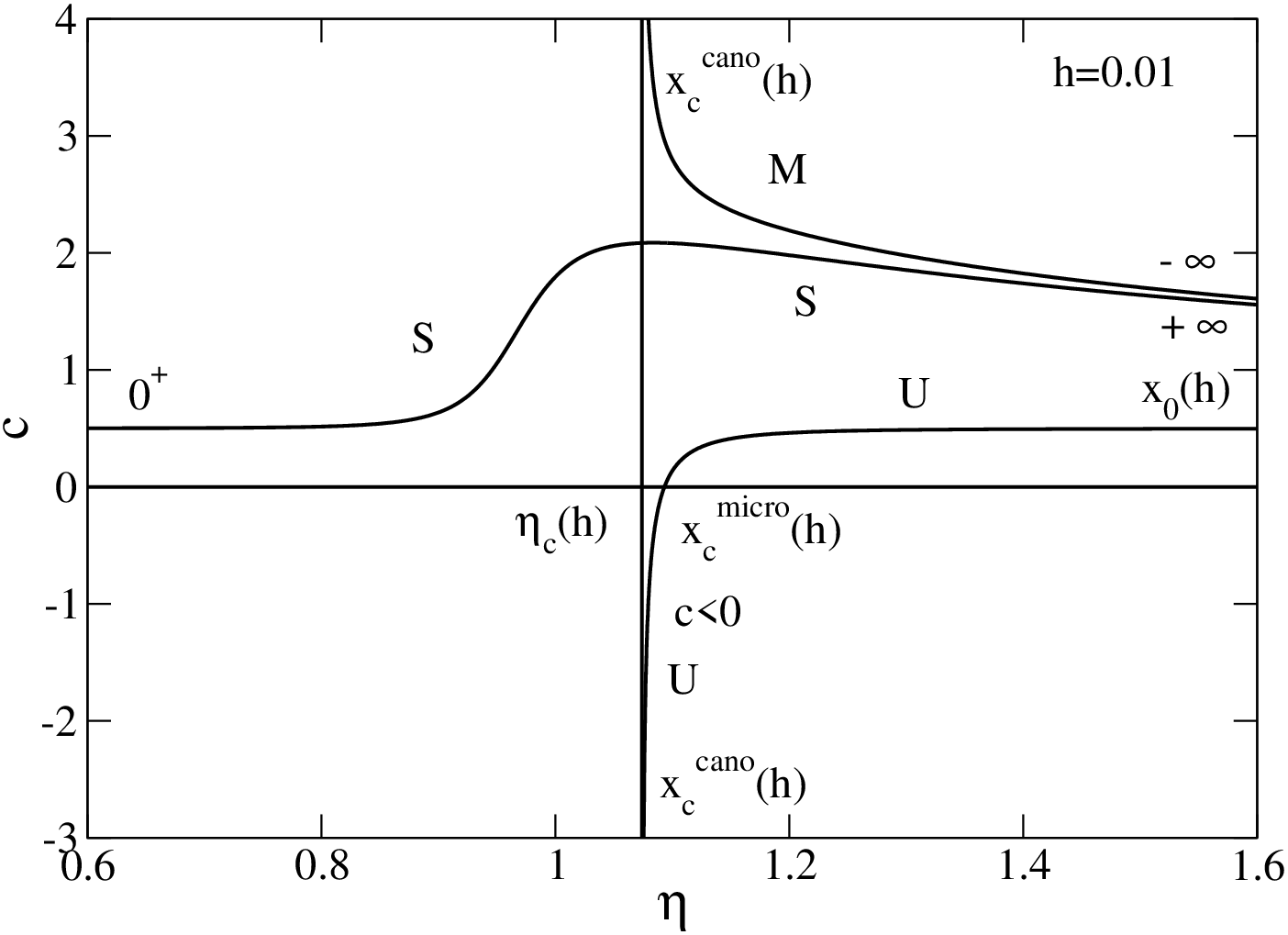}
\caption{Specific heat $c$  as a function of the inverse temperature $\eta$ for $h=0.01$. We find that $c\rightarrow 1/2$ for $\eta\rightarrow 0$ (corresponding to $x\rightarrow 0$), $c\rightarrow 1$ for $\eta\rightarrow +\infty$ (corresponding to $x\rightarrow \pm\infty$) and $c\rightarrow 1/2$ for $\eta\rightarrow +\infty$ (corresponding to $x\rightarrow x_0(h)$). In the anti-aligned phase ($-\infty<x\le x_0(h)$), the specific heat diverges like $c\propto \pm (\eta-\eta_c(h))^{-1/2}\rightarrow \pm\infty$ at the turning point of temperature $\eta_c(h)$ (corresponding to $x=x_c^{cano}(h)$). Close to $\eta_c(h)$, there is a region of negative specific heat (corresponding to $x_c^{cano}(h)<x<x_c^{micro}(h)$). The study of Section \ref{sec_poincare} shows that the condition of instability in the canonical ensemble corresponds to $x_c^{cano}(h)<x\le x_0(h)$. In particular, the states with negative specific heats are unstable in the canonical ensemble in agreement with general theorems of statistical mechanics.}
\label{chaleur-h0.01}
\end{center}
\end{figure}

\begin{figure}
\begin{center}
\includegraphics[clip,scale=0.3]{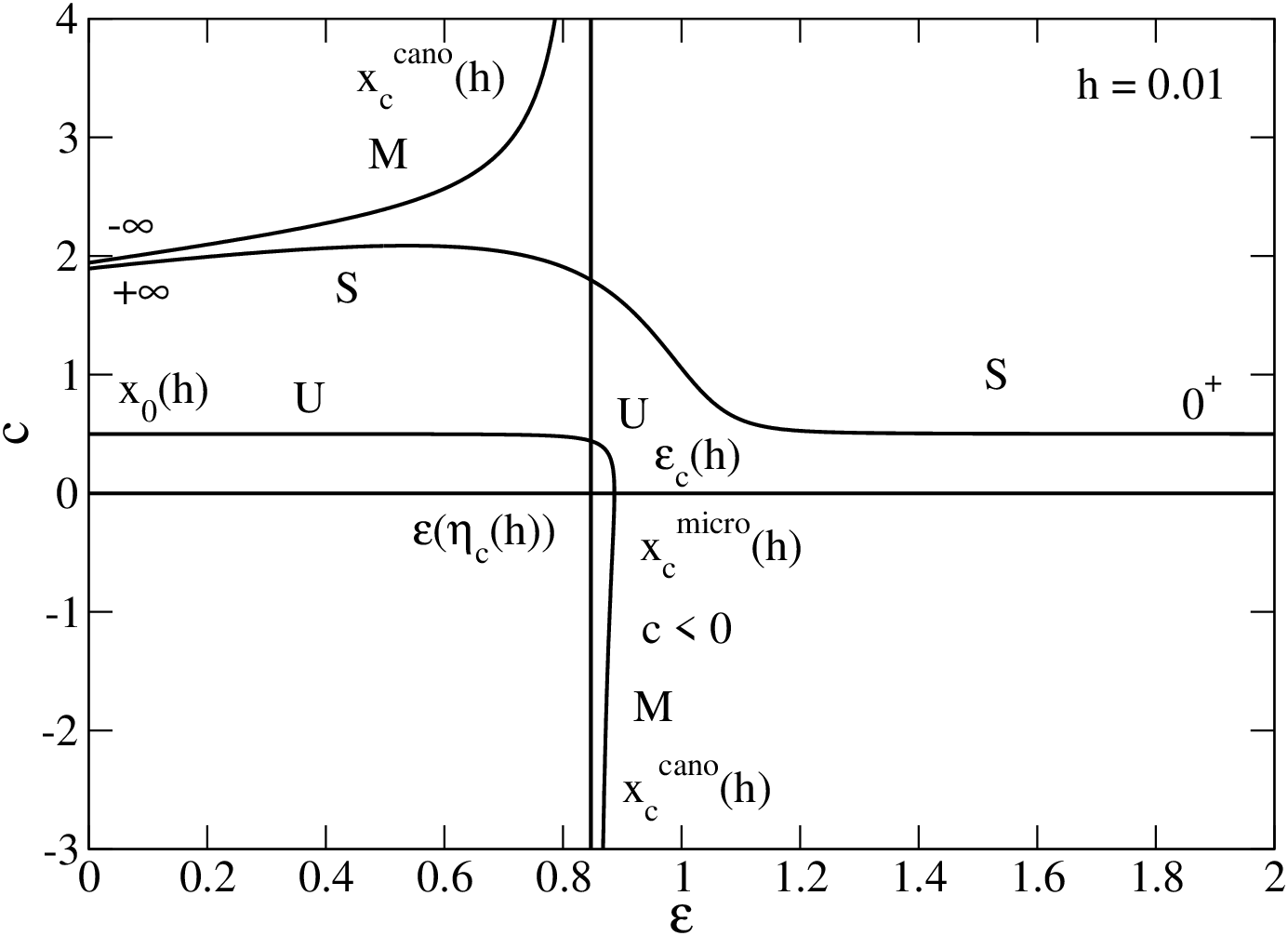}
\caption{Specific heat $c$ as a function of energy $\epsilon$ for $h=0.01$. We find that $c\rightarrow 1/2$ for $\epsilon\rightarrow +\infty$ (corresponding to $x\rightarrow 0$), $c\rightarrow 1$ for $\epsilon\rightarrow \epsilon_{min}(h)$ and $\epsilon\rightarrow \epsilon'_{min}(h)$ (corresponding to $x\rightarrow \pm\infty$) and $c\rightarrow 1/2$ for $\epsilon\rightarrow \epsilon_0(h)$ (corresponding to $x\rightarrow x_0(h)$). In the anti-aligned phase ($-\infty<x\le x_0(h)$), the specific heat diverges like $c\propto - (\epsilon-\epsilon(\eta_c(h)))^{-1}\rightarrow \pm\infty$ at the energy $\epsilon(\eta_c(h))$ associated with the turning point of temperature $\eta_c(h)$ (corresponding to $x=x_c^{cano}(h)$). Close to $\epsilon(\eta_c(h))$, there is a region of negative specific heat (corresponding to $x_c^{cano}(h)<x<x_c^{micro}(h)$). The specific heat becomes zero at the turning point of energy $\epsilon_c(h)$ (corresponding to $x=x_c^{micro}(h)$). The study of Section \ref{sec_poincare} shows that the condition of instability in the microcanonical ensemble corresponds to $x_c^{micro}(h)<x\le x_0(h)$. Therefore, the states with negative specific heats are stable in the microcanonical ensemble.}
\label{chaleurEnergie-h0.01}
\end{center}
\end{figure}

\begin{figure}
\begin{center}
\includegraphics[clip,scale=0.3]{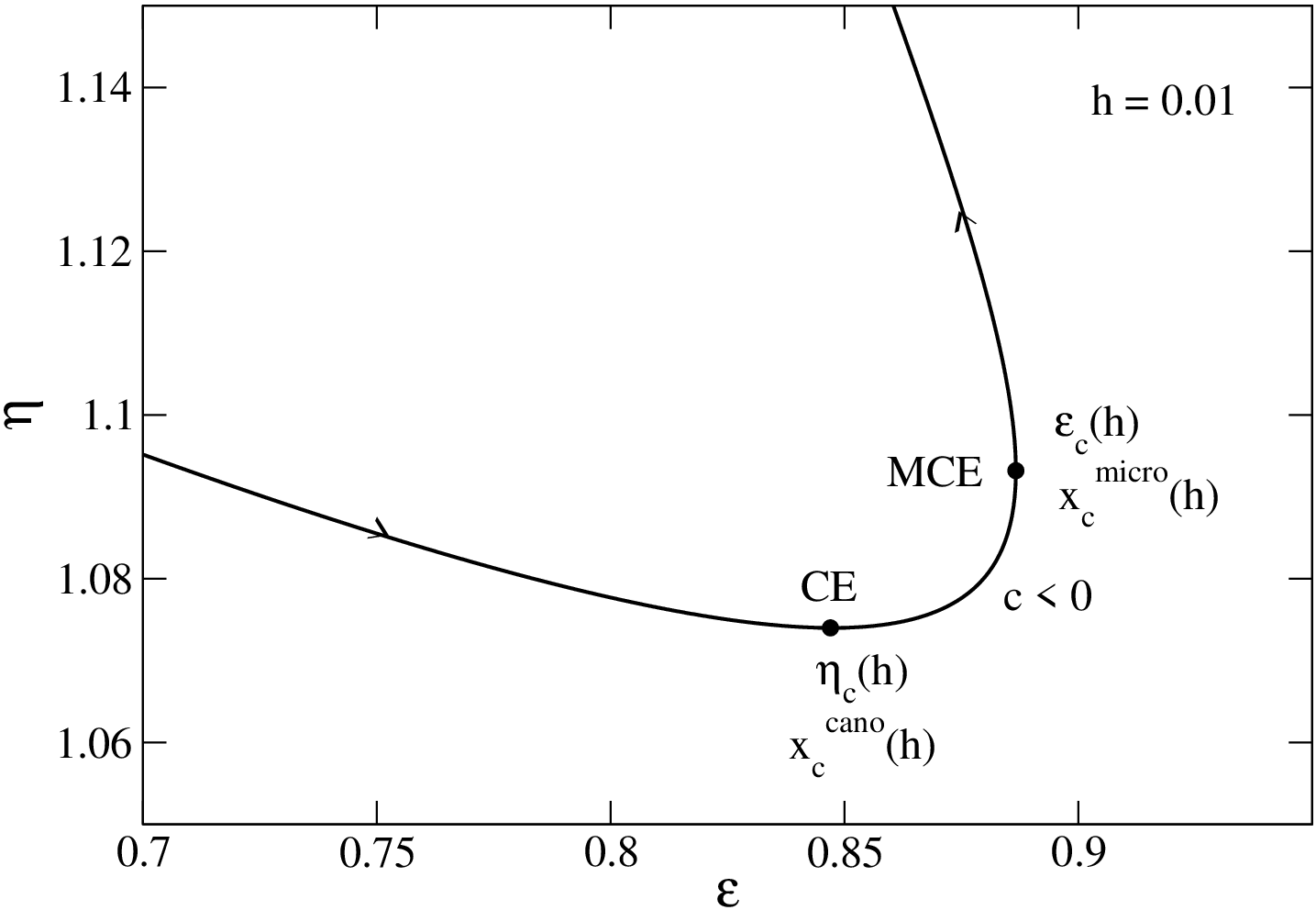}
\caption{Enlargement of the series of equilibria $\eta(\epsilon)$ in the region of negative specific heats for $h=0.01$. In the canonical ensemble, the system becomes unstable after the turning point of temperature ($c=\infty$)  and in the microcanonical ensemble, the system becomes unstable after the turning point of energy ($c=0$). In particular, in the region of negative specific heats (i.e. between CE and MCE) the system is stable in the microcanonical ensemble but unstable in the canonical ensemble. This can be viewed as a situation of ensembles inequivalence. Note, however, that it concerns metastable states (see Section \ref{sec_poincare}).}
\label{eec}
\end{center}
\end{figure}

The specific heat is defined by
$C=\partial E/\partial T$ in both ensembles. Using dimensionless variables, the specific heat per  particle $c=C/N$ can be written
\begin{eqnarray}
\label{sh1}
c=\frac{1}{2}\frac{d\epsilon}{d(1/\eta)}.
\end{eqnarray}
From equations (\ref{t7}) and (\ref{t8}) we easily obtain
\begin{eqnarray}
\label{sh3}
\frac{d(1/\eta)}{dx}=-\frac{1}{x\eta(x)}+\frac{2}{x}b'(x),
\end{eqnarray}
and
\begin{eqnarray}
\label{sh2}
\frac{d\epsilon}{dx}=-\frac{1}{x\eta(x)}+2b'(x)\left \lbrack\frac{1}{x}-\frac{x}{\eta(x)}\right \rbrack,
\end{eqnarray}
where $b'(x)$ is given by
\begin{eqnarray}
\label{msc5}
b'(x)=1-\frac{b(x)}{x}-b(x)^2.
\end{eqnarray}
To obtain this expression, we have used the identities $I'_0(x)=I_1(x)$ and
\begin{eqnarray}
\label{msc5b}
I'_n(x)=I_{n-1}(x)-\frac{n}{x}I_n(x).
\end{eqnarray}
Therefore, the specific heat can be written
\begin{eqnarray}
\label{sh4}
c=\frac{1-2b'(x)\left \lbrack \eta(x)-x^2\right \rbrack}{2\left\lbrack 1-2b'(x)\eta(x)\right\rbrack}.
\end{eqnarray}
The specific heat is infinite at the turning point of temperature
$\eta_c(h)$ corresponding to $x_c^{cano}(h)$ and is zero at the
turning point of energy $\epsilon_c(h)$ corresponding to
$x_c^{micro}(h)$.  It is represented as a function of the inverse
temperature $\eta$ (appropriate to the canonical ensemble) in Figure
\ref{chaleur-h0.01} and as a function of the energy $\epsilon$
(appropriate to the microcanonical ensemble) in Figure
\ref{chaleurEnergie-h0.01}. The series of equilibria with positive
magnetization has positive specific heat for any $0\le
x<+\infty$. Alternatively, there exists a region of negative specific
heat in the series of equilibria with negative magnetization between
$x_c^{cano}(h)$ and $x_c^{micro}(h)$ (while the specific heat is
positive in the rest of the curve).  This region of negative specific
heat can be seen directly on the series of equilibria $\eta(\epsilon)$
of Figure \ref{epsilon-etaNEW} (see a zoom in Figure \ref{eec}). In
this region, the temperature decreases when the energy increases! This
is the first occurrence of negative specific heats for the
Boltzmannian statistical equilibrium state of the HMF
model\footnote{Out-of-equilibrium distributions with negative specific
heats have been found in \cite{prlneg,cc}.}. Indeed, in the absence of
magnetic field $h=0$, the specific heat is always positive but
undergoes a discontinuity $\Delta c=2$ at $\eta=\eta_c$ or
$\epsilon=\epsilon_c$ \cite{cvb}.

It is a general result of statistical mechanics that stable
states in the canonical ensemble have positive specific heat since the specific heat $C=\beta^2\langle (\Delta E)^2\rangle\ge 0$ measures the variance of the fluctuations of energy.
Therefore, we can already conclude that the states between points CE and
MCE in the series of equilibria (see Figure \ref{eec}) are thermodynamically unstable in the canonical ensemble. However, negative specific heat is not a necessary condition of instability in the canonical ensemble. On the other hand, stable states  in the microcanonical ensemble can have negative specific heats \cite{cdr}. Therefore, the study of the specific heat is not sufficient to settle the stability/instability of the system.

\subsection{Poincar\'e theorem}
\label{sec_poincare}

\begin{figure}
\begin{center}
\includegraphics[clip,scale=0.3]{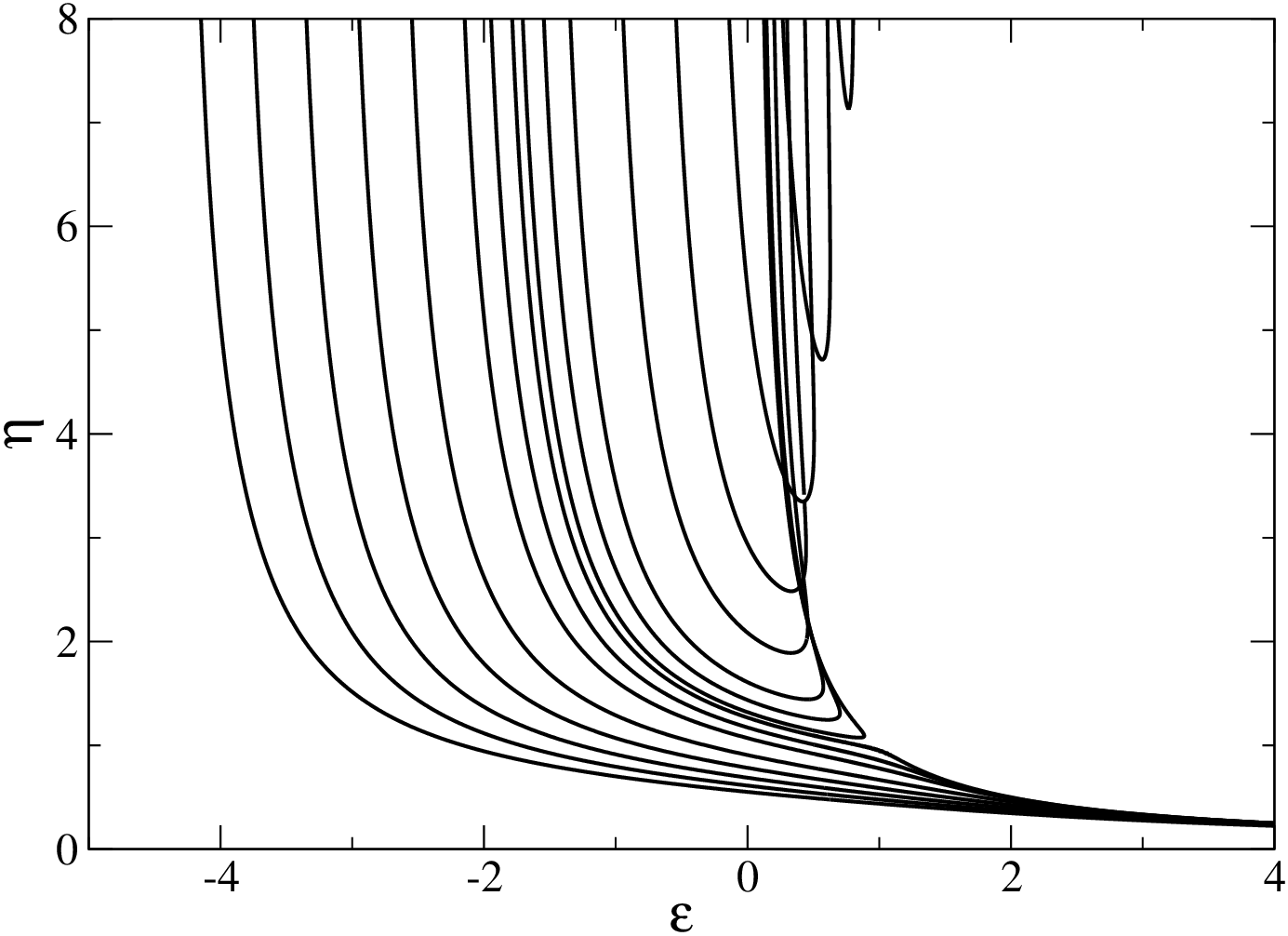}
\caption{Series of equilibria (caloric curve) giving the inverse temperature $\eta$ as a function of the energy $\epsilon$  for different values of $h= 0.01, 0.05, 0.1, 0.2, 0.3, 0.4, 0.5, 0.6$.}
\label{epsilon-eta}
\end{center}
\end{figure}

\begin{figure}
\begin{center}
\includegraphics[clip,scale=0.3]{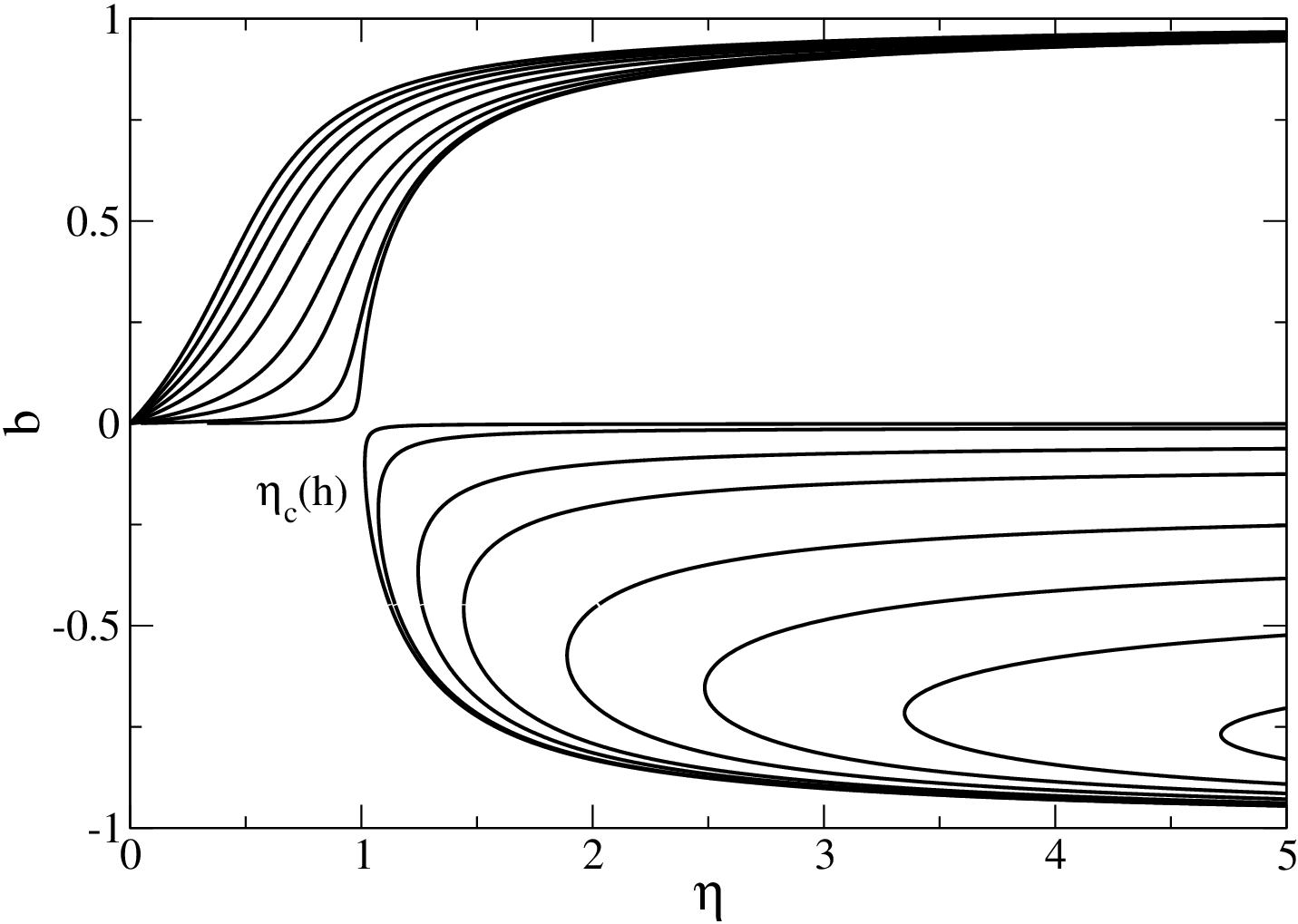}
\caption{Magnetization $b$ (order parameter) as a function of the inverse temperature $\eta$ for $h=0.001, 0.01, 0.05, 0.1, 0.2, 0.3, 0.4, 0.5, 0.6$.}
\label{eta-b}
\end{center}
\end{figure}

\begin{figure}
\begin{center}
\includegraphics[clip,scale=0.3]{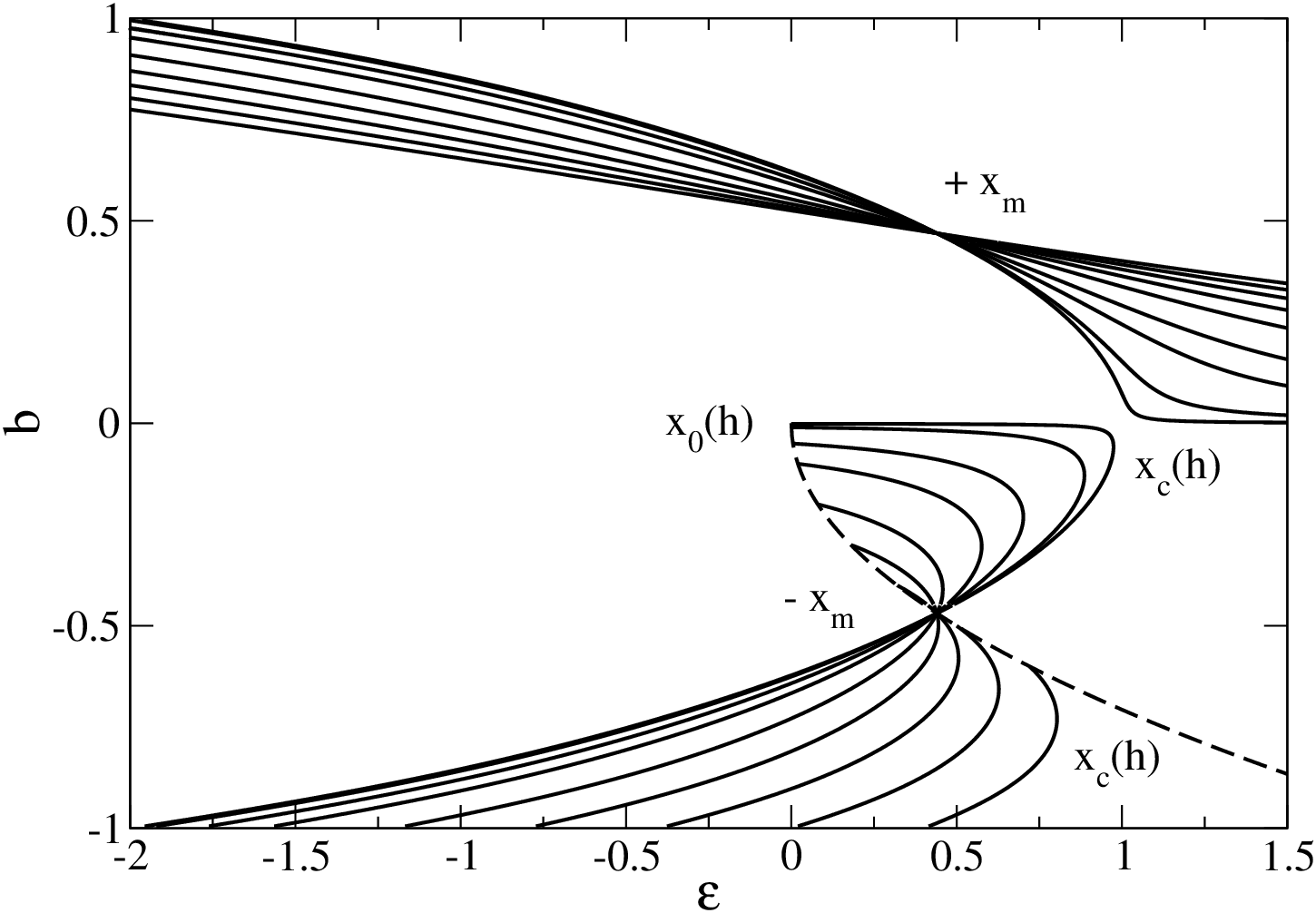}
\caption{Magnetization $b$ (order parameter) as a function of the energy $\epsilon$ for $h=0.001, 0.01, 0.05, 0.1, 0.2, 0.3, 0.4, 0.5, 0.6$. The dashed line corresponds to $b=-(\epsilon/2)^{1/2}$. The ``magic'' points $\pm x_m$ where all the curves cross each other are defined in Section \ref{sec_ms}.}
\label{epsilon-b}
\end{center}
\end{figure}

The stability of the critical points in each ensemble can be easily obtained
by applying the Poincar\'e theory of linear series of equilibria. This method has been extensively used in astrophysical problems (see, e.g.,
\cite{katzrev,ijmpb}). Linear series of equilibria are plotted in Figs. \ref{epsilon-etaNEW}-\ref{fZOOM} for $h=0.01$ and in Figs. \ref{epsilon-eta}-\ref{epsilon-b} for different values of $h$.

Let us first consider the canonical ensemble. We have to determine the minima of free energy at fixed mass. A global minimum will be called fully stable (S), a local minimum will be called metastable (M) and a maximum or a saddle point will be called unstable (U).  To apply the Poincar\'e
theorem, we just have to plot $\epsilon$ as a function of $\eta$ (see Figure \ref{epsilon-etaNEW} rotated by $90^o$). The
series of equilibria with positive magnetization exists for any inverse temperature
$\eta>0$. We know that the system is stable at high temperatures since it becomes equivalent to a classical gas without interaction (see Section \ref{sec_se}). Since
the series of equilibria with positive magnetization does not present
turning point of temperature, nor bifurcation, we conclude that the whole branch  is stable. The series of equilibria with negative magnetization
exists for $\eta>\eta_c(h)$. We know that the states with  inverse temperature $\eta\rightarrow +\infty$ and energy $\epsilon'_{min}(h)$  are stable since they are local energy minima at fixed mass (see Section \ref{sec_se}). On the other hand, the series of equilibria with negative magnetization
presents a unique turning point of temperature at
$\eta_c(h)$. Therefore, starting from $(\epsilon'_{min}(h),+\infty)$,
the series of equilibria with negative
magnetization is stable before the turning point of temperature
CE  and it becomes, and remains, unstable afterwards.
By comparing the free energies of the stable solutions in competition (see Figure \ref{f}), we see that the states with
positive magnetization always have a lower free energy than the states
with negative magnetization. Therefore, the states with positive
magnetization (i.e. $0\le x<+\infty$) are fully stable (S), the states with large negative
magnetization (i.e. $-\infty<x< x_c^{cano}(h)$) are metastable (M) and the states with small negative
magnetization (i.e. $x_c^{cano}(h)<x<x_0(h)$) are unstable (U). In conclusion: (i) for
$\eta<\eta_c(h)$ there exists a unique equilibrium state that is fully
stable (S); (ii) for $\eta>\eta_c(h)$ there exists one fully stable state
(S), one metastable state (M) and one unstable state (U).

Let us now consider the microcanonical ensemble. We have to determine the maxima  of entropy at fixed mass and energy. A global maximum will be called fully stable (S), a local maximum will be called metastable (M) and a minimum or saddle point will be called unstable (U). To apply the
Poincar\'e theorem, we have to plot $\eta$ as a function of
$\epsilon$ (see Figure \ref{epsilon-etaNEW}). The series of equilibria with positive magnetization
exists for any $\epsilon\ge \epsilon_{min}(h)$. We know that the system is
stable at high energies since it becomes equivalent to a classical gas without interaction (see Section \ref{sec_se}). Since the series of equilibria with
positive magnetization does not present turning point of energy, nor bifurcation, it
follows that the whole branch is stable. The series of equilibria with
negative magnetization exists for $\epsilon_{min}'(h)\le \epsilon\le
\epsilon_{c}(h)$. We know that the states with energy
$\epsilon'_{min}(h)$ and inverse temperature $\eta\rightarrow +\infty$
are stable since they correspond to local energy minima (see Section \ref{sec_se}). On the other hand, the series of equilibria with negative magnetization
presents a unique turning point of energy at
$\epsilon_c(h)$. Therefore, starting from $(\epsilon'_{min}(h),+\infty)$, the series of equilibria with negative
magnetization is stable before the turning point of energy MCE
and it becomes, and remains, unstable afterwards. By comparing the
entropies of the stable solutions in competition (see Figure \ref{entropie}), we see that the states with
positive magnetization always have a higher entropy than the states
with negative magnetization. Therefore, the states with positive
magnetization (i.e. $0\le x<+\infty$) are full stable (S), the states with large negative
magnetization (i.e. $-\infty<x< x_c^{micro}(h)$) are metastable (M) and the states with small negative
magnetization (i.e. $x_c^{micro}(h)<x<x_0(h)$) are unstable (U). In conclusion: (i) for
$\epsilon>\epsilon_c(h)$, there exists a unique equilibrium state that is fully
stable (S); (ii) for $\epsilon_0(h)<\epsilon<\epsilon_c(h)$, there exists one fully stable state
(S), one metastable state (M) and one unstable state (U); (iii) for $\epsilon_{min}'(h)<\epsilon<\epsilon_0(h)$, there exists one fully stable state
(S) and one metastable state (M); (iv) for $\epsilon_{min}(h)<\epsilon<\epsilon'_{min}(h)$, there exists only one fully stable state (S).

If we only consider fully stable states (S), we conclude that the
ensembles are equivalent and that the specific heat is always
positive. Furthermore, there is no phase transition when $h\neq 0$, contrary to the case $h=0$ which displays a second order phase transition at $(\epsilon_c,\eta_c)=(1,1)$. However, if we take into account metastable
states\footnote{Metastable
states are very important in systems with long-range interactions because they have tremendously
long lifetimes, scaling like $e^N$. Therefore, they can be considered as  stable states in
practice \cite{art,metastable}.}  we find a small region of ensembles inequivalence. Indeed,
we note that the states situated in the region between points CE and
MCE have negative specific heats (see Figure \ref{eec}).  These states are unstable in the
canonical ensemble while they are stable in the microcanonical
ensemble.  We recall that negative specific heat is a sufficient but
not necessary condition of canonical instability. In particular, the
states past point MCE are unstable (in both ensembles) while they have
positive specific heats. We note that the series of equilibria
becomes unstable in the canonical ensemble when the specific heat
passes from positive to negative values (the point CE has infinite
specific heat) while the series of equilibria becomes unstable in the
microcanonical ensemble when the specific heat passes from negative to
positive values (the point MCE has zero specific heat).

\begin{figure}
\begin{center}
\includegraphics[clip,scale=0.3]{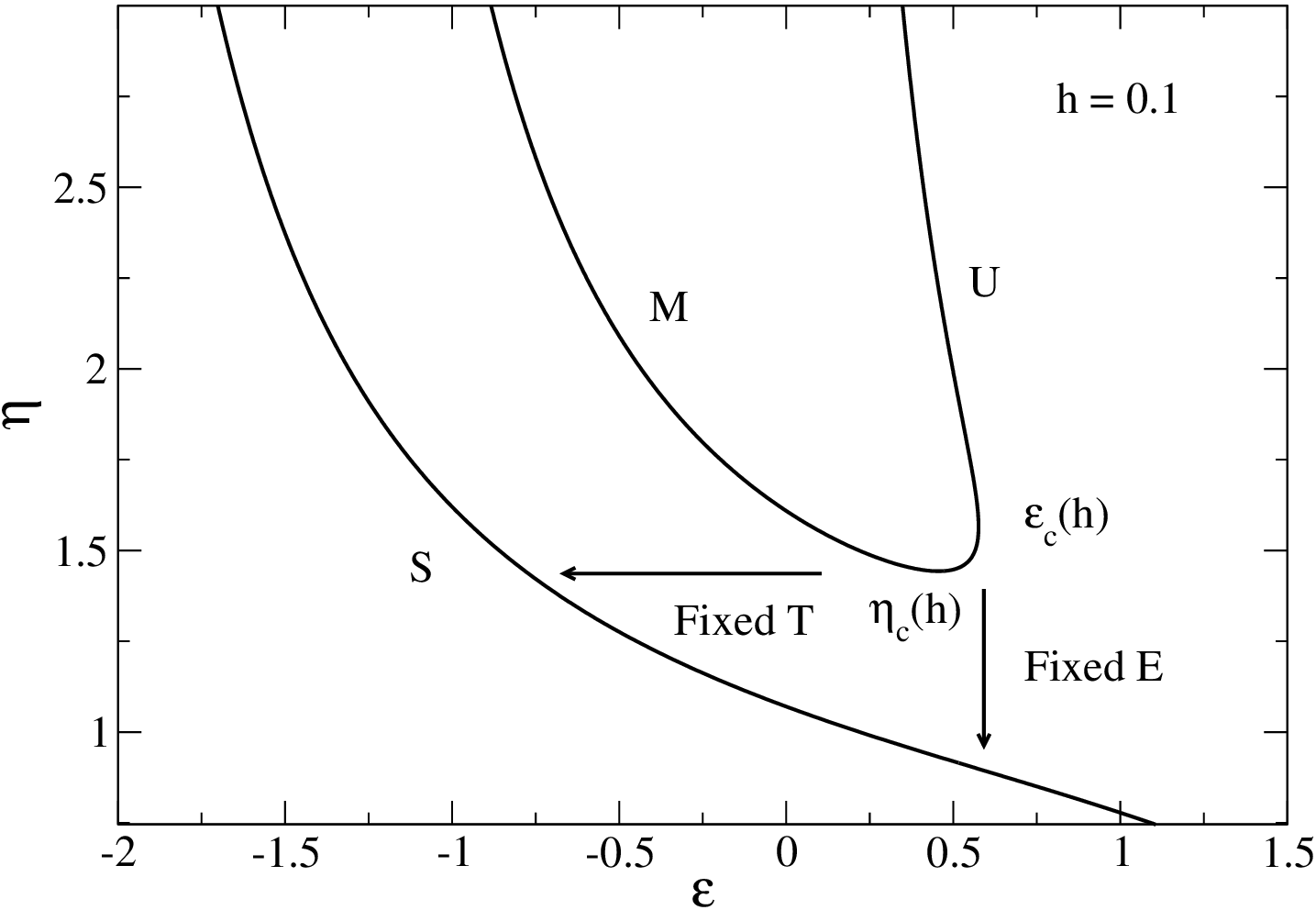}
\caption{Zeroth order phase transitions corresponding to the reorganization of the system from the anti-aligned phase to the aligned phase. This is similar to the gravothermal catastrophe and isothermal collapse of self-gravitating systems \cite{ijmpb}.}
\label{reorganization}
\end{center}
\end{figure}

These results are very similar to those obtained in the thermodynamics
of self-gravitating systems (see \cite{ijmpb} for a review). For
self-gravitating systems, there also exists a region of negative
specific heat in the microcanonical ensemble in which the temperature
decreases when the energy increases. On the other hand, when the
system reaches the turning point of temperature or energy (called
spinodal points), the metastable branch disappears and the system
undergoes an ``isothermal collapse'' (in the canonical ensemble) or a
``gravothermal catastrophe'' (in the microcanonical ensemble) until an
equilibrium state is reached (for systems with a small-scale
cut-off). Similar phenomena occur in the present problem (see Figure
\ref{reorganization}). In the canonical ensemble, when we reach the
spinodal point $\eta_c(h)$, the metastable branch made of states with
negative magnetization disappears and the system undergoes a sort of
instability similar to the isothermal collapse (fixed $T$). Similarly,
in the microcanonical ensemble, when we reach the spinodal point
$\epsilon_c(h)$, the metastable branch made of states with negative
magnetization disappears and the system undergoes a sort of
instability similar to the gravothermal catastrophe (fixed $E$). The
system reorganizes itself and finally reaches an equilibrium state
with positive magnetization. This corresponds to a zeroth order phase
transition marked by the discontinuity of entropy and free energy (see
Figs. \ref{entropie}-\ref{fZOOM}). There is no first order phase
transition in the HMF model with a magnetic field contrary to the case
of self-gravitating systems \cite{ijmpb}. Furthermore, in the
gravitational case, the system undergoes a transition from a
homogeneous phase to a clustered phase while for the HMF model with a
magnetic field, it undergoes a reorganization from an anti-aligned
phase (magnetization pointing in a direction opposed to the field) to
an aligned phase (magnetization pointing in the same direction as the
field).

{\it Important remark:} a more detailed stability analysis (see Appendix \ref{sec_lm}) reveals that the metastable states are, in fact, unstable with respect to perturbations that change the phase of the magnetization (i.e. $\delta B_y\neq 0$). Therefore, if we allow for these perturbations, the anti-aligned phase becomes unstable and the richness of the problem disappears. In the following, we shall consider only perturbations for which $\delta B_y=0$. However, one should keep in mind this remark in the interpretation of the results.

\subsection{Phase diagrams}
\label{sec_cmpd}

\begin{figure}
\begin{center}
\includegraphics[clip,scale=0.3]{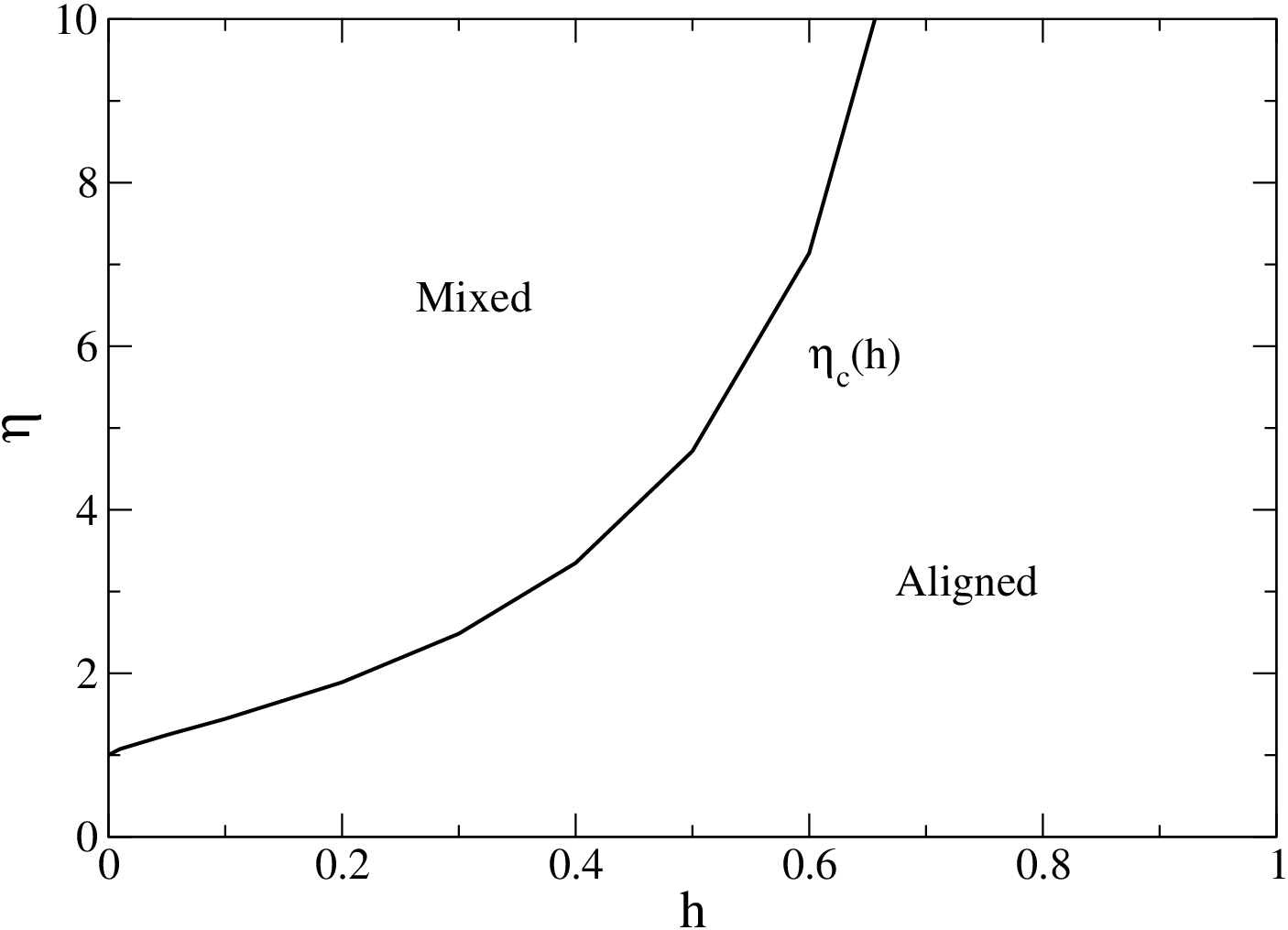}
\caption{Canonical phase diagram in the $(h,\eta)$ plane where $h$ is the external parameter (magnetic field) and $\eta$ is the control parameter (inverse temperature).}
\label{phasediagcano}
\end{center}
\end{figure}

\begin{figure}
\begin{center}
\includegraphics[clip,scale=0.3]{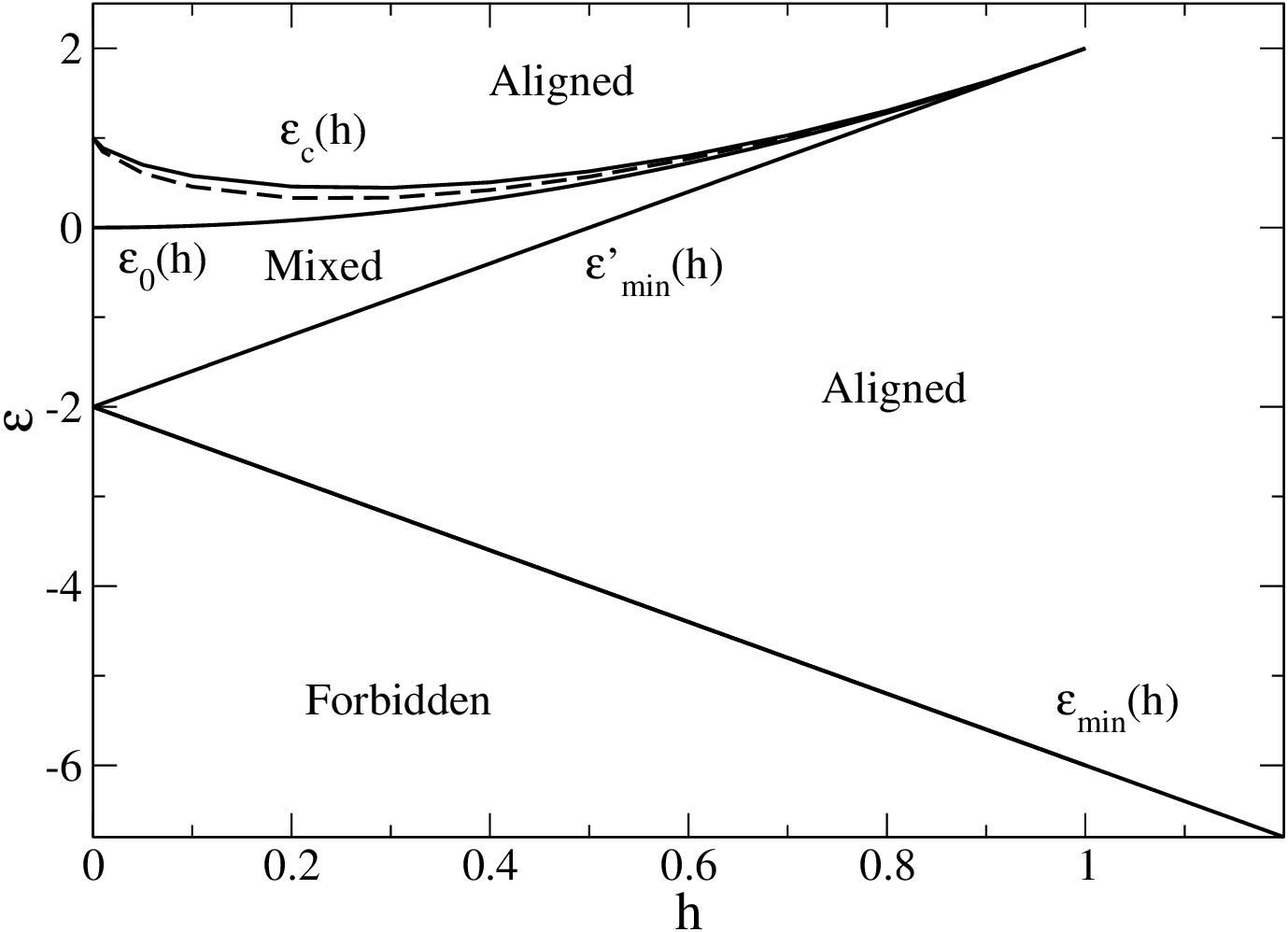}
\caption{Microcanonical phase diagram in the $(h,\epsilon)$ plane where $h$ is the external parameter (magnetic field) and $\epsilon$ is the control parameter (energy).}
\label{phasediagmicro}
\end{center}
\end{figure}

The preceding results can be summarized by drawing appropriate canonical and microcanonical phase diagrams (see Figs. \ref{phasediagcano} and \ref{phasediagmicro}).

Let us first describe the canonical phase diagram in the $(h,\eta)$
plane represented in Figure \ref{phasediagcano}. The solid line
corresponds to the critical inverse temperature $\eta_c(h)$ that
exists only for $h<1$ and tends to $+\infty$ when $h\rightarrow
1^-$. This curve divides the parameter space in two regions. In the
region denoted ``aligned'', the magnetization has the same sign as the
magnetic field. This is the case for any temperature when $h>1$ or for
inverse temperatures $\eta<\eta_c(h)$ when $h<1$. The region denoted
``mixed'' corresponds to a mixed zone in which the magnetization can
be aligned or anti-aligned with the magnetic field. In the
$N\rightarrow +\infty$ limit, the system can be observed in only one
of these two phases depending on the way it has been initially
prepared. For finite $N$, the system undergoes random transitions from
the aligned phase (fully stable) to the anti-aligned phase
(metastable) as described in Section \ref{sec_cs}. As explained in
Section \ref{sec_poincare}, all the stable states in the canonical
ensemble have positive specific heat $c>0$.

The microcanonical phase diagram in the $(h,\epsilon)$ plane is
represented in Figure \ref{phasediagmicro}. The upper solid line
corresponds to the critical energy $\epsilon_c(h)$ that exists only
for $h<1$ and tends to $2$ when $h\rightarrow 1^-$. We have also
represented the characteristic energies $\epsilon_0(h)$,
$\epsilon'_{min}(h)$ and $\epsilon_{min}(h)$.  These curves divide the
parameter space in two regions (of course, the region below the
minimum energy $\epsilon_{min}(h)$ is forbidden). In the region
denoted ``aligned'', the magnetization has the same sign as the
magnetic field. This is the case for any energy $\epsilon\ge
\epsilon_{min}(h)$ when $h>1$ and for energies
$\epsilon>\epsilon_c(h)$ or $\epsilon_{min}(h)\le
\epsilon<\epsilon'_{min}(h)$ when $h<1$. The region denoted ``mixed'',
delimited by the curves $\epsilon'_{min}(h)$ and $\epsilon_c(h)$,
corresponds to a mixed zone in which the magnetization can be aligned
or anti-aligned with the magnetic field. In the $N\rightarrow +\infty$
limit, the system can be observed in only one of these two phases
depending on the way it has been initially prepared. The subpart of
the mixed region delimited by the curves $\epsilon'_{min}(h)$ and
$\epsilon_0(h)$ exhibits phase space gaps responsible for ergodicity
breaking as described in Section \ref{sec_mst}. Therefore, for finite
$N$, the system undergoes random transitions between the aligned phase
(fully stable) and the anti-aligned phase (metastable) when
$\epsilon_0(h)\le \epsilon\le \epsilon_c(h)$ while it remains blocked
in one of these two phases when $\epsilon'_{min}(h)\le \epsilon\le
\epsilon_0(h)$ even for small $N$. On the other hand, the dashed line
corresponds to the energy $\epsilon(\eta_c(h))$ associated with the
turning point of temperature. As explained in Section \ref{sec_sh},
the region between the curves $\epsilon(\eta_c(h))$ and
$\epsilon_c(h)$ corresponds to states with negative specific heats
$c<0$ which are stable in the microcanonical ensemble but unstable in
the canonical ensemble. This is therefore a region of ensembles
inequivalence.

\section{Stability analysis in the canonical ensemble}
\label{sec_cano}

In this section, we analytically study the thermodynamical stability of the system in the canonical ensemble, using a procedure similar to the one developed in \cite{isostab} for the case $h=0$.

\subsection{The free energy $F(B)$}
\label{sec_f}

The minimization problem (\ref{m14}) determines the statistical equilibrium state of the HMF model in the canonical ensemble.
It is shown in Appendix A.2. of \cite{yukawa} that the solution of (\ref{m14}) is given by
\begin{eqnarray}
\label{f1}
f(\theta,v)=\left (\frac{\beta}{2\pi}\right )^{1/2} \, \rho({\theta})\, e^{-\beta\frac{v^2}{2}},
\end{eqnarray}
where $\rho(\theta)$ is the solution of
\begin{eqnarray}
\label{f2}
\min_\rho\left\lbrace F\lbrack \rho\rbrack\, |\, M\lbrack \rho\rbrack=M\right\rbrace,
\end{eqnarray}
where
\begin{eqnarray}
\label{f3}
F[\rho]=\frac{1}{2}\int\rho\Phi\, d\theta+\int\rho\Phi_{ext}\, d\theta+{T}\int \rho\ln\rho\, d\theta,
\end{eqnarray}
is the configurational free energy. Therefore, the minimization
problems (\ref{m14}) and (\ref{f2}) are equivalent:
\begin{eqnarray}
\label{f4}
(\ref{m14}) \Leftrightarrow (\ref{f2}).
\end{eqnarray}
This equivalence holds for global and local minimization \cite{yukawa}: (i) $f(\theta,v)$ is the global minimum of (\ref{m14}) iff $\rho(\theta)$ is the global minimum of (\ref{f2}) and (ii) $f(\theta,v)$ is a local minimum of (\ref{m14}) iff $\rho(\theta)$ is a local minimum of (\ref{f2}). We are therefore led to considering the minimization problem (\ref{f2}) which is simpler to study since it involves the density $\rho(\theta)$ instead of the distribution function $f(\theta,v)$.

For the HMF model, the potential energy is given by equation (\ref{m10}) so that the free energy (\ref{f3}) can be rewritten
\begin{eqnarray}
\label{f5}
F[\rho]=-\frac{\pi B^2}{k}-\frac{2\pi}{k}B_x H+{T}\int \rho\ln\rho\, d\theta.
\end{eqnarray}
Let us determine the {\it global} minimum of free energy at fixed mass. To that purpose, we shall reduce the minimization problem (\ref{f2})  to an equivalent but simpler minimization problem. To solve the minimization problem (\ref{f2}), we proceed in two steps: we first minimize $F[\rho]$ at fixed $M$ {\it and} $B_x$ and $B_y$. Writing the variational principle as
\begin{eqnarray}
\label{f6}
\delta \left (\int \rho\ln\rho\, d\theta\right )+\alpha\delta M+\mu_x\delta B_x+\mu_y\delta B_y=0,
\end{eqnarray}
we obtain
\begin{eqnarray}
\label{f7}
\rho_1(\theta)=Ae^{\lambda\cos\theta}.
\end{eqnarray}
We have anticipated the fact that the magnetization of the global minimum of free energy is parallel to the magnetic field so that $B_y=0$, implying $\mu_y=0$. The Lagrange multipliers $A=e^{-1-\alpha}$ and  $\lambda=-\frac{k}{2\pi}\mu_x$  are determined by the constraints $M$ and $B_x$ (denoted $B$ in the following)  through the equations
\begin{equation}
A=\frac{M}{2\pi I_{0}(\lambda)},
\label{f8}
\end{equation}
and
\begin{equation}
b\equiv \frac{2\pi B}{kM}=\frac{I_{1}(\lambda)}{I_{0}(\lambda)}.
\label{f9}
\end{equation}
Equation (\ref{f7}) is the (unique) global minimum of $F[\rho]$ with the previous constraints since $\delta^2 F=\frac{1}{2}T\int \frac{(\delta \rho)^2}{\rho}\, d\theta> 0$ (the constraints are linear in $\rho$ so that their second variations vanish). Then, we can express the free energy $F[\rho]$ as a function of $B$ by writing $F(B)\equiv F[\rho_1]$. After straightforward calculations, we obtain
\begin{eqnarray}
F(B)=-\frac{\pi B^2}{k}-\frac{2\pi}{k}BH+T \lambda \frac{2\pi B}{k}-MT\ln I_0(\lambda),\nonumber\\
\label{f10}
\end{eqnarray}
where $\lambda(B)$ is given by equation (\ref{f9}).   Finally, the minimization problem (\ref{f2})  is equivalent to the minimization problem
\begin{eqnarray}
\label{f11}
\min_B\left\lbrace F(B)\right\rbrace,
\end{eqnarray}
in the sense that the solution of (\ref{f2}) is given by equations (\ref{f7}), (\ref{f8}) and (\ref{f9})  where $B$ is the solution of (\ref{f11}). Note that the mass constraint is taken into account implicitly in the variational problem (\ref{f11}). Therefore, (\ref{f2}) and (\ref{f11}) are equivalent for global minimization:
\begin{eqnarray}
\label{f12}
(\ref{f2}) \Leftrightarrow (\ref{f11}).
\end{eqnarray}
Furthermore, we show in Appendix \ref{sec_lmc} that they are also equivalent for local minimization provided that we impose the constraint $\delta B_y=0$ to the perturbations (otherwise the metastable states are always unstable). Under these conditions, the  equivalence (\ref{f12}) holds for global and local minimization: (i) $\rho(\theta)$ is the global minimum of (\ref{f2}) iff $B$ is the global minimum of (\ref{f11}) and (ii) $\rho(\theta)$ is a local minimum of (\ref{f2}) iff $B$ is a local minimum of (\ref{f11}). We are therefore led to considering the minimization problem (\ref{f11}) which is simpler to study since, for given $T$ and $M$, we just have to determine the minimum of a {\it function} $F(B)$ instead of the minimum of a functional $F[\rho]$  at fixed mass.

\subsection{The condition of canonical stability}
\label{sec_cs}

Let us therefore study the function $F(B)$ defined by equations
(\ref{f10}) and (\ref{f9}) for given $T$ and $M$. Introducing the dimensionless
variables of Section \ref{sec_c}, we have to study the function
\begin{eqnarray}
f(b)=-2b^2-4bh+\frac{2}{\eta}\lambda b-\frac{2}{\eta}\ln I_{0}(\lambda),
\label{cs1}
\end{eqnarray}
where $\lambda(b)$ is given by equation (\ref{f9}) and $\eta$ is prescribed. Its first
derivative is
\begin{eqnarray}
f'(b)=-4(b+h)+\frac{2}{\eta}\lambda+\frac{2}{\eta}\left (b-\frac{I_0'(\lambda)}{I_0(\lambda)}\right )\frac{d\lambda}{db}.
\label{cs2}
\end{eqnarray}
Using the identity $I'_0(\lambda)=I_1(\lambda)$ and the relation (\ref{f9}), we see that the term in parenthesis vanishes. Then, we get
\begin{eqnarray}
f'(b)=-4(b+h)+\frac{2}{\eta}\lambda.
\label{cs3}
\end{eqnarray}
The critical points of $f(b)$, satisfying $f'(b)=0$, correspond therefore to
\begin{eqnarray}
\lambda=x\equiv 2\eta (b+h).
\label{cs4}
\end{eqnarray}
Substituting this result in equation (\ref{f9}), we obtain the self-consistency relation
\begin{equation}
b=\frac{I_{1}(2\eta(b+h))}{I_{0}(2\eta (b+h))},
\label{cs5}
\end{equation}
which determines the magnetization $b$ as a function of the inverse temperature $\eta$. This returns the equilibrium results of Section \ref{sec_c}.

Now, a critical point of $f(b)$ is a minimum if $f''(b)> 0$ and a maximum if $f''(b)<0$. Differentiating  equation (\ref{cs3}) with respect to $b$, we find that
\begin{eqnarray}
f''(b)=\frac{2}{\eta}\frac{d\lambda}{db}-4.
\label{cs6}
\end{eqnarray}
Therefore, a critical point $\lambda=x$ is a minimum if
\begin{eqnarray}
b'(x)<\frac{1}{2\eta},
\label{cs7}
\end{eqnarray}
and a maximum if the inequality is reversed. According to equation (\ref{sh3}), the turning point of temperature
($d\eta/dx=0$) corresponds to
\begin{eqnarray}
b'(x)=\frac{1}{2\eta(x)}.
\label{cs8}
\end{eqnarray}
Comparing equation (\ref{cs8}) with equation (\ref{cs7}), we see that
the change of stability in the canonical ensemble corresponds to the
turning point of temperature $\eta_c(h)$ occuring at $x_c^{cano}(h)$
in agreement with the Poincar\'e theorem. Using equation (\ref{sh4}),
it is also easy to establish that when the specific heat is negative,
the converse of inequality (\ref{cs7}) is always fulfilled so that the
system is unstable in agreement with general theorems of statistical
mechanics.

\begin{figure}
\begin{center}
\includegraphics[clip,scale=0.3]{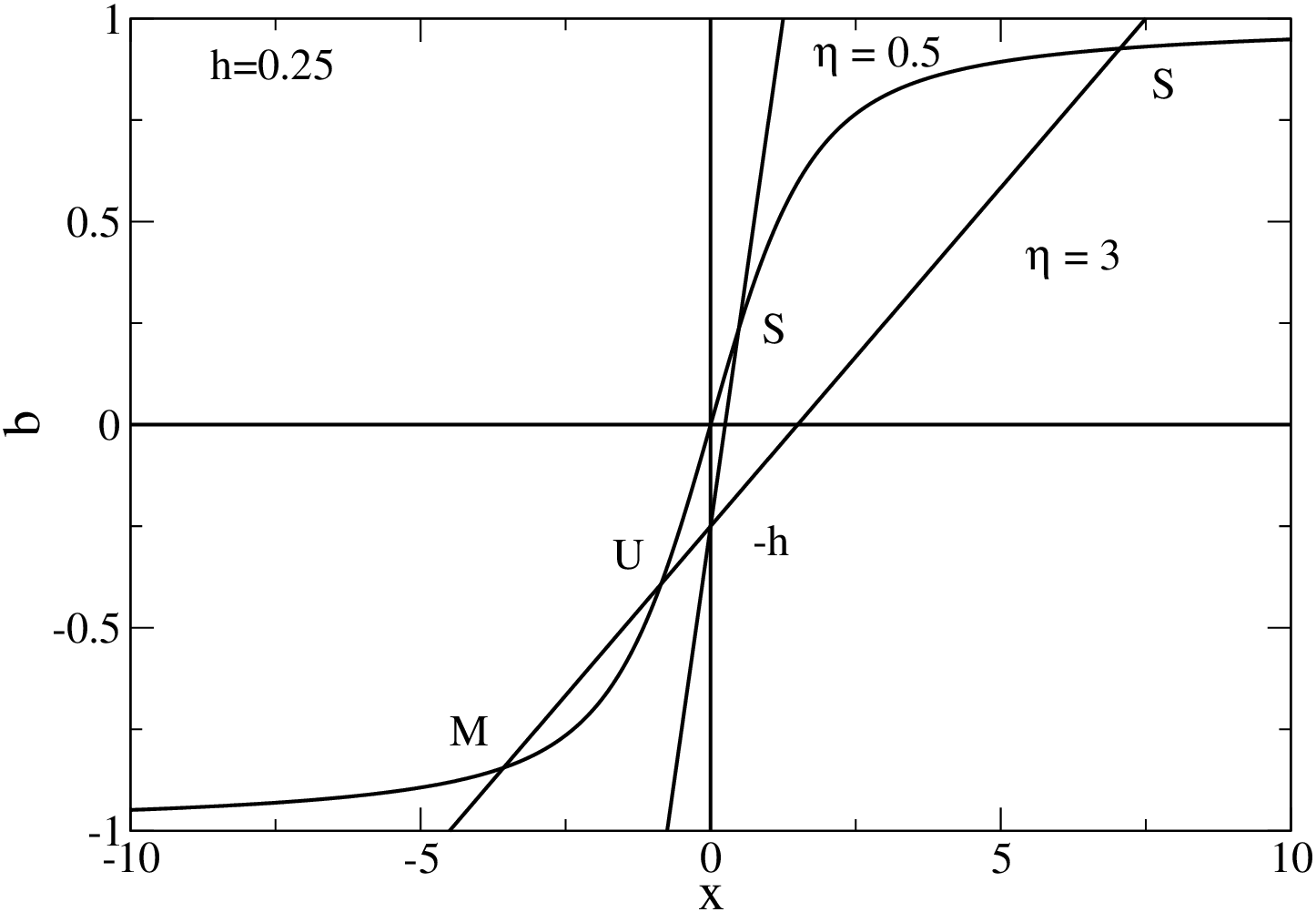}
\caption{Graphical construction determining the critical points of $f(b)$ and their stability. The critical points are determined by the intersection(s) between the curve $b=b(\lambda)$ defined by equation (\ref{f9}) and the straight line $b=\lambda/(2\eta)-h$. The critical point is a minimum (resp. maximum) of $f(b)$ if the slope of the curve $b(\lambda)$ at that point is smaller (resp. larger) than the slope of the straight line $b=\lambda/(2\eta)-h$.}
\label{blambda}
\end{center}
\end{figure}

We can determine the minima and maxima of the function $f(b)$ by a
simple graphical construction. To that purpose, we plot $b$ as a
function of $\lambda$ according to
$b(\lambda)={I_1(\lambda)}/{I_0(\lambda)}$. This function is
represented in Figure \ref{blambda}. We note that $b\rightarrow \pm 1$
for $\lambda\rightarrow
\pm\infty$ and that $b\sim \lambda/2$ for $\lambda\rightarrow 0$.  According
to equation (\ref{cs4}), the critical points of $f(b)$ are determined
by the intersection of this curve with the straight line $b=\lambda/(2\eta)-h$.
For given $\eta$, this determines $x(\eta)$ and $b(\eta)$. For
$\eta<\eta_c(h)$, there is a unique solution $b_{S}>0$ that has the same
sign as the imposed magnetic field. For $\eta\rightarrow 0$, we see that $x_S\rightarrow 0^+$ and $b_S\rightarrow 0^+$. For $\eta>\eta_c(h)$, there are three
solutions: a solution $b_{S}>0$ that has the same sign as the imposed
magnetic field and two solutions $b_{U}<0$ and $b_{M}<b_{U}$ whose
sign is opposite to the sign of the imposed magnetic field. For $\eta\rightarrow +\infty$, we see that $x_S\rightarrow +\infty$, $b_S\rightarrow 1$, $x_M\rightarrow -\infty$, $b_M\rightarrow -1$, $x_U\rightarrow x_0(h)$ and $b_U\rightarrow b_0(h)=-h$.  According
to inequality (\ref{cs7}), a solution is a minimum of $f(b)$ if
$b'(x)<1/(2\eta)$ and a maximum if $b'(x)>1/(2\eta)$. Therefore, {\it
a critical point of free energy $f(b)$ is a minimum (resp. maximum) if
the slope of the main curve is lower (resp. higher) than the slope of
the straight line at the point of intersection}. From this criterion,
we directly conclude that the solutions $b_S$ and $b_M$ are minima of
free energy while the solution $b_{U}$ is a maximum of free
energy.

\begin{figure}
\begin{center}
\includegraphics[clip,scale=0.3]{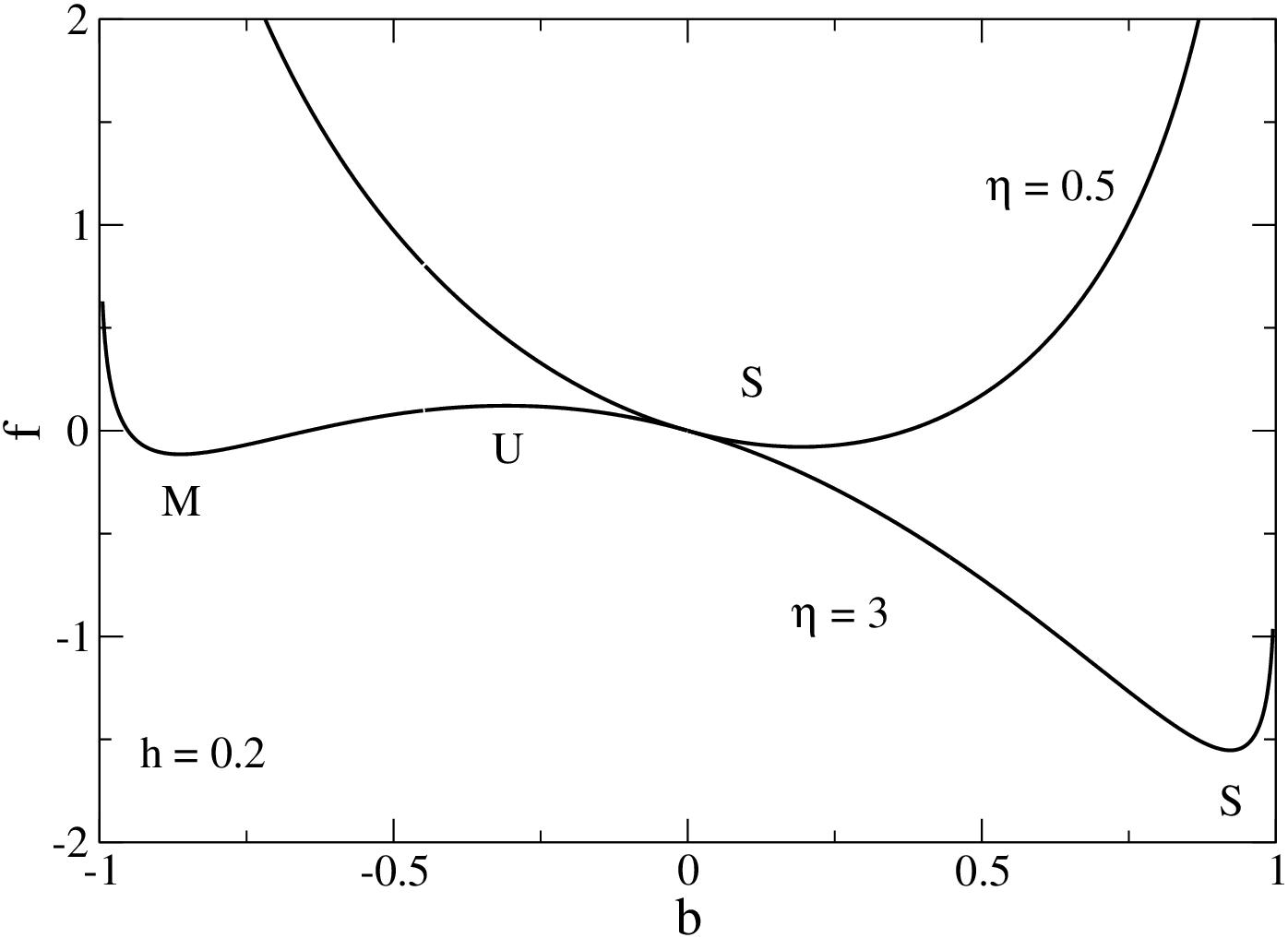}
\caption{Free energy $f(b)$ as a function of the magnetization $b$ for a given value of the inverse temperature $\eta$ and magnetic field $h$ (specifically $h=0.2$). For $\eta<\eta_c(h)\simeq 1.89$, this curve has a (unique) global  minimum at $b_S>0$. For $\eta>\eta_c(h)$, this curve has a  global  minimum at $b_S>0$, a local  minimum at $b_M<0$ and a local maximum at $b_U<0$.}
\label{b-f-h0.2}
\end{center}
\end{figure}

To complete our analysis, it can be useful to plot the function $f(b)$ for a prescribed inverse temperature $\eta$. It is given in parametric form by
\begin{eqnarray}
f(\lambda)=-2b(\lambda)^2-4b(\lambda)h+\frac{2}{\eta}\lambda b(\lambda)-\frac{2}{\eta}\ln I_{0}(\lambda),
\label{cs11}
\end{eqnarray}
together with equation (\ref{f9}).  Eliminating $\lambda$ between the expressions (\ref{cs11}) and (\ref{f9}), we obtain the free energy $f(b)$ as a function of the magnetization $b$ for a fixed value of the inverse temperature $\eta$. For $\eta<\eta_c(h)$ and $\eta>\eta_c(h)$, this function displays the two behaviors described above, as illustrated in Figure \ref{b-f-h0.2}.

It can be shown that the probability of a fluctuation with magnetization $b$ in the canonical ensemble is given by $P_{CE}(b)=\frac{1}{Z(\eta)} e^{-\frac{1}{2}\eta Nf(b)}$ (see Appendix \ref{sec_magnetization}). When there exists only one (global) minimum of free energy $b_S$, as in the case $h>1$ or in the case $h<1$ and $\eta<\eta_c(h)$, the situation is simple. For $N\rightarrow +\infty$, the distribution of magnetization is strongly peaked around the optimal value $b=b_S$ so that, after a transient regime, the system will be found in that state. For $h<1$ and $\eta>\eta_c(h)$, the free energy $f(b)$ presents a global minimum at $b_S$ and a local minimum at $b_M$. For  $N\rightarrow +\infty$, these states have infinite lifetime and the system will be found in one of them depending on how it has been initially prepared. For finite $N$, the system will jump from one state to the other. Of course, it will spend more time in the global minimum of free energy (fully stable) than in the local one (metastable). To pass from one state to the other, the system has to overcome an entropy barrier which is played by the unstable solution $b_U$. This barrier scales like $\frac{1}{2}\eta N|\Delta f|$. For finite $N$, this barrier is finite so that random transitions from the fully stable state $b_S$ to the metastable state $b_M$ are possible. For $N\rightarrow +\infty$, the barrier is too hard to cross and the system remains in one of these two states as previously indicated (according to the Kramers formula, the lifetime of these states scales like $t_{life}\sim e^{\frac{1}{2}\eta N|\Delta f|}\rightarrow +\infty)$ \cite{art,metastable}.

\section{Stability analysis in the microcanonical ensemble}
\label{sec_micro}

In this section, we analytically study the thermodynamical stability of the system in the microcanonical ensemble, using a procedure similar to the one developed in \cite{isostab} for the case $h=0$.

\subsection{The entropy $S(B)$}
\label{sec_sb}

The maximization problem (\ref{m13}) determines the statistical equilibrium state of the HMF model in the microcanonical ensemble. It is shown in Appendix A.1. of \cite{yukawa} that the solution of (\ref{m13}) is given by
\begin{eqnarray}
\label{sb1}
f(\theta,v)=\left (\frac{\beta}{2\pi}\right )^{1/2} \, \rho({\theta})\, e^{-\beta\frac{v^2}{2}},
\end{eqnarray}
where the inverse temperature $\beta=1/T$ is determined by the energy constraint
\begin{eqnarray}
\label{sb2}
E=\frac{1}{2}MT+W,
\end{eqnarray}
and $\rho(\theta)$ is the solution of
\begin{eqnarray}
\label{sb3}
\max_\rho\left\lbrace S\lbrack \rho\rbrack\, |\, M\lbrack \rho\rbrack=M\right\rbrace,
\end{eqnarray}
where
\begin{eqnarray}
\label{sb4}
S[\rho]=\frac{1}{2}M\ln T-\int \rho\ln\rho\, d\theta,
\end{eqnarray}
is the configurational entropy. Eliminating the temperature thanks to
the constraint (\ref{sb2}), we can write the entropy in terms of
$\rho$ alone as
\begin{eqnarray}
\label{sb5}
S[\rho]=-\int \rho\ln\rho\, d\theta+\frac{1}{2}M\ln (E-W[\rho]).
\end{eqnarray}
Therefore, the maximization problems (\ref{m13}) and (\ref{sb3}) are equivalent:
\begin{eqnarray}
\label{sb6}
(\ref{m13}) \Leftrightarrow (\ref{sb3}).
\end{eqnarray}
This equivalence holds for global and local maximization \cite{yukawa}: (i) $f(\theta,v)$ is the global maximum of (\ref{m13}) iff $\rho(\theta)$ is the global maximum of (\ref{sb3}) and (ii) $f(\theta,v)$ is a local maximum of (\ref{m13}) iff $\rho(\theta)$ is a local maximum of (\ref{sb3}). We are led therefore to considering the maximization problem (\ref{sb3}) which is simpler to study since it involves the density $\rho(\theta)$ instead of the distribution function $f(\theta,v)$.

For the HMF model, the potential energy is given by equation (\ref{m10}) so that  the energy (\ref{sb2}) and the entropy (\ref{sb5}) can be rewritten
\begin{eqnarray}
\label{sb7}
E=\frac{1}{2}MT-\frac{\pi B^2}{k}-\frac{2\pi}{k}B_x H,
\end{eqnarray}
\begin{eqnarray}
\label{sb8}
S[\rho]=-\int \rho\ln\rho\, d\theta+\frac{1}{2}M\ln \left (E+\frac{\pi B^2}{k}+\frac{2\pi}{k}B_x H\right ).\nonumber\\
\end{eqnarray}
Let us determine the {\it global} maximum of entropy at fixed mass (the conservation of energy is automatically taken into account in equation (\ref{sb8})). To that purpose, we shall reduce the maximization problem (\ref{sb3})  to an equivalent but simpler maximization problem. To solve the maximization problem  (\ref{sb3}), we proceed in two steps: we first maximize $S[\rho]$ at fixed $M$ {\it and} $B_x$ and $B_y$. Writing the variational problem as
\begin{eqnarray}
\label{sb9}
-\delta \left (\int \rho\ln\rho\, d\theta\right )-\alpha\delta M-\mu_x\delta B_x-\mu_y \delta B_y=0,
\end{eqnarray}
we obtain
\begin{eqnarray}
\label{sb10}
\rho_1(\theta)=Ae^{\lambda\cos\theta},
\end{eqnarray}
where $A$ and $\lambda$ are determined by the constraints $M$ and $B$ through the equations
\begin{equation}
A=\frac{M}{2\pi I_{0}(\lambda)},
\label{sb11}
\end{equation}
and
\begin{equation}
b\equiv \frac{2\pi B}{kM}=\frac{I_{1}(\lambda)}{I_{0}(\lambda)}.
\label{sb12}
\end{equation}
We have anticipated the fact that the magnetization of the global maximum  of entropy is parallel to the magnetic field so that $B_y=0$, implying $\mu_y=0$.  Equation (\ref{sb10}) is the (unique) global maximum of $S[\rho]$ with the previous constraints since $\delta^2 S=-\frac{1}{2}\int \frac{(\delta \rho)^2}{\rho}\, d\theta< 0$ (the constraints are linear in $\rho$ so that their second order variations vanish). Then, we can express the entropy $S$ as a function of $B$ by writing $S(B)\equiv S[\rho_1]$. After straightforward calculations, we obtain
\begin{eqnarray}
S(B)=M\ln I_0(\lambda)-\frac{2\pi B}{k}\lambda\nonumber\\
+\frac{M}{2}\ln \left (E+\frac{\pi B^2}{k}+\frac{2\pi}{k}BH\right ),
\label{sb13}
\end{eqnarray}
where $\lambda(B)$ is given by equation (\ref{sb12}). Finally, the maximization problem (\ref{sb3}) is equivalent to the maximization problem
\begin{eqnarray}
\label{sb14}
\max_B\left\lbrace S(B)\right\rbrace,
\end{eqnarray}
in the sense that the solution of (\ref{sb3}) is given by equations (\ref{sb10})-(\ref{sb12})  where $B$ is the solution of (\ref{sb14}). Note that the energy and mass constraints are taken into account implicitly in the variational problem (\ref{sb14}). Therefore, (\ref{sb3}) and (\ref{sb14}) are equivalent for global maximization:
\begin{eqnarray}
\label{sb15}
(\ref{sb3}) \Leftrightarrow (\ref{sb14}).
\end{eqnarray}
Furthermore, we show in Appendix \ref{sec_lmm} that they are also equivalent for local maximization provided that we impose the constraint $\delta B_y=0$ to the perturbations (otherwise the metastable states are always unstable). Under these conditions, the  equivalence (\ref{sb15}) holds for global and local maximization: (i) $\rho(\theta)$ is the global maximum of (\ref{sb3}) iff $B$ is the global maximum of (\ref{sb14}) and (ii) $\rho(\theta)$ is a local maximum of (\ref{sb3}) iff $B$ is a local maximum of (\ref{sb14}). We are therefore led to considering the maximization problem (\ref{sb14}) which is simpler to study since, for given $E$ and $M$, we just have to determine the maximum of a {\it function} $S(B)$ instead of the maximum of a functional $S[\rho]$  at fixed mass.

\subsection{The condition of microcanonical stability}
\label{sec_mst}

Let us therefore study the function $S(B)$ defined by equations (\ref{sb13}) and (\ref{sb12}). Introducing the  dimensionless variables of  Section \ref{sec_c}, we have to study the function
\begin{eqnarray}
s(b)=\ln I_0(\lambda)-b\lambda
+\frac{1}{2}\ln \left (\epsilon+2b^2+4bh\right ),
\label{ms1}
\end{eqnarray}
where $\lambda(b)$ is given by equation (\ref{sb12}).
Its first derivative is
\begin{eqnarray}
s'(b)=\left (\frac{I_0'(\lambda)}{I_0(\lambda)}-b\right )\frac{d\lambda}{db}-\lambda
+\frac{2(b+h)}{\epsilon+2b^2+4bh},
\label{ms2}
\end{eqnarray}
Using the identity $I'_0(x)=I_1(x)$  and the relation (\ref{sb12}), we see that the term in parenthesis vanishes. Then, we get
\begin{eqnarray}
s'(b)=-\lambda
+2\eta(b+h),
\label{ms3}
\end{eqnarray}
where the inverse temperature $\eta$ is determined by the energy constraint (\ref{sb7}) which can be rewritten in dimensionless form
\begin{eqnarray}
\epsilon=\frac{1}{\eta}-2b^2-4hb.
\label{ms4}
\end{eqnarray}
The critical points of $s(b)$, satisfying $s'(b)=0$, correspond to
\begin{eqnarray}
\lambda=x\equiv 2\eta (b+h).
\label{ms5}
\end{eqnarray}
Substituting this result in equation (\ref{sb12}), we obtain the self-consistency relation
\begin{equation}
b=\frac{I_{1}(2\eta (b+h))}{I_{0}(2\eta (b+h))},
\label{ms6}
\end{equation}
which, together with equation (\ref{ms4}),  determines the magnetization as a function of the energy. This returns the equilibrium relationships of Section \ref{sec_c}.

Now, a critical point of $s(b)$ is a maximum if $s''(b)<0$ and a minimum if $s''(b)>0$. Differentiating  equation (\ref{ms3}) with respect to $b$, and recalling that the inverse temperature  $\eta$ is a function of $b$ given by equation (\ref{ms4}), we find that
\begin{eqnarray}
s''(b)=2\eta-\frac{d\lambda}{db}-8\eta^2(b+h)^2.
\label{ms7}
\end{eqnarray}
Therefore, a critical point of $s(b)$ is a maximum if
\begin{eqnarray}
\frac{1}{b'(x)}> 2\eta-8\eta^2 (b+h)^2,
\label{ms8}
\end{eqnarray}
and a minimum if the inequality is reversed. Using equation (\ref{ms5}), inequality (\ref{ms8}) can be rewritten
\begin{eqnarray}
\frac{1}{b'(x)}> 2(\eta-x^2).
\label{ms9}
\end{eqnarray}
According to equation (\ref{sh2}), the turning point of energy ($d\epsilon/dx=0$) corresponds to
\begin{eqnarray}
\frac{1}{b'(x)}= 2(\eta(x)-x^2).
\label{ms10}
\end{eqnarray}
Comparing equation (\ref{ms10}) with equation (\ref{ms9}), we see that
the change of stability in the microcanonical ensemble corresponds to
the turning point of energy $\epsilon_c(h)$ in agreement with the
Poincar\'e theorem. On the other hand, using equation (\ref{cs6}), we
note that
\begin{eqnarray}
s''(b)=-\frac{1}{2}\eta f''(b)-8\eta^2(b+h)^2.
\label{ms11}
\end{eqnarray}
Since the last term in equation (\ref{ms11}) is negative, we recover the fact that canonical stability implies microcanonical stability \cite{ellis}. Indeed, if the critical point is a minimum of free energy ($f''(b)>0$), then it is a fortiori a maximum of entropy ($s''(b)<0$).

\begin{figure}
\begin{center}
\includegraphics[clip,scale=0.3]{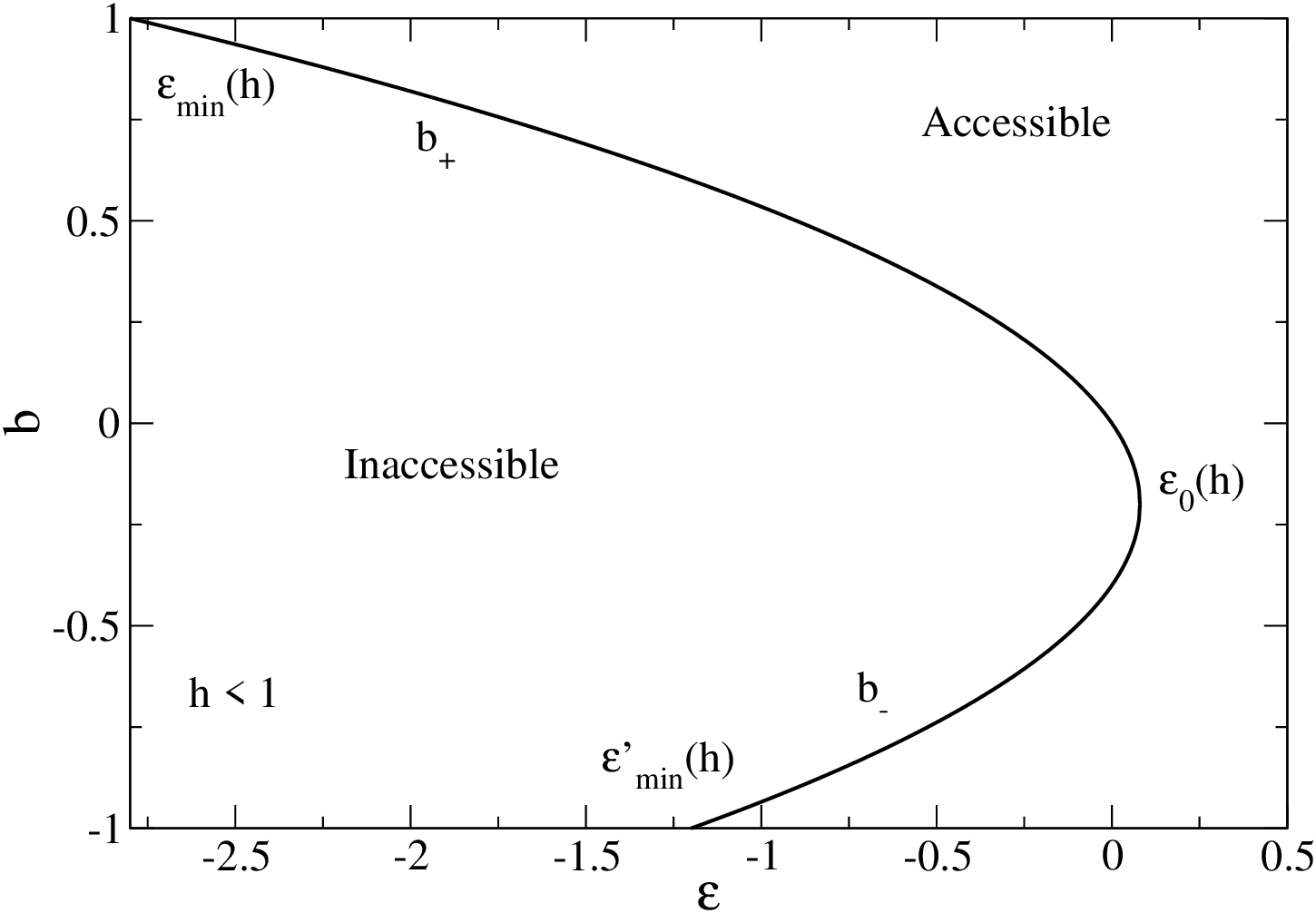}
\caption{Accessible ranges of magnetizations when $h<1$ (specifically $h=0.2$).}
\label{gaph0.2}
\end{center}
\end{figure}

\begin{figure}
\begin{center}
\includegraphics[clip,scale=0.3]{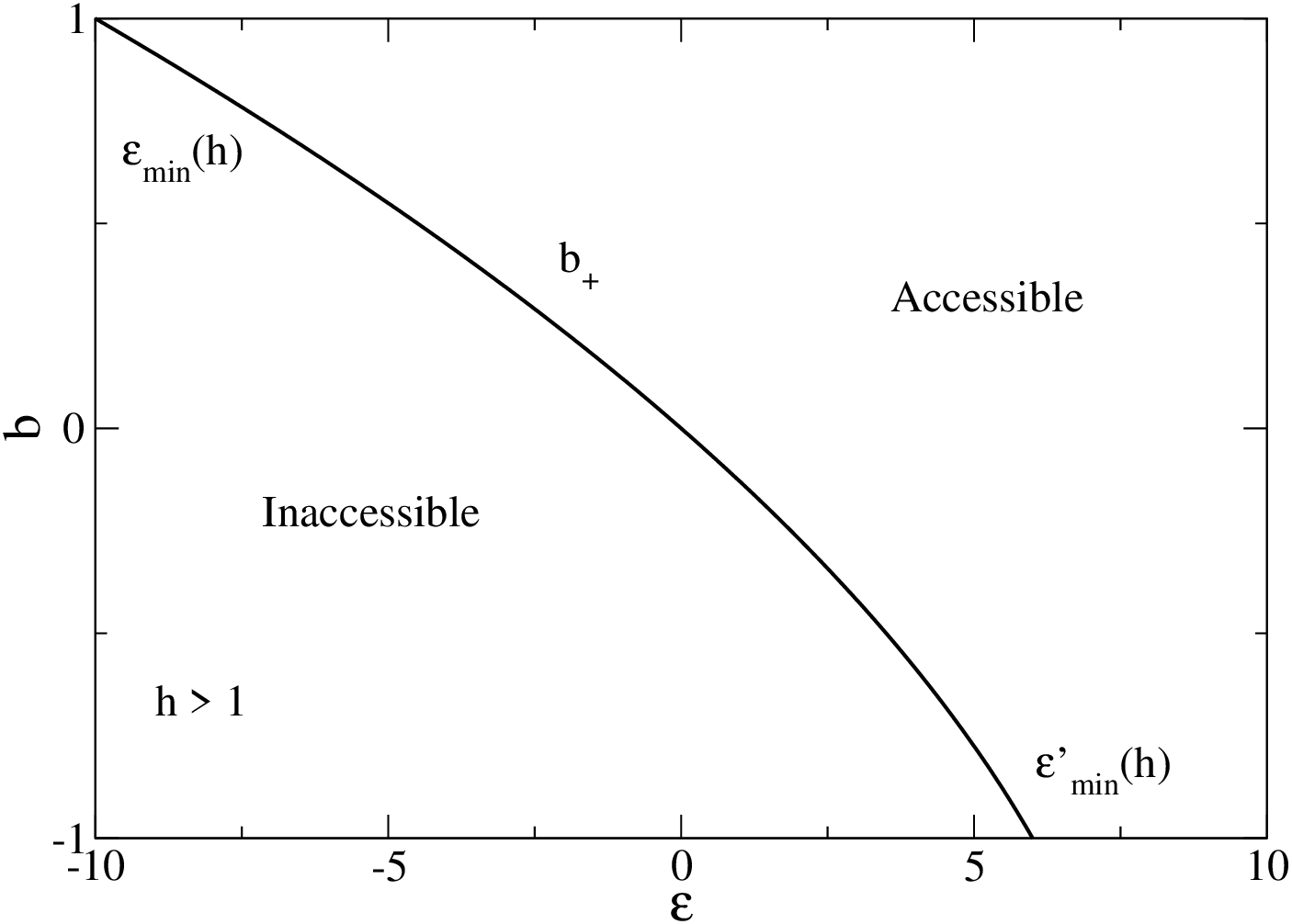}
\caption{Accessible ranges of magnetizations when $h>1$ (specifically $h=2$).}
\label{gaph2}
\end{center}
\end{figure}

To complete our analysis, it can be useful to plot the function $s(b)$ for a prescribed energy $\epsilon$. It is given in parametric form by
\begin{eqnarray}
\label{ms12}
s(\lambda)=\ln I_0(\lambda)-b(\lambda)\lambda+\frac{1}{2}\ln\left (\epsilon+2b(\lambda)^2+4hb(\lambda)\right ),\nonumber\\
\end{eqnarray}
together with equation (\ref{sb12}). Eliminating $\lambda$ between equations (\ref{ms12}) and (\ref{sb12}), we obtain the entropy $s(b)$ as a function of the magnetization $b$ for a fixed value of the energy $\epsilon$. Before going further, we must take into account the possibility of phase space gaps in the system. Indeed, since the temperature is positive, the energy equation (\ref{ms4}) implies that $\epsilon+2b^2+4hb>0$. Depending on the value of the energy, this constraint may restrict the range of accessible magnetizations (see Figs. \ref{gaph0.2} and \ref{gaph2}). We must distinguish several cases.

\begin{figure}
\begin{center}
\includegraphics[clip,scale=0.3]{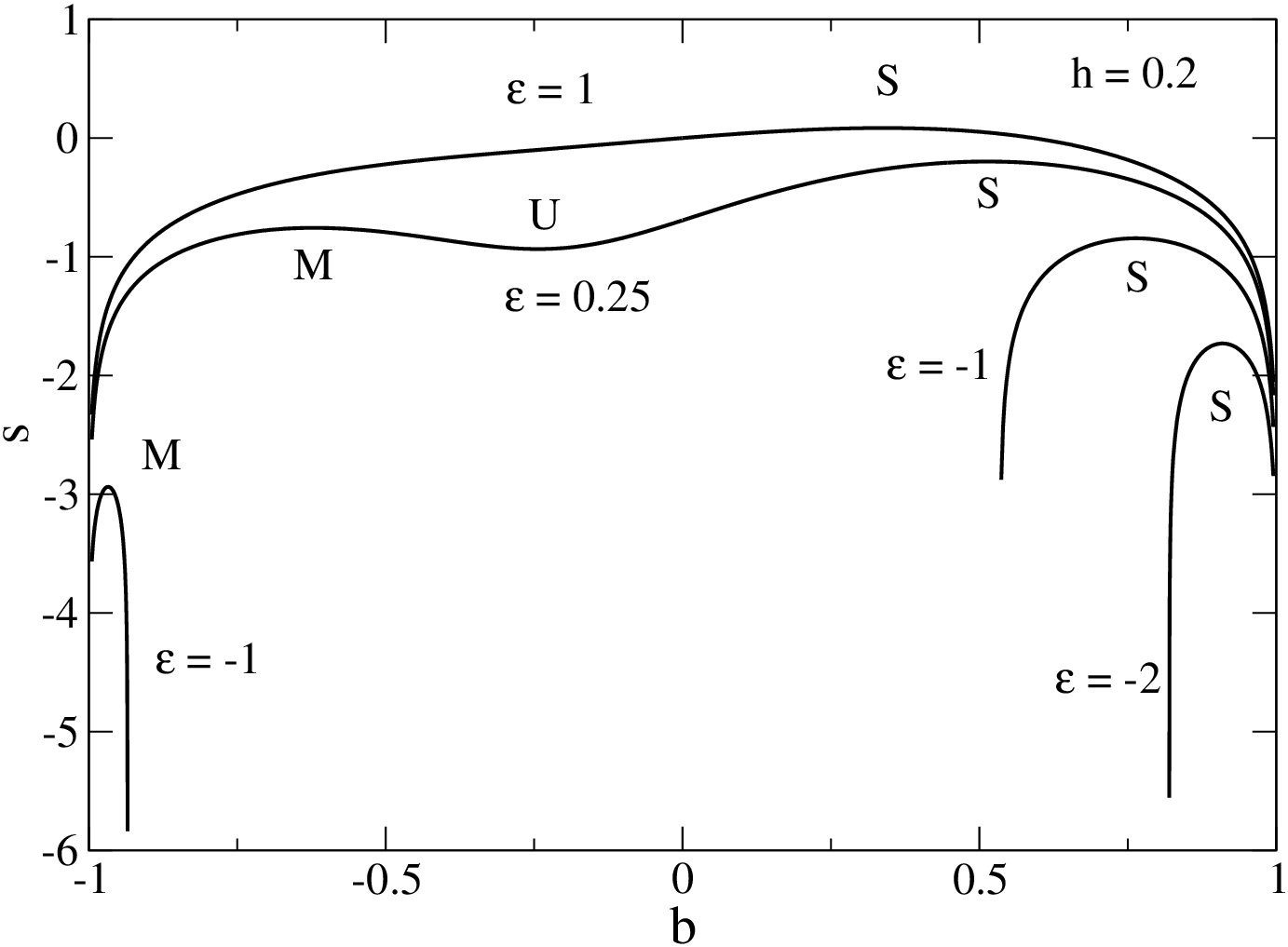}
\caption{Entropy $s(b)$ as a function of the magnetization for a given value of the energy $\epsilon$ when $h<1$ (specifically $h=0.2$). For $\epsilon>\epsilon_c(h)\simeq 0.458$, this curve has a unique (global) maximum at $b_S>0$. For $\epsilon_0(h)=0.08<\epsilon<\epsilon_c(h)$, it has a global maximum at $b_S>0$, a local maximum at $b_M<0$ and a local minimum at $b_U<0$. For $\epsilon_{min}'(h)=-1.2<\epsilon<\epsilon_0(h)$, it has a global maximum at $b_S>0$ and  a local maximum at $b_M<0$. For $\epsilon_{min}(h)=-2.8<\epsilon<\epsilon_{min}'(h)$, it has a unique (global) maximum at $b_S>0$. For $\epsilon_{min}'(h)<\epsilon<\epsilon_0(h)$, there is a gap of magnetization corresponding to the interval $[b_{-}(h,\epsilon),b_{+}(h,\epsilon)]$ and for $\epsilon_{min}(h)<\epsilon<\epsilon_{min}'(h)$, only the interval $]b_{+}(h,\epsilon),1]$ is accessible.}
\label{b-s-h0.2}
\end{center}
\end{figure}

Let us first assume that $h<1$. (i) For $\epsilon>\epsilon_c(h)$, all magnetizations are allowed and the entropy $s(b)$ has a unique (global) maximum at $b_S>0$. (ii) For $\epsilon_0(h)<\epsilon<\epsilon_c(h)$, all magnetizations are allowed and the entropy $s(b)$ has a global maximum at $b_S>0$, a local maximum at $b_M<0$ and a local minimum at $b_U<0$. (iii) For $\epsilon_{min}'(h)<\epsilon<\epsilon_0(h)$, only magnetizations in the intervals  $[-1, b_{-}(h)[$ and $]b_+(h),1]$ with
\begin{eqnarray}
\label{ms13}
b_{\pm}(h,\epsilon)=-h\pm h\sqrt{1-\frac{\epsilon}{\epsilon_0(h)}},
\end{eqnarray}
are allowed and the entropy $s(b)$ has a global maximum at $b_S>0$ and  a local maximum at $b_M<0$. (iv) For $\epsilon_{min}(h)<\epsilon<\epsilon_{min}'(h)$, only magnetizations in the range $]b_+(h),1]$ are allowed and the entropy $s(b)$ has a unique (global) maximum at $b_S>0$. These results are illustrated in Figure \ref{b-s-h0.2}.

\begin{figure}
\begin{center}
\includegraphics[clip,scale=0.3]{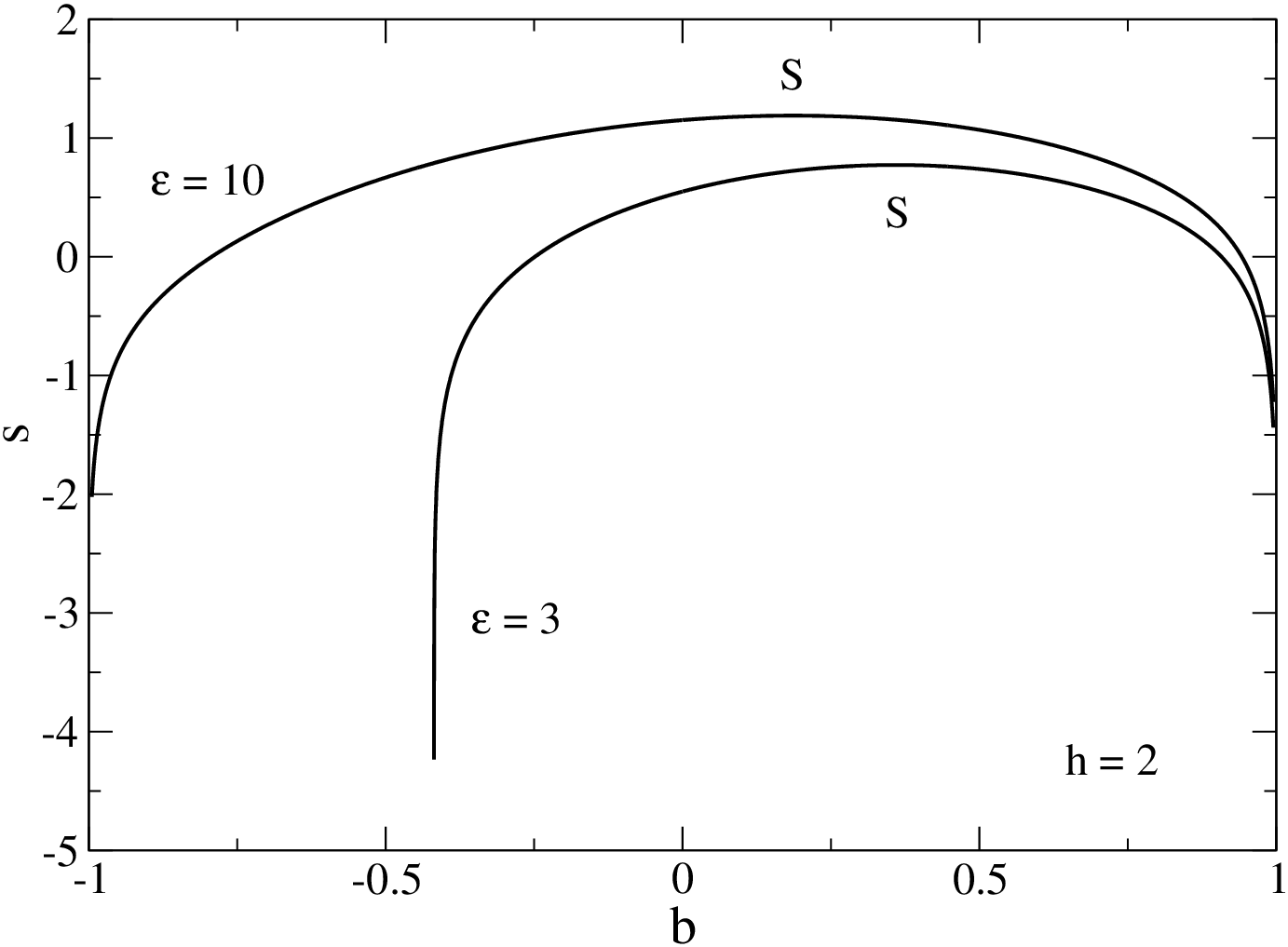}
\caption{Entropy $s(b)$ as a function of the magnetization for a given value of the energy $\epsilon$ when $h>1$ (specifically $h=2$). This curve has a unique (global) maximum at $b_S>0$.  For $\epsilon_{min}(h)=-10<\epsilon<\epsilon'_{min}(h)=6$, only the interval $]b_{+}(h,\epsilon),1]$ is accessible.}
\label{b-s-h2}
\end{center}
\end{figure}

Let us now assume that $h>1$. (i) For $\epsilon>\epsilon'_{min}(h)$, all magnetizations are allowed and the entropy $s(b)$ has a unique (global) maximum at $b_S>0$. (ii) For $\epsilon_{min}(h)<\epsilon<\epsilon'_{min}(h)$,  only magnetizations in the range $]b_+(h),1]$  are allowed and the entropy $s(b)$ has a global maximum at $b_S>0$. These results are illustrated in Figure \ref{b-s-h2}.

It can be shown that the probability of a fluctuation with
magnetization $b$ in the microcanonical ensemble is given by
$P_{MCE}(b)=\frac{1}{g(\epsilon)} e^{Ns(b)}$ (see Appendix
\ref{sec_magnetization}). When there exists only one (global) entropy
maximum $b_S$, as in the case $h>1$ or in the case $h<1$ and
$\epsilon>\epsilon_c(h)$ or $\epsilon_{min}(h)\le
\epsilon<\epsilon'_{min}(h)$, the situation is simple. For
$N\rightarrow +\infty$, the distribution of magnetization is strongly
peaked around the optimal value $b=b_S$ so that, after a transient
regime, the system will be found in that state. For $h<1$ and
$\epsilon_0(h)<\epsilon<\epsilon_c(h)$, the entropy curve presents a
global maximum at $b_S$ and a local maximum at $b_M$ separated by a
local minimum ar $b_U$. For $N\rightarrow +\infty$, these states have
infinite lifetime and the system will be found in one of them
depending on how it has been initially prepared. For finite $N$, the
system will jump from one state to the other. Of course, it will spend
more time in the global entropy maximum (fully stable) than in the
local one (metastable). To pass from one state to the other, the
system has to overcome an entropy barrier which is played by the
unstable solution $b_U$. This barrier scales like $N|\Delta s|$. For
finite $N$, this barrier is finite so that random transitions from the
fully stable state $b_S$ to the metastable state $b_M$ are
possible. For $N\rightarrow +\infty$, the barrier is too hard to cross
and the system remains in one of these two states as previously
indicated (according to the Kramers formula, the lifetime of these
states scales like $t_{life}\sim e^{N\Delta s}\rightarrow
+\infty)$ \cite{art,metastable}. Interestingly, for
$\epsilon_{min}'(h)<\epsilon<\epsilon_0(h)$, the entropy curve still
possesses two entropy maxima $b_S$ and $b_M$ but the model exhibits a
gap in the magnetization corresponding to the interval
$[b_{-}(h,\epsilon),b_{+}(h,\epsilon)]$. Such phase space gaps have
been previously observed in other systems with long range interaction
and they are responsible for ergodicity breaking
\cite{mukamel,bouchet}. Indeed, in the presence of such a gap, the
system cannot jump from one state to the other {\it even when $N$ is
small} (the gap is equivalent to having an infinite entropy barrier
between the two states). In that case, the system remains blocked in
one of these two states for ever and ergodicity is broken (time
averages do not coincide with ensemble averages). This type of
ergodicity breaking has been illustrated in \cite{mukamel,bouchet} for
a generalized isotropic $XY$ model with two and four mean-field
interactions. The HMF model with a magnetic field is another system
(maybe simpler) where ergodicity breaking should be observed in some
range of parameters. Note, in contrast, that these features do not
arise in the canonical ensemble (see Section \ref{sec_cs}) since all
the values of the magnetization are accessible.

{\it Remark:} in the usual HMF model ($h=0$), the entropy maximum is degenerate due to the rotational $U(1)$ symmetry. Therefore, there exists an infinity of equilibrium states that only differ by their phase $\phi$ or equivalently by the position of the density maximum. For finite $N$, the system will explore these different maxima randomly. For $N\rightarrow +\infty$, it will remain blocked in one of them.

\section{Magnetic susceptibility}
\label{sec_ms}

\subsection{Canonical ensemble}
\label{sec_msc}

\begin{figure}
\begin{center}
\includegraphics[clip,scale=0.3]{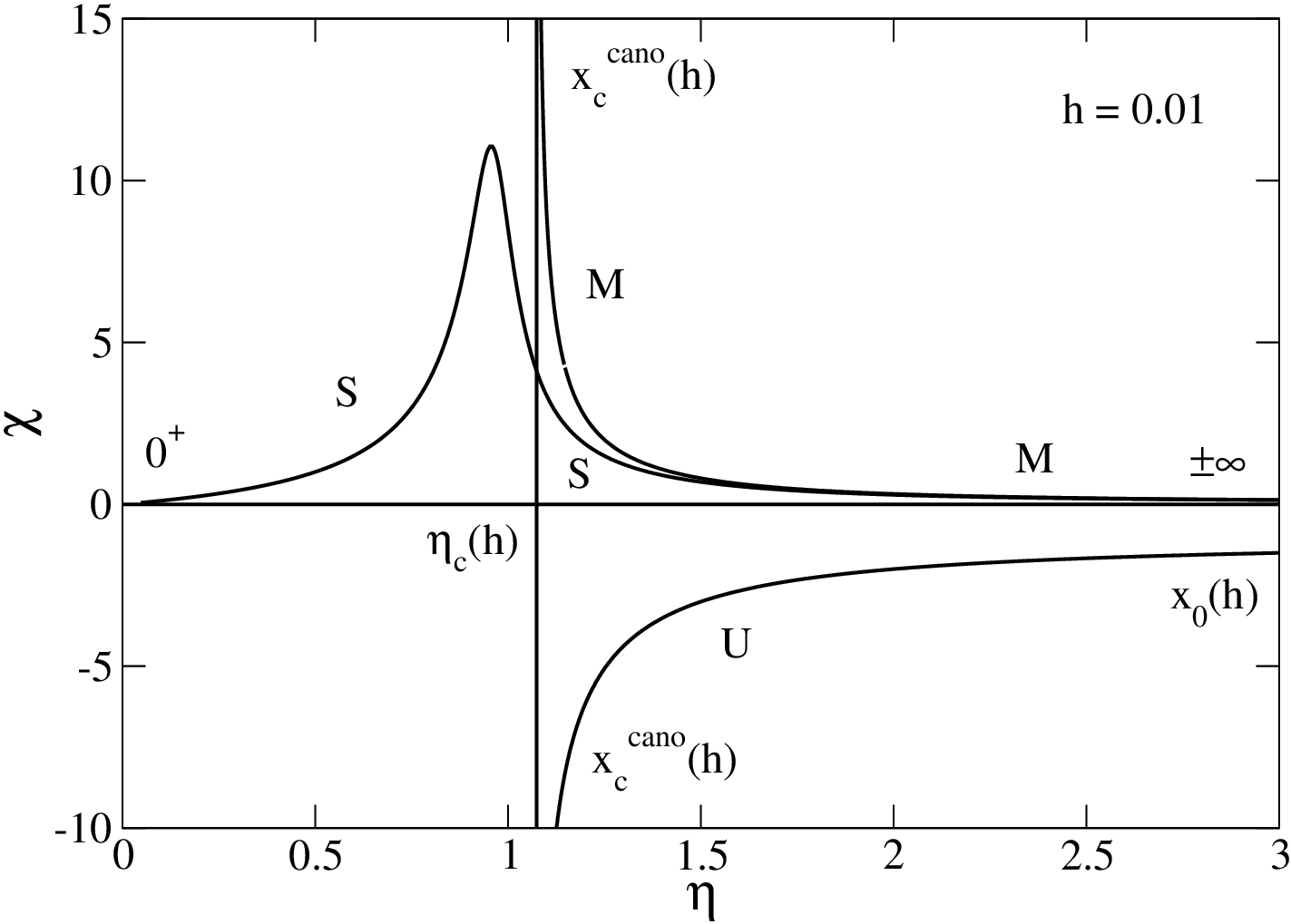}
\caption{Magnetic susceptibility $\chi$ versus inverse temperature $\eta$ for a magnetic field $h=0.01$. Using the results of Appendix \ref{sec_asy}, we find that $\chi\sim \eta$ for $\eta\rightarrow 0$ (corresponding to $x\rightarrow 0$), $\chi\sim 1/(4(1+h)^2\eta)$ for $\eta\rightarrow +\infty$ (corresponding to $x\rightarrow +\infty$), $\chi\sim 1/(4(1-h)^2\eta)$ for $\eta\rightarrow +\infty$ (corresponding to $x\rightarrow -\infty$) and $\chi\rightarrow -1$ for $\eta\rightarrow +\infty$ (corresponding to $x\rightarrow x_0(h)$). On the other hand, $\chi\propto \pm (\eta-\eta_c(h))^{-1/2}$ when $\eta\rightarrow \eta_c(h)^+$ (corresponding to $x\rightarrow x_c^{cano}(h)$). The stable (S) and metastable (M) states have $\chi>0$ while the unstable (U) states have $\chi<0$. Note that the susceptibility increases rapidly close to $\eta_c(h)$. This is related to the second order phase transition at $\eta_c=1$ when $h=0$ (see Section \ref{sec_hzero}).}
\label{eta-chi-h0.01}
\end{center}
\end{figure}

The curve giving the magnetization $b$ as a function of the magnetic
field $h$ at fixed inverse temperature $\eta$ (canonical ensemble) is given in
parametric form by
\begin{eqnarray}
\label{msc1}
b\equiv \frac{2\pi B}{kM}=\frac{I_1(x)}{I_0(x)},
\end{eqnarray}
\begin{eqnarray}
\label{msc2}
h\equiv \frac{2\pi H}{kM}=\frac{x}{2\eta}-b(x).
\end{eqnarray}
The magnetic susceptibility in the canonical ensemble is defined by $\chi=({\partial B}/{\partial H})_{T}$. Introducing dimensionless variables, it can be written
\begin{eqnarray}
\label{msc3}
\chi=\left (\frac{\partial b}{\partial h}\right )_{\eta}.
\end{eqnarray}
Using equations (\ref{msc1}) and (\ref{msc2}), we obtain
\begin{eqnarray}
\label{msc4}
\chi(x)=\frac{2\eta}{\frac{1}{b'(x)}-2\eta},
\end{eqnarray}
where $b'(x)$ is given by equation (\ref{msc5}).  We observe that the magnetic susceptibility (\ref{msc4})  diverges when the condition (\ref{cs8}) is fulfilled, i.e. at the turning point of temperature $\eta_c(h)$. Furthermore, we have the following results: (i) the stable branch (S) has positive magnetic susceptibility for all $x\ge 0$, i.e. for all inverse temperatures $\eta$. (ii) The metastable branch (M) has positive magnetic susceptibility for all $x<x_c^{cano}(h)$, i.e. for all $\eta>\eta_c(h)$. (iii) The unstable branch (U) has negative magnetic susceptibility for all $x_c^{cano}(h)<x\le x_0(h)$, i.e. for all $\eta>\eta_c(h)$. These results are illustrated on Figure \ref{eta-chi-h0.01}.

It is a general result of statistical mechanics that the magnetic
susceptibility is positive in the canonical ensemble
($\chi\ge 0$) since it is a measure of the variance of the fluctuations of the magnetization $\chi=\frac{2\pi}{k}\beta\langle (\Delta B)^2\rangle$. Similarly, the specific heat is positive in the canonical ensemble since it is a measure of the variance of the fluctuations of energy $C=\beta^2\langle (\Delta E)^2\rangle$. For the HMF model with a magnetic field, our study shows that, in the canonical ensemble, all unstable states ($x_c^{cano}(h)<x\le x_0(h)$) have negative magnetic susceptibility but only a fraction of them ($x_c^{cano}(h)<x<x_c^{micro}(h)$) has negative specific heats (see Section \ref{sec_sh}).

\begin{figure}
\begin{center}
\includegraphics[clip,scale=0.3]{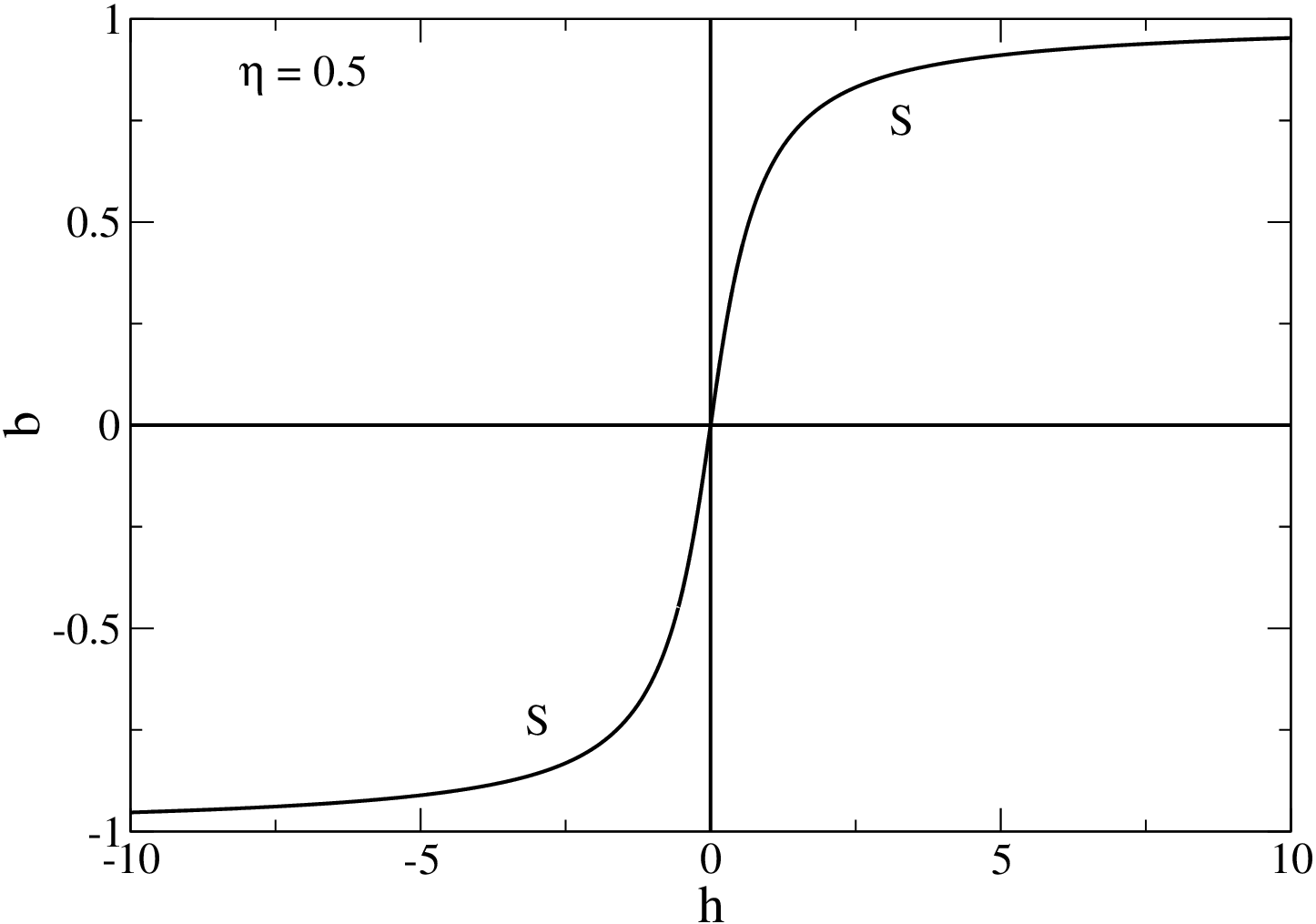}
\caption{Magnetization $b$ versus magnetic field $h$ in the canonical ensemble for $\eta<\eta_c=1$.}
\label{h-b-eta0.5}
\end{center}
\end{figure}

\begin{figure}
\begin{center}
\includegraphics[clip,scale=0.3]{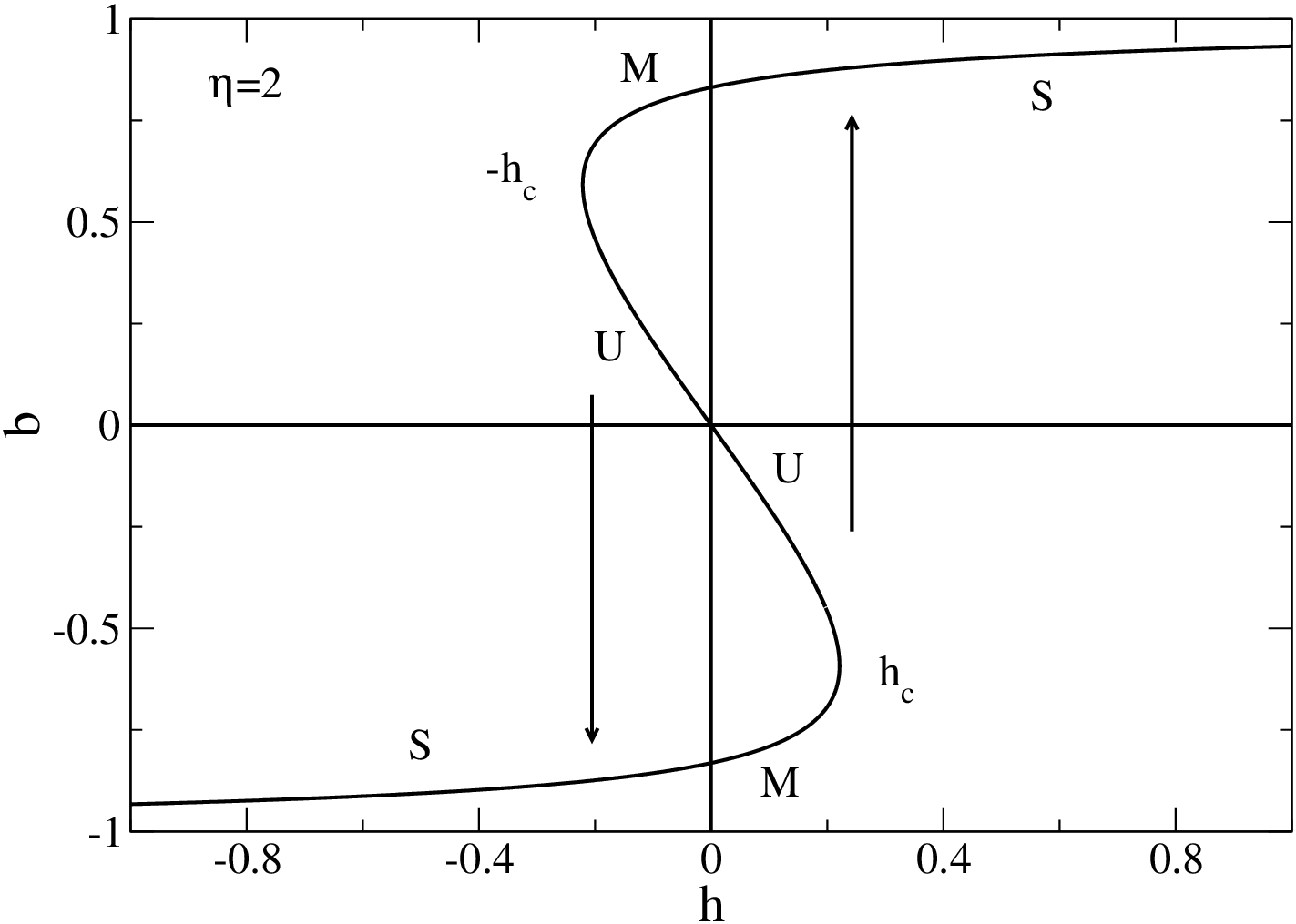}
\caption{Magnetization $b$ versus magnetic field $h$ in the canonical ensemble for $\eta>\eta_c=1$. This curve displays a classical hysteretic cycle similar to the Ising model in a magnetic field.}
\label{h-b-eta2}
\end{center}
\end{figure}

We shall now study the magnetic curve $b(h)$ for a fixed inverse temperature $\eta$. To understand the following discussion, it can be useful to consider  Figure \ref{eta-b} in parallel. On this figure, we fix the value of $\eta$ and progressively increase the value of $h$, starting from $h=0$. Let us introduce the critical magnetic field $h_c(\eta)$ such that $\eta_c(h_c)=\eta$. For $\eta<\eta_c=1$, this equation has no solution. For $\eta>\eta_c=1$, this equation has one solution. We are now ready to study the curves $b(h)$ for different values of the inverse temperature $\eta$.

If $\eta<\eta_c=1$: there is only one stable state $b_S>0$ with $\chi_S>0$.

If $\eta>\eta_c=1$: (i) For  $0<h< h_c(\eta)$, there is one stable state $b_S>0$ with $\chi_S>0$, one metastable state $b_M<0$ with $\chi_M>0$ and one unstable state $b_U<0$ with $\chi_U<0$. (ii)  For $h>h_c(\eta)$, there is only one stable state $b_S>0$ with $\chi_S>0$.

Of course, the magnetic  curves $b(h)$ are antisymmetric with respect to a change of sign $h\rightarrow -h$ of the magnetic field. They are represented in Figures \ref{h-b-eta0.5} and \ref{h-b-eta2} for the different cases described above. For $\eta<\eta_c=1$ (see Figure \ref{h-b-eta0.5}), the curve is univalued, going from $b=-1$ for $h\rightarrow -\infty$ to $b=+1$ for $h\rightarrow +\infty$. For $\eta>\eta_c=1$ (see Figure \ref{h-b-eta2}),  the curve $b(h)$ displays an hysteretic cycle similar to the one observed for the Ising model in a magnetic field. For large positive $h$, the magnetization tends to its maximum value $b=+1$. When we reduce the magnetic field $h$, the magnetization decreases since $\chi>0$. For $h>0$, the upper branch (S) is fully stable and corresponds to states whose magnetization $b>0$ has the same direction as the magnetic field. When $h<0$, this branch becomes metastable (M) since these states have a magnetization opposite  to the magnetic field. However, since metastable states have tremendously long lifetimes for long-range interactions \cite{art,metastable}, the system is expected to  remain on this branch. For $h<-h_c(\eta)$, the metastable branch disappears and the system jumps on the fully stable branch (S) with a magnetization $b<0$ having the same direction as the magnetic field. This transition corresponds to the reorganization of the system from an anti-aligned phase to an aligned phase.  If we keep decreasing the magnetic field, the magnetization decreases since $\chi>0$ until the minimum value $b=-1$ reached for $h\rightarrow -\infty$. If we now increase the magnetic field, the magnetization increases. For $h<0$, the system is on the stable branch (S) corresponding to states with negative  magnetization $b<0$ aligned with the magnetic field. For $h>0$, this branch becomes metastable (M) since these states have a magnetization opposite to the magnetic field. However, the system remains on this branch for the reason given previously. For  $h>h_c(\eta)$ the metastable branch disappears and the system jumps on the fully stable branch with positive magnetization. We have thus described a classical hysteretic cycle. In the canonical ensemble, the stable and metastable states have positive magnetic susceptibilities and unstable states have negative susceptibilities like for the classical Ising model.

\subsection{Microcanonical ensemble}
\label{sec_msm}

The curve giving the magnetization $b$ as a function of the magnetic
field $h$ at fixed energy $\epsilon$ (microcanonical ensemble) is given in
parametric form by
\begin{eqnarray}
\label{msm1}
b\equiv \frac{2\pi B}{kM}=\frac{I_1(x)}{I_0(x)},
\end{eqnarray}
\begin{eqnarray}
\label{msm2}
h(x)=\frac{\epsilon-\frac{2}{x}b(x)+2b(x)^2}{\frac{2}{x}-4b(x)}.
\end{eqnarray}
The magnetic susceptibility in the microcanonical ensemble is defined by $\chi=({\partial B}/{\partial H})_{E}$. Introducing dimensionless variables, it can be written
\begin{eqnarray}
\label{msm3}
\chi=\left (\frac{\partial b}{\partial h}\right )_{\epsilon}.
\end{eqnarray}
Using equations (\ref{msm1}), (\ref{msm2}) and (\ref{t7}), we obtain
\begin{eqnarray}
\label{msm4}
\chi(x)=\frac{2\eta(x)\left (1-2b(x)x\right )}{\frac{1}{b'(x)}-2\left (\eta(x)-x^2\right )},
\end{eqnarray}
where $b'(x)$ is given by equation (\ref{msc5}). In the microcanonical
ensemble, the magnetic susceptibility $\chi$ of stable states can be
either positive or negative \cite{cdr}. Indeed, it can be shown that
$\chi=\frac{2\pi}{k}(\beta \langle (\Delta B)^2\rangle+B(\partial
B/\partial E))$ \cite{vg} (see also Appendix
\ref{sec_magnetization}). We first observe that the magnetic
susceptibility diverges when the condition (\ref{ms10}) is fulfilled,
i.e. at the turning point of energy. We also observe that the magnetic
susceptibility vanishes when $b(x)=1/(2x)$. Together with equation
(\ref{msm1}) this gives $x=\pm x_m\simeq \pm 1.0657$. Then, using
equations (\ref{t6})-(\ref{t8}), we obtain $b=\pm b_m\simeq \pm
0.46918$ and $\epsilon=\epsilon_m=1/(2x_m^2)\simeq 0.44025$. We stress
that {\it these values are independent of $h$!} (by contrast, the
corresponding inverse temperature $\eta=\eta_m(h)\simeq
0.53285/(0.46918+h)$ depends on $h$). Therefore, all the series of
equilibria $b(\epsilon)$ with positive magnetization pass by the
``magic'' point $(\epsilon_m, b_m)$. On the other hand, all the series
of equilibria $b(\epsilon)$ with negative magnetization pass by the
``magic'' point $(\epsilon_m, -b_m)$ provided that $0\le h\le b_m$
(corresponding to $x_0(h)\ge -x_m$).  Let us finally determine the
relative position of the turning point of energy $x_c^{micro}(h)$ with
respect to the point $-x_m$ at which the magnetic susceptibility
vanishes. A simple calculation shows that they coincide for
$h=h_m\equiv 1/(x_m(4x_m^2-1))\simeq 0.2649$. More precisely,
$x_c^{micro}(h)>-x_m$ for $0\le h<h_m$ while $x_c^{micro}(h)<-x_m$ for
$h_m<h<b_m$. These results can be checked on Figure \ref{epsilon-b}.

\begin{figure}
\begin{center}
\includegraphics[clip,scale=0.3]{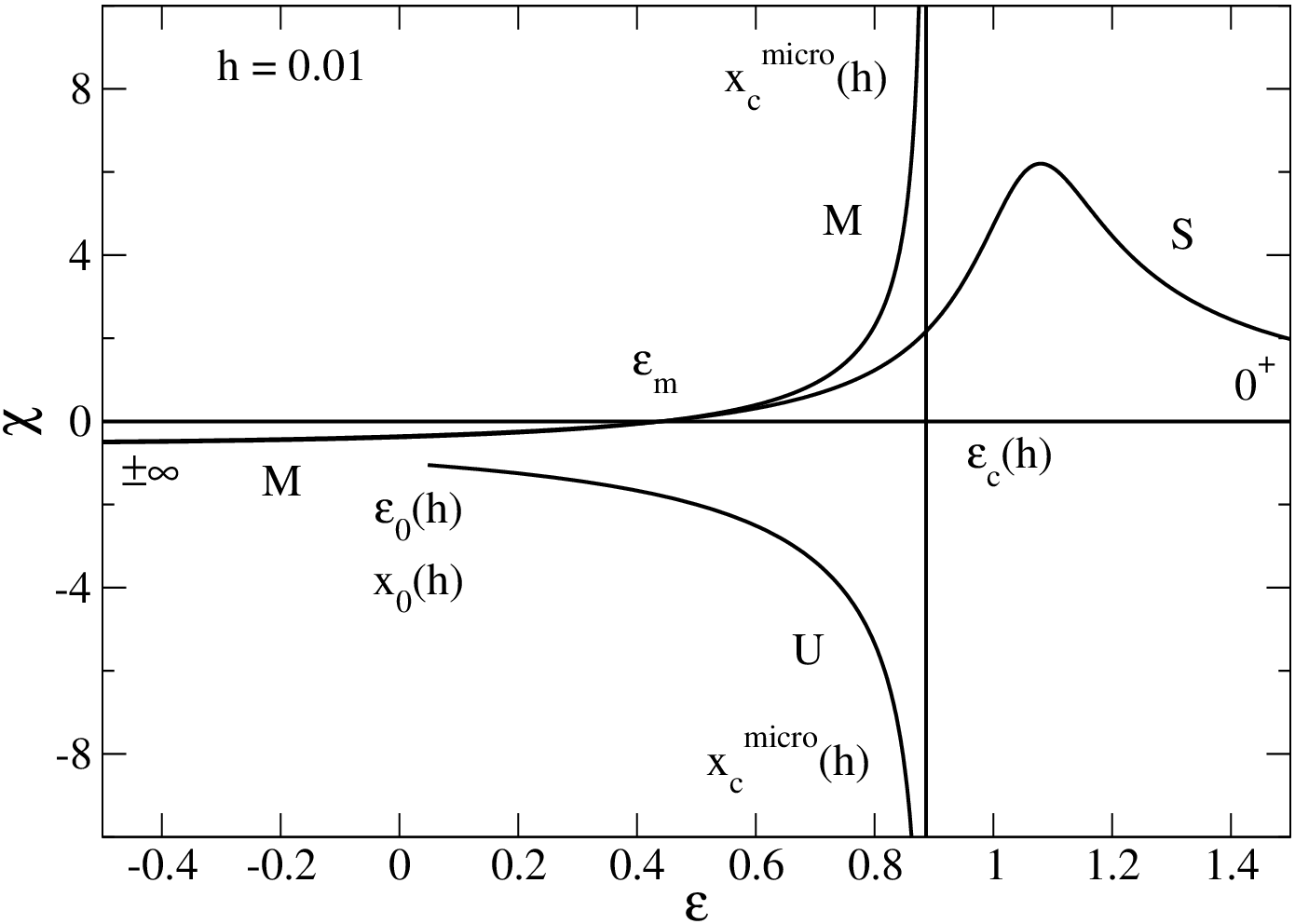}
\caption{Magnetic susceptibility $\chi$ versus energy $\epsilon$ for a magnetic field $h=0.01<h_m$. We find that $\chi\sim 1/\epsilon$ for $\epsilon\rightarrow +\infty$ (corresponding to $x\rightarrow 0$), $\chi\rightarrow -1/(2(1+h))$ for $\epsilon\rightarrow \epsilon_{min}(h)$ (corresponding to $x\rightarrow +\infty$), $\chi\rightarrow -1/(2(1-h))$ for $\epsilon\rightarrow \epsilon'_{min}(h)$ (corresponding to $x\rightarrow -\infty$) and $\chi\rightarrow -1-2h x_0(h)$ for $\epsilon\rightarrow \epsilon_{0}(h)$ (corresponding to $x\rightarrow x_0(h)$). On the other hand, $\chi\propto \pm (\epsilon_c(h)-\epsilon)^{-1/2}$ when $\epsilon\rightarrow \epsilon_c(h)^-$ (corresponding to $x\rightarrow x_c^{micro}(h)$). The stable and metastable states have $\chi>0$ for $\epsilon>\epsilon_m$ and $\chi<0$ for $\epsilon<\epsilon_m$.  Note that the susceptibility increases rapidly close to $\epsilon_c(h)$. This is related to the existence of the second order phase transition at $\epsilon_c=1$ when $h=0$ (see Section \ref{sec_hzero}).}
\label{epsilon-chi-h0.01}
\end{center}
\end{figure}

\begin{figure}
\begin{center}
\includegraphics[clip,scale=0.3]{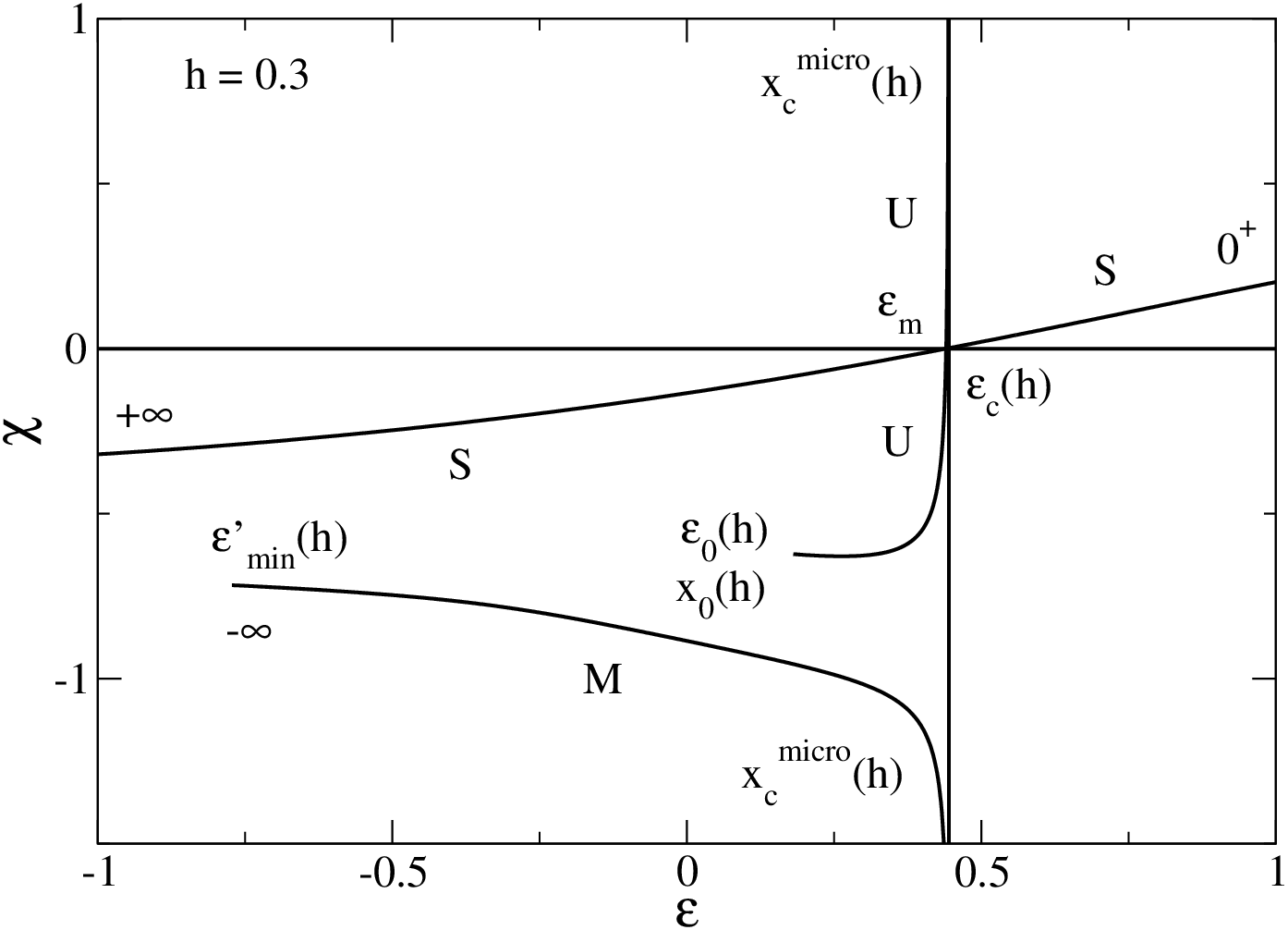}
\caption{Magnetic susceptibility $\chi$ versus energy $\epsilon$ for a magnetic field $h_m<h=0.3<b_m$. The metastable states have $\chi<0$. The stable states have $\chi>0$ for $\epsilon>\epsilon_m$ and $\chi<0$ for $\epsilon<\epsilon_m$.}
\label{epsilon-chi-h0.3}
\end{center}
\end{figure}

\begin{figure}
\begin{center}
\includegraphics[clip,scale=0.3]{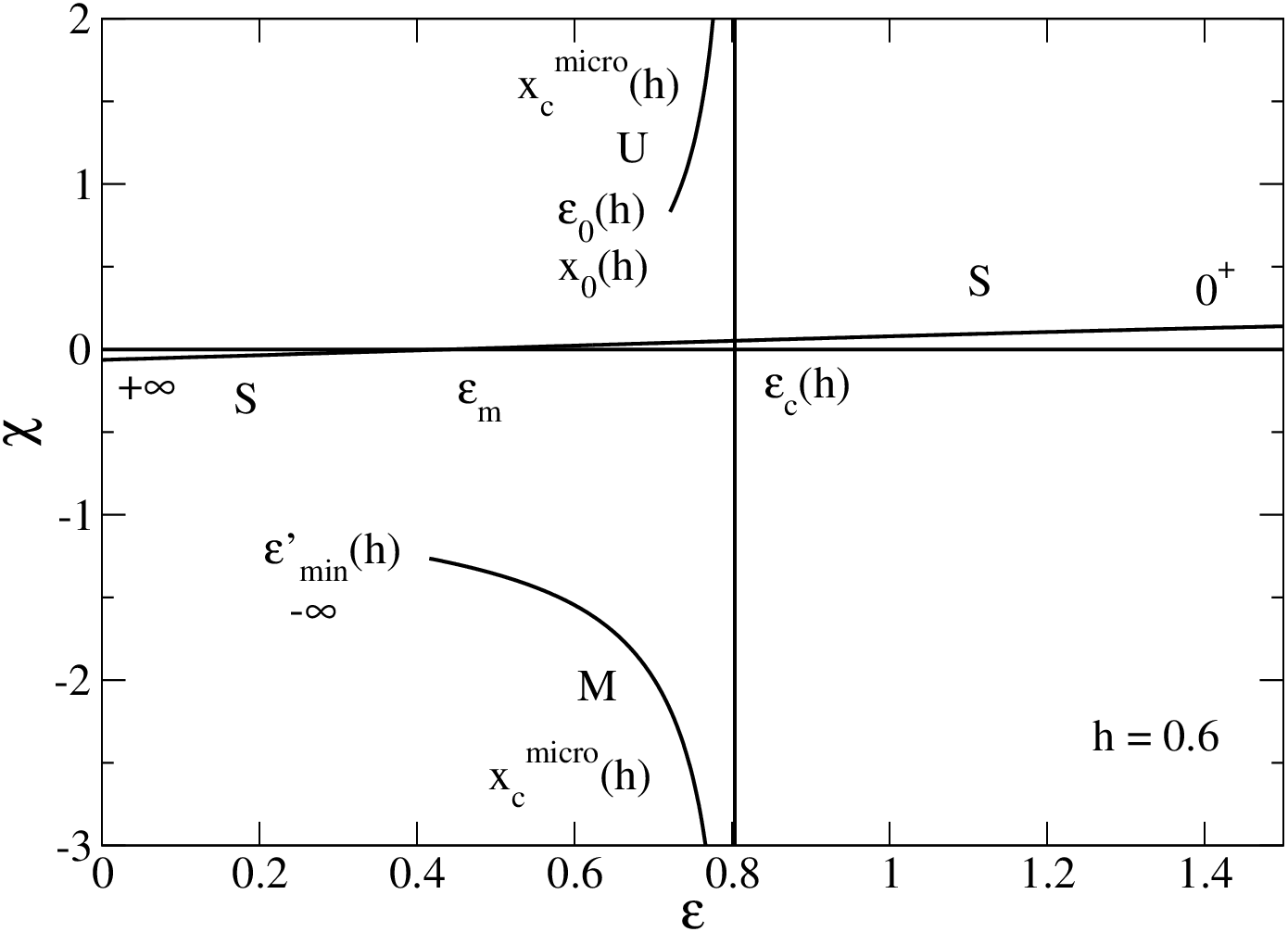}
\caption{Magnetic susceptibility $\chi$ versus energy $\epsilon$ for a magnetic field $h=0.6>b_m$. The metastable states have $\chi<0$. The stable states have $\chi>0$ for $\epsilon>\epsilon_m$ and $\chi<0$ for $\epsilon<\epsilon_m$.}
\label{epsilon-chi-h0.6}
\end{center}
\end{figure}

We can now state the main results (it is recommended to consider Figure \ref{epsilon-b} in parallel): (i) for any $h>0$, the stable branch (S) has positive magnetic susceptibility for $0\le x\le x_m$ (i.e. $\epsilon\ge \epsilon_m$) and negative magnetic susceptibility for $x\ge x_m$ (i.e. $\epsilon_{min}(h)\le \epsilon\le \epsilon_m$); (ii) for $h<h_m$,  the metastable branch (M) has negative magnetic susceptibility for $x\le -x_m$ (i.e. $\epsilon_{min}'(h)\le \epsilon\le \epsilon_m$) and positive magnetic susceptibility for $-x_m\le x<x_c^{micro}(h)$ (i.e. $\epsilon_m\le \epsilon<\epsilon_c(h)$). The unstable branch (U) has negative magnetic susceptibility for $x_c^{micro}(h)<x\le x_0(h)$ (i.e. $\epsilon_0(h)\le \epsilon<\epsilon_c(h)$); (iii) for $h_m<h<b_m$, the metastable branch (M) has negative magnetic susceptibility for $x<x_c^{micro}(h)$ (i.e. $\epsilon_{min}'(h)\le \epsilon<\epsilon_c(h)$). The unstable branch (U) has positive magnetic susceptibility for $x_c^{micro}(h)<x\le -x_m$ (i.e. $\epsilon_m\le \epsilon<\epsilon_c(h)$) and negative magnetic susceptibility for $-x_m\le x\le x_0(h)$ (i.e. $\epsilon_0(h)\le \epsilon\le \epsilon_m$);  (iv) for $b_m<h<1$,  the metastable branch (M) has negative magnetic susceptibility for $x<x_c^{micro}(h)$ (i.e. $\epsilon_{min}'(h)\le \epsilon<\epsilon_c(h)$).  The unstable branch (U) has positive magnetic susceptibility for $x_c^{micro}(h)<x\le x_0(h)$ (i.e. $\epsilon_0(h)\le \epsilon<\epsilon_c(h)$). These results are illustrated on Figures \ref{epsilon-chi-h0.01}-\ref{epsilon-chi-h0.6}.

We shall now study the magnetic curve $b(h)$ for a fixed energy
$\epsilon$. To understand the following discussion, it can be useful
to consider Figure \ref{epsilon-b} in parallel. On this figure, we fix
the value of $\epsilon$ and progressively increases the value of $h$,
starting from $h=0$. To prepare the following discussion, we introduce
some critical magnetic fields. Let $h_c(\epsilon)$ be the magnetic
field such that $\epsilon_c(h_c)=\epsilon$. For $\epsilon>2$, this
equation has no solution. For $\epsilon_c=1<\epsilon<2$, this equation
has a unique solution. For $\epsilon_m<\epsilon<\epsilon_c=1$, this
equation has two solutions denoted $h_c^{(1)}(\epsilon)$ and
$h_c^{(2)}(\epsilon)$. For $\epsilon<\epsilon_m$, this equation has no
solution. Let $h_0(\epsilon)$ be the magnetic field such that
$\epsilon_0(h_0)=\epsilon$. For $\epsilon>2$, this equation has no
solution. For $0<\epsilon<2$, this equation has a unique solution
explicitly given by $h_0(\epsilon)=(\epsilon/2)^{1/2}$. For
$\epsilon<0$ it has no solution. Let $h'_{min}(\epsilon)$ be the
magnetic field such that $\epsilon'_{min}(h'_{min})=\epsilon$. For
$\epsilon>2$, this equation has no solution.  For $-2<\epsilon<2$,
this equation has a unique solution explicitly given by
$h'_{min}(\epsilon)=(\epsilon+2)/4$. For $\epsilon<-2$, this equation
has no solution.  Let $h_{min}(\epsilon)$ be the magnetic field such
that $\epsilon_{min}(h_{min})=\epsilon$. For $\epsilon<-2$, this
equation has a unique solution explicitly given by
$h_{min}(\epsilon)=-(\epsilon+2)/4$. For $\epsilon>-2$, this equation
has no solution. We are now ready to study the curves $b(h)$ for
different values of energy $\epsilon$.

If $\epsilon>2$: there is only one stable state $b_S>0$ with $\chi_S>0$. The magnetic curve $b(h)$ is similar to the one of Figure \ref{h-b-eta0.5} in the canonical ensemble.

\begin{figure}
\begin{center}
\includegraphics[clip,scale=0.3]{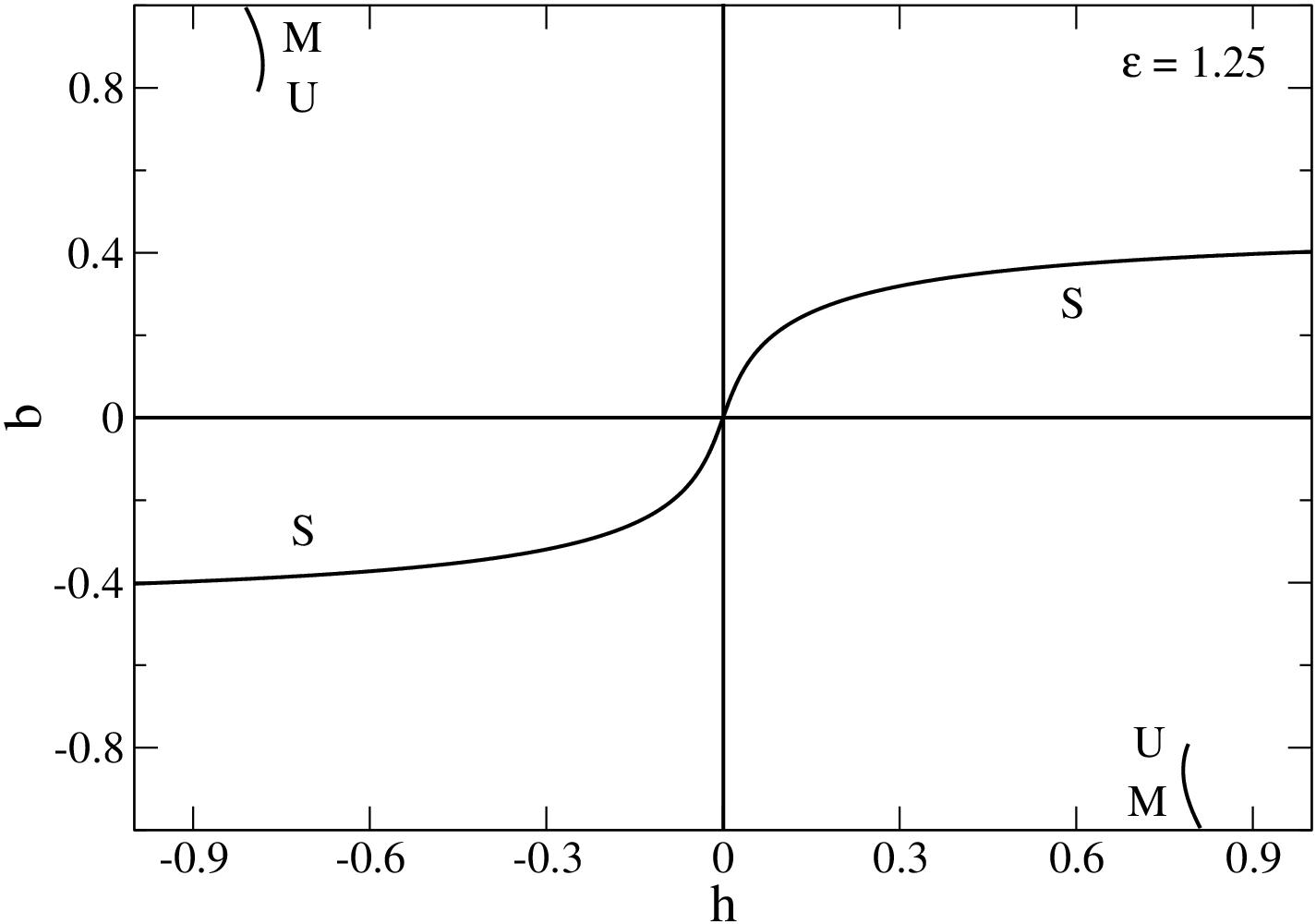}
\caption{Magnetization $b$ versus magnetic field $h$ in the microcanonical ensemble for $\epsilon_c=1<\epsilon=1.25<2$.}
\label{h-b-epsilon1.25}
\end{center}
\end{figure}

If $\epsilon_c=1<\epsilon<2$ (see Figure \ref{h-b-epsilon1.25}): (i) For  $0<h< h_c(\epsilon)$, there is only one stable state $b_S>0$ with $\chi_S>0$. (ii) For $h_c(\epsilon)<h<h_0(\epsilon)$, there is one stable state $b_S>0$ with $\chi_S>0$, one metastable state $b_M<0$ with $\chi_M<0$ and one unstable state $b_U<0$ with $\chi_U>0$. (iii) For $h_0(\epsilon)<h<h'_{min}(\epsilon)$, there is one stable state $b_S>0$ with $\chi_S>0$ and one metastable state $b_M<0$ with $\chi_M<0$. (iv) For $h>h'_{min}(\epsilon)$, there is only one stable state $b_S>0$ with $\chi_S>0$. If we start from the stable branch (S), we remain on this branch for all $h$. Furthermore, the magnetization increases with $h$ since $\chi>0$. If we start from the metastable branch (M), made of anti-aligned states, then we jump on the stable branch, made of aligned states, for $|h|>h_{min}'(\epsilon)$ or $|h|<h_c(\epsilon)$. Note that the metastable branch has negative susceptibility so that $b$ decreases with $h$.

\begin{figure}
\begin{center}
\includegraphics[clip,scale=0.3]{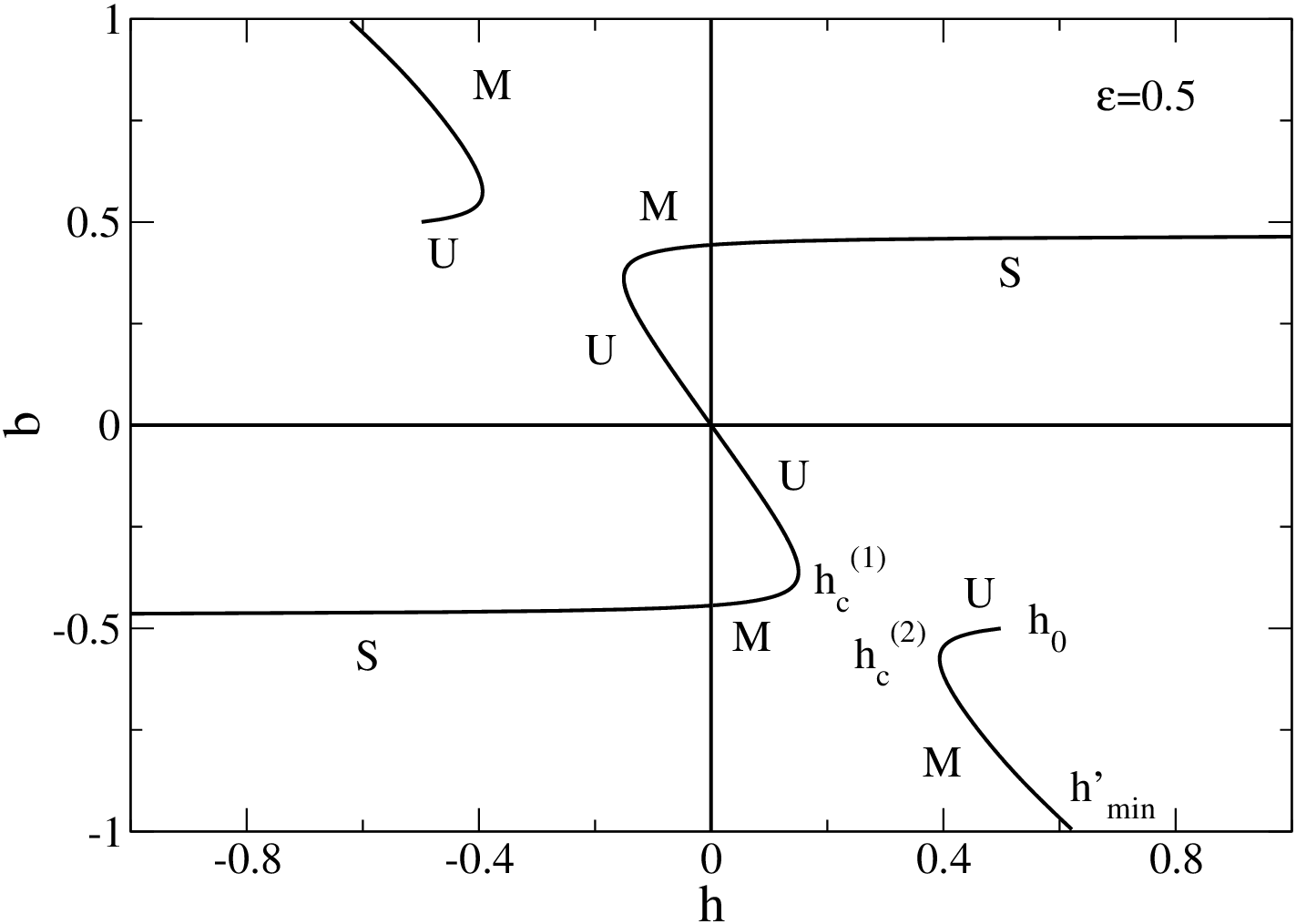}
\caption{Magnetization $b$ versus magnetic field $h$ in the microcanonical ensemble for $\epsilon_m<\epsilon=0.5<\epsilon_c=1$.}
\label{h-b-epsilon0.5}
\end{center}
\end{figure}

If $\epsilon_m<\epsilon<\epsilon_c=1$ (see Figure \ref{h-b-epsilon0.5}): (i) For  $0<h< h_c^{(1)}(\epsilon)$, there is one stable state $b_S>0$ with $\chi_S>0$, one metastable state $b_M<0$ with $\chi_M>0$ and one unstable state $b_U<0$ with $\chi_U<0$. (ii)  For $h_c^{(1)}(\epsilon)<h<h_c^{(2)}(\epsilon)$, there is only one stable state $b_S>0$ with $\chi_S>0$. (iii) For $h_c^{(2)}(\epsilon)<h<h_0(\epsilon)$, there is one stable state $b_S>0$ with $\chi_S>0$, one metastable state $b_M<0$ with $\chi_M<0$ and one unstable state $b_U<0$ with $\chi_U>0$. (iv) For $h_0(\epsilon)<h<h'_{min}(\epsilon)$, there is one stable state $b_S>0$ with $\chi_S>0$ and  one metastable state $b_M<0$ with $\chi_M<0$. (iv) For $h>h'_{min}(\epsilon)$, there is only one stable state $b_S>0$ with $\chi_S>0$. The central curve displays a classical hysteretic cycle similar to the one described in Section \ref{sec_msc} in the canonical ensemble. On the other hand, if we start from the  metastable branch on the periphery, then the system undergoes a transition to the stable branch for  $|h|>h_{min}'(\epsilon)$ or $|h|<h_c^{(2)}(\epsilon)$ as in the previous case.

\begin{figure}
\begin{center}
\includegraphics[clip,scale=0.3]{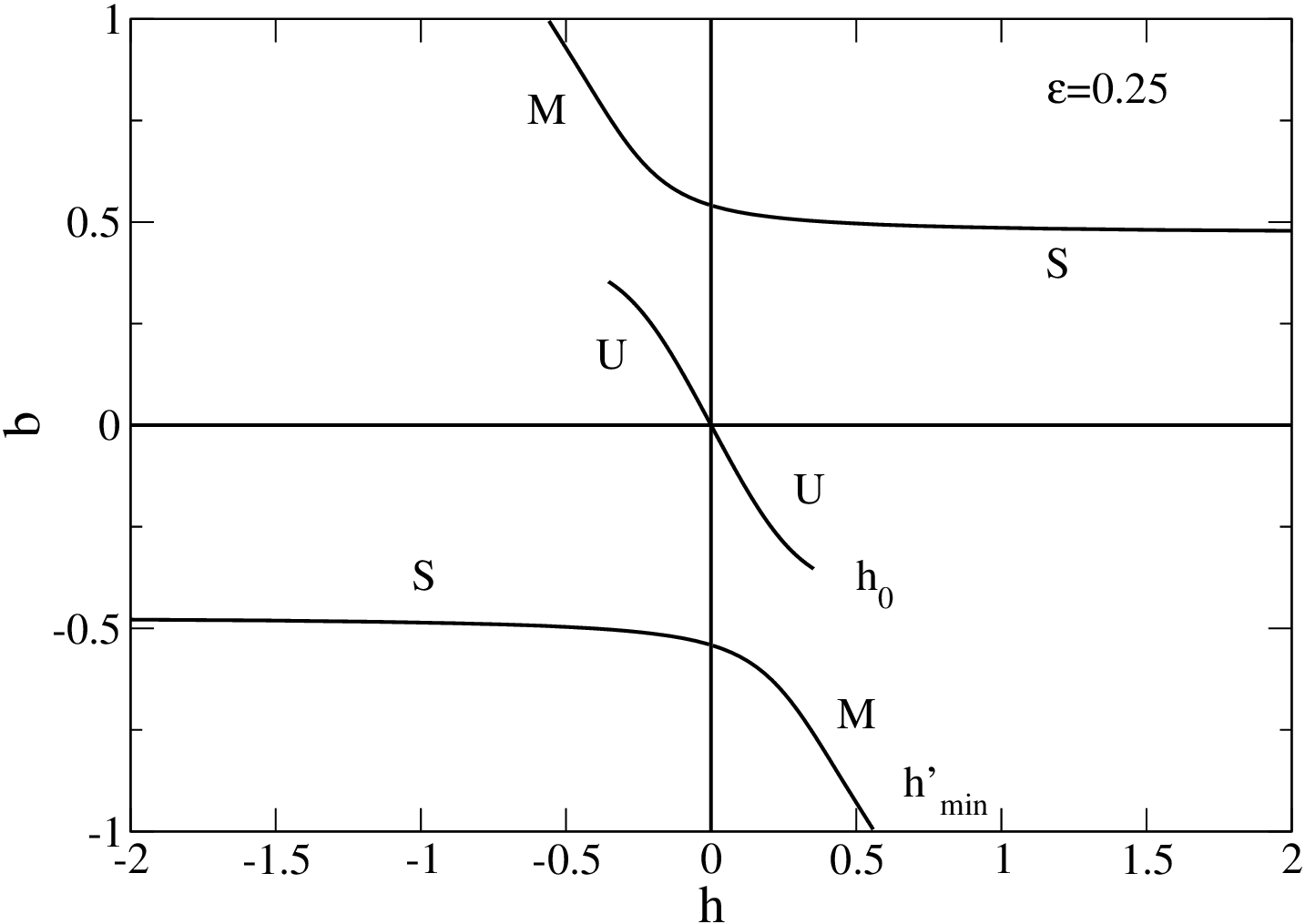}
\caption{Magnetization $b$ versus magnetic field $h$ in the microcanonical ensemble for $0<\epsilon=0.25<\epsilon_m$.}
\label{h-b-epsilon0.25}
\end{center}
\end{figure}

If $0<\epsilon<\epsilon_m$ (see Figure \ref{h-b-epsilon0.25}): (i) For  $0<h<h_0(\epsilon)$, there is one stable state $b_S>0$ with $\chi_S<0$, one metastable state $b_M<0$ with $\chi_M<0$ and one unstable state $b_U<0$ with $\chi_U<0$. (ii)  For $h_0(\epsilon)<h<h'_{min}(\epsilon)$, there is one stable state $b_S>0$ with $\chi_S<0$ and one metastable state $b_M<0$ with $\chi_M<0$. (iii) For $h>h'_{min}(\epsilon)$, there is only one stable state $b_S>0$ with $\chi_S<0$.This case is interesting since it yields a new type of hysteretic cycle related to the fact that the stable and metastable states have negative magnetic susceptibility. For large positive $h$, the magnetization tends to $b=b_s$. When we reduce the magnetic field $h$, the magnetization increases since $\chi<0$. For $h>0$, the upper branch (S) is fully stable and its magnetization $b>0$ has the same direction as the magnetic field. When $h<0$, this branch becomes metastable (M) since its magnetization is now opposite to the magnetic field.  However, since metastable states have tremendously long lifetimes for long-range interactions \cite{art,metastable}, the system remains on this branch. The magnetization increases until the maximum value $b=1$ so that the system is in its most anti-aligned configuration. For $h<-h'_{min}(\epsilon)$, the metastable branch disappears and the system jumps on the fully stable branch (S) with a magnetization $b<0$ having the same direction as the magnetic field. This transition corresponds to the reorganization of the system from an anti-aligned phase to an aligned phase. If we decrease the magnetic field, the magnetization increases to $-b_s$ since $\chi<0$. If we now increase the magnetic field, the magnetization decreases. For $h<0$, the system is on the stable branch (S) corresponding to states with negative magnetization $b<0$ aligned with the magnetic field. For $h>0$, this branch becomes metastable (M) since these states become anti-aligned with the magnetic field. However, the system remains on this branch for the reason given previously. For  $h>h'_{min}(\epsilon)$ the metastable branch disappears and the system jumps on the fully stable branch with positive magnetization. We have thus described a new type of hysteretic cycle. In the microcanonical ensemble, the stable and metastable states have negative magnetic susceptibilities so that, increasing the magnetic field, the magnetization decreases.

\begin{figure}
\begin{center}
\includegraphics[clip,scale=0.3]{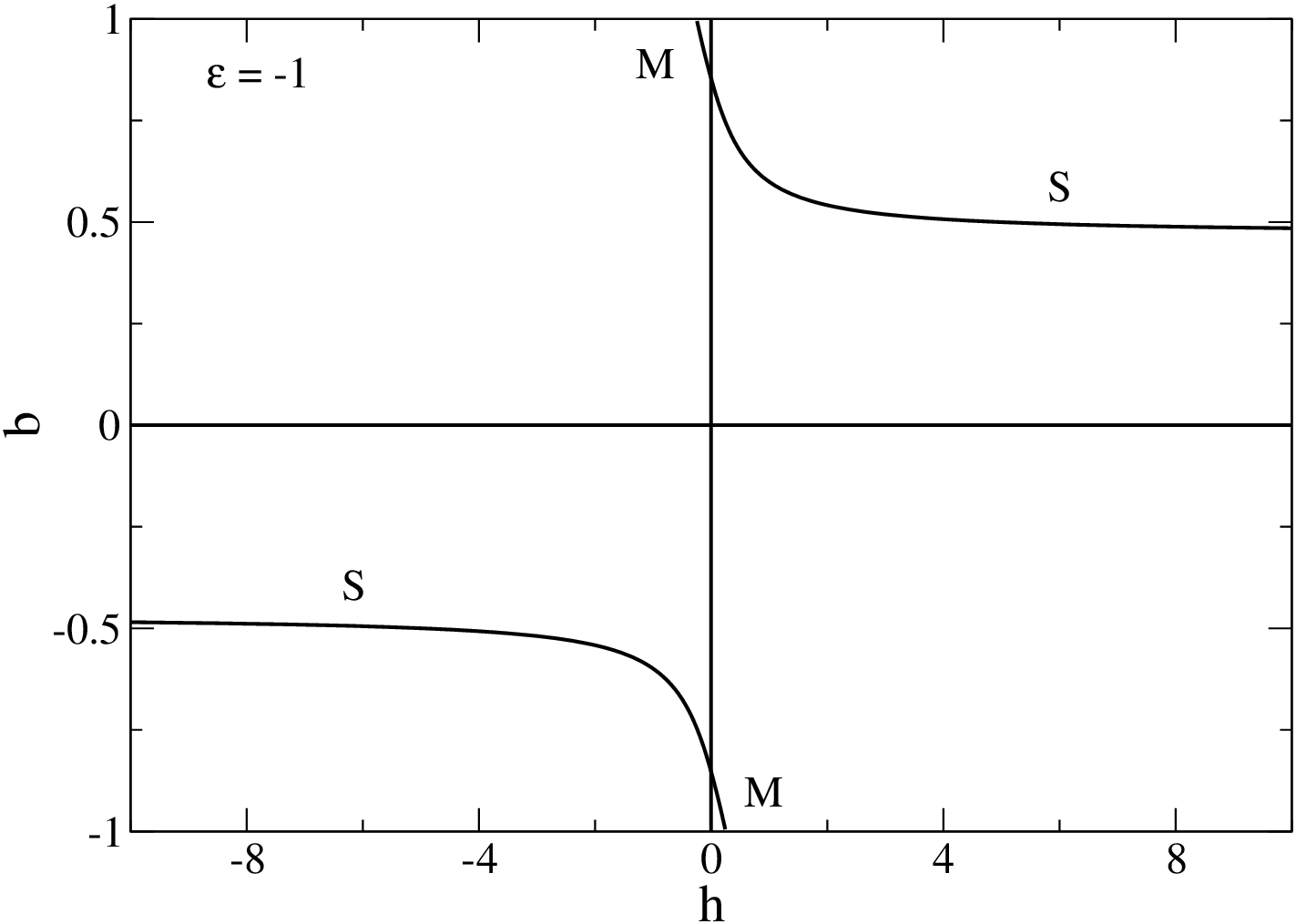}
\caption{Magnetization $b$ versus magnetic field $h$ in the microcanonical ensemble for $-2<\epsilon=-1<0$.}
\label{h-b-epsilonMINUS1}
\end{center}
\end{figure}

If $-2<\epsilon<0$ (see Figure \ref{h-b-epsilonMINUS1}): (i)  For $0<h<h'_{min}(\epsilon)$, there is one stable state $b_S>0$ with $\chi_S<0$ and one metastable state $b_M<0$ with $\chi_M<0$. (ii) For $h>h'_{min}(\epsilon)$, there is only one stable state $b_S>0$ with $\chi_S<0$. The discussion is essentially the same as the one given previously since the only difference is the nonexistence of unstable states (which play no role) in the present case.

\begin{figure}
\begin{center}
\includegraphics[clip,scale=0.3]{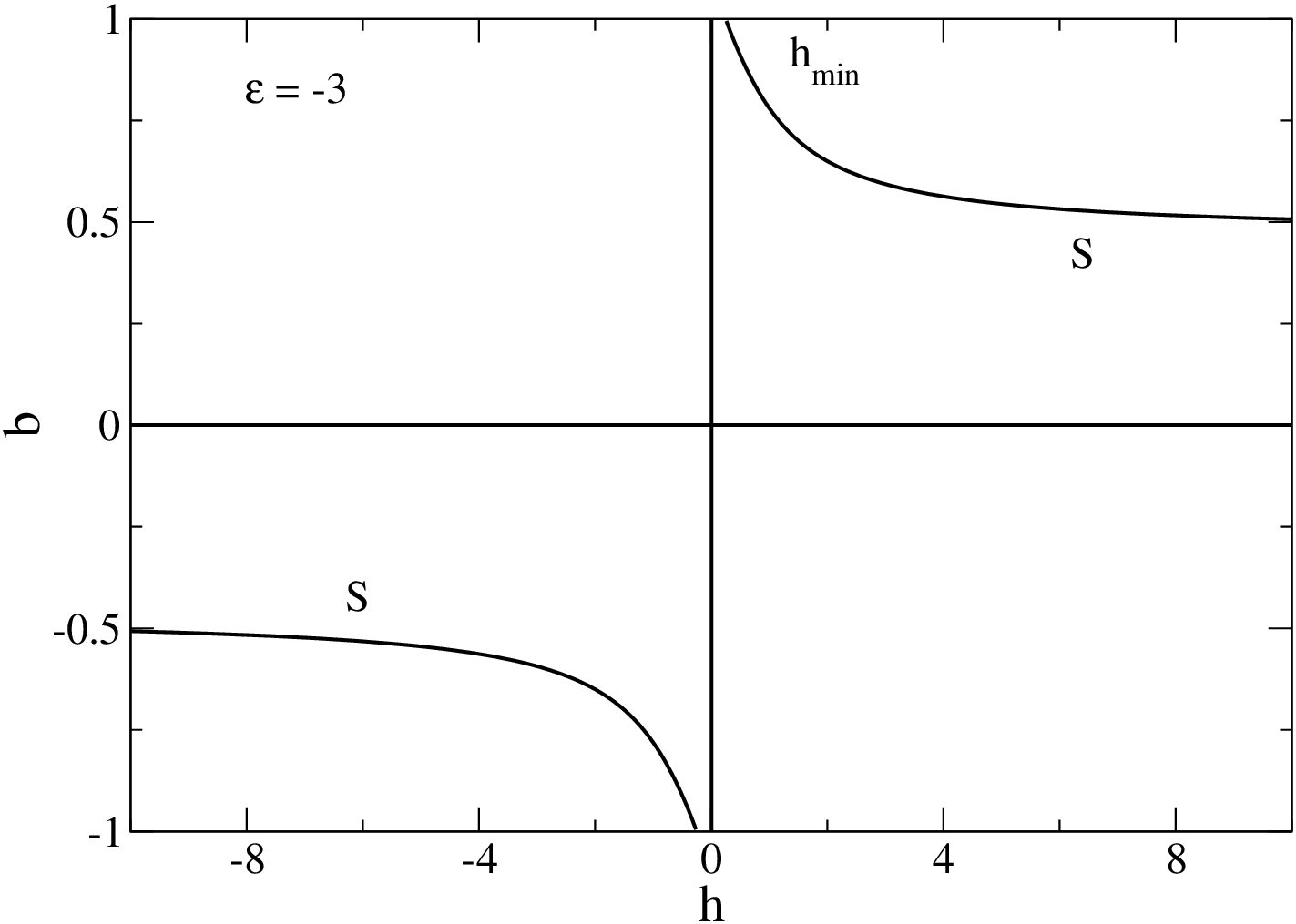}
\caption{Magnetization $b$ versus magnetic field $h$ in the microcanonical ensemble for $\epsilon=-3<-2$.}
\label{h-b-epsilonMINUS3}
\end{center}
\end{figure}

If $\epsilon<-2$ (see Figure \ref{h-b-epsilonMINUS3}): (i)  For $0<h<h_{min}(\epsilon)$, there is no solution (this region is inaccessible). (ii) For $h>h_{min}(\epsilon)$, there is only one stable state $b_S>0$ with $\chi_S<0$.

\subsection{The limit $h\rightarrow 0$}
\label{sec_hzero}

In this section, we study the magnetic susceptibility in the limit $h\rightarrow 0$ and investigate its divergence at the critical inverse temperature $\eta_c=1$ or critical energy $\epsilon_c=1$ (see also \cite{vg} with another presentation).

\begin{figure}
\begin{center}
\includegraphics[clip,scale=0.3]{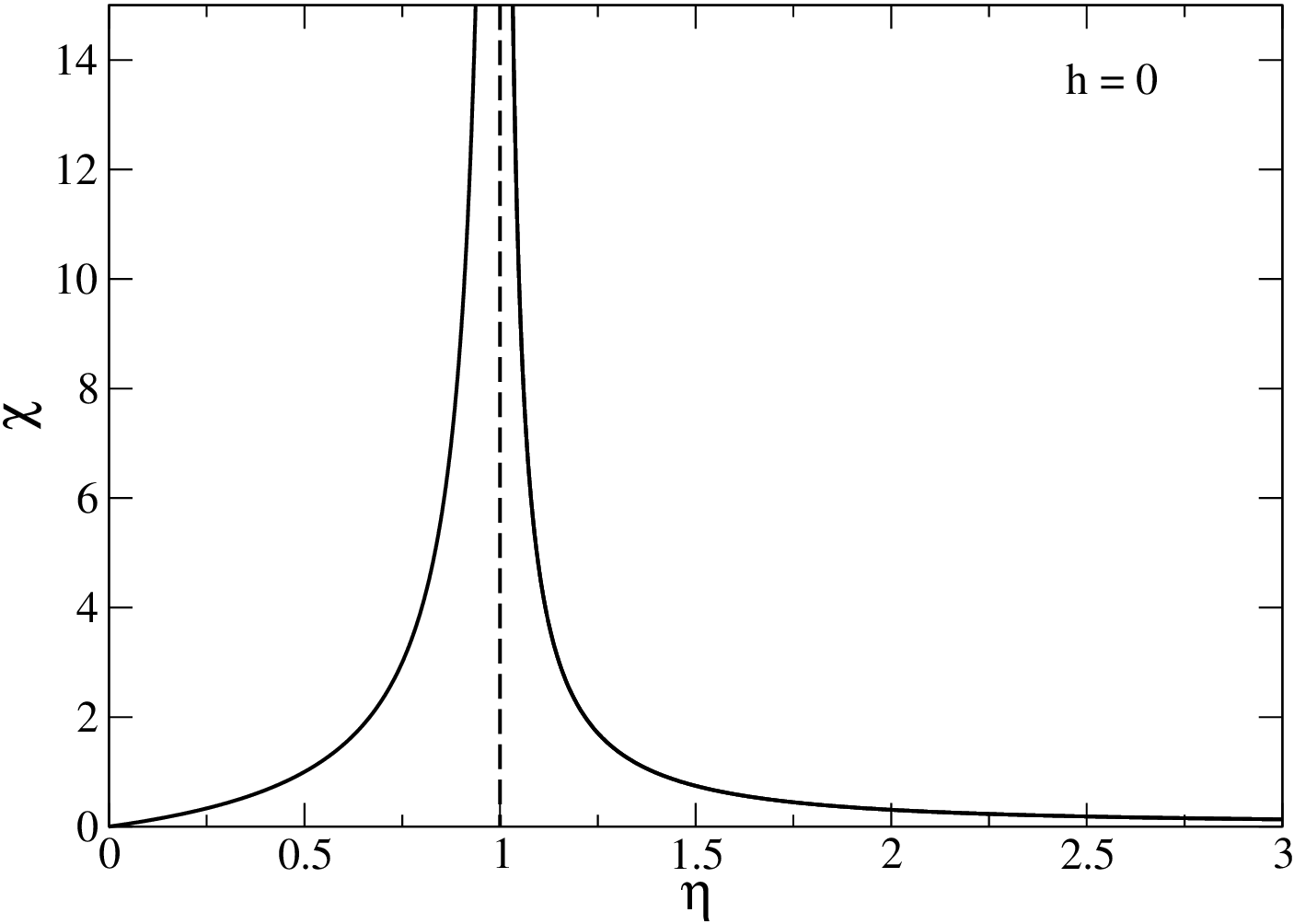}
\caption{Magnetic susceptibility $\chi$ as a function of the inverse temperature $\eta$ in the canonical ensemble for $h\rightarrow 0$.}
\label{eta-chi-h0}
\end{center}
\end{figure}

Let us first consider the canonical ensemble. For any inverse temperature $\eta<\eta_c=1$, we know that $b\rightarrow 0$ when $h\rightarrow 0$ \cite{cvb}. Therefore, we can expand equations (\ref{msc1}) and (\ref{msc2}) in powers of $x\rightarrow 0$ at fixed $\eta$. We obtain the equivalents
\begin{eqnarray}
\label{hzero1}
b(x)\sim \frac{x}{2},\qquad h(x)\sim \left (\frac{1}{\eta}-1\right )\frac{x}{2},
\end{eqnarray}
from which we deduce the expression of the magnetic susceptibility
\begin{eqnarray}
\label{hzero2}
\chi=\frac{\eta}{1-\eta}.
\end{eqnarray}
This equation is valid  for any $\eta<\eta_c=1$. In particular, the magnetic susceptibility diverges at the critical point ($\eta\rightarrow \eta_c^-=1^-$) like
\begin{eqnarray}
\label{hzero3}
\chi\sim \frac{1}{1-\eta}.
\end{eqnarray}
For $\eta>\eta_c=1$ and $h\rightarrow 0$, the magnetic susceptibility $\chi(\eta)$ is given by equations (\ref{msc4}) and (\ref{t7}) with $h=0$. These equations can be written
\begin{eqnarray}
\label{hzero4}
\chi(x)=\frac{1}{\frac{1}{2\eta(x) b'(x)}-1},
\end{eqnarray}
\begin{eqnarray}
\label{hzero5}
\eta(x)=\frac{x}{2b(x)}.
\end{eqnarray}
For $\eta\rightarrow \eta_c=1$, we can expand the previous equations in powers of $x\rightarrow 0$. We obtain the equivalents
\begin{eqnarray}
\label{hzero6}
\eta(x)-1\sim\frac{x^2}{8},\qquad \chi(x)\sim\frac{4}{x^2},
\end{eqnarray}
so that the magnetic susceptibility diverges at the critical point ($\eta\rightarrow \eta_c^+=1^+$) like
\begin{eqnarray}
\label{hzero7}
\chi\sim \frac{1}{2(\eta-1)}.
\end{eqnarray}
Note that equations (\ref{hzero3}) and (\ref{hzero7}) differ by a factor $2$. There is the same difference for the Ising model in a magnetic field. The magnetic susceptibility is plotted as a function of $\eta$  in Figure \ref{eta-chi-h0}.

\begin{figure}
\begin{center}
\includegraphics[clip,scale=0.3]{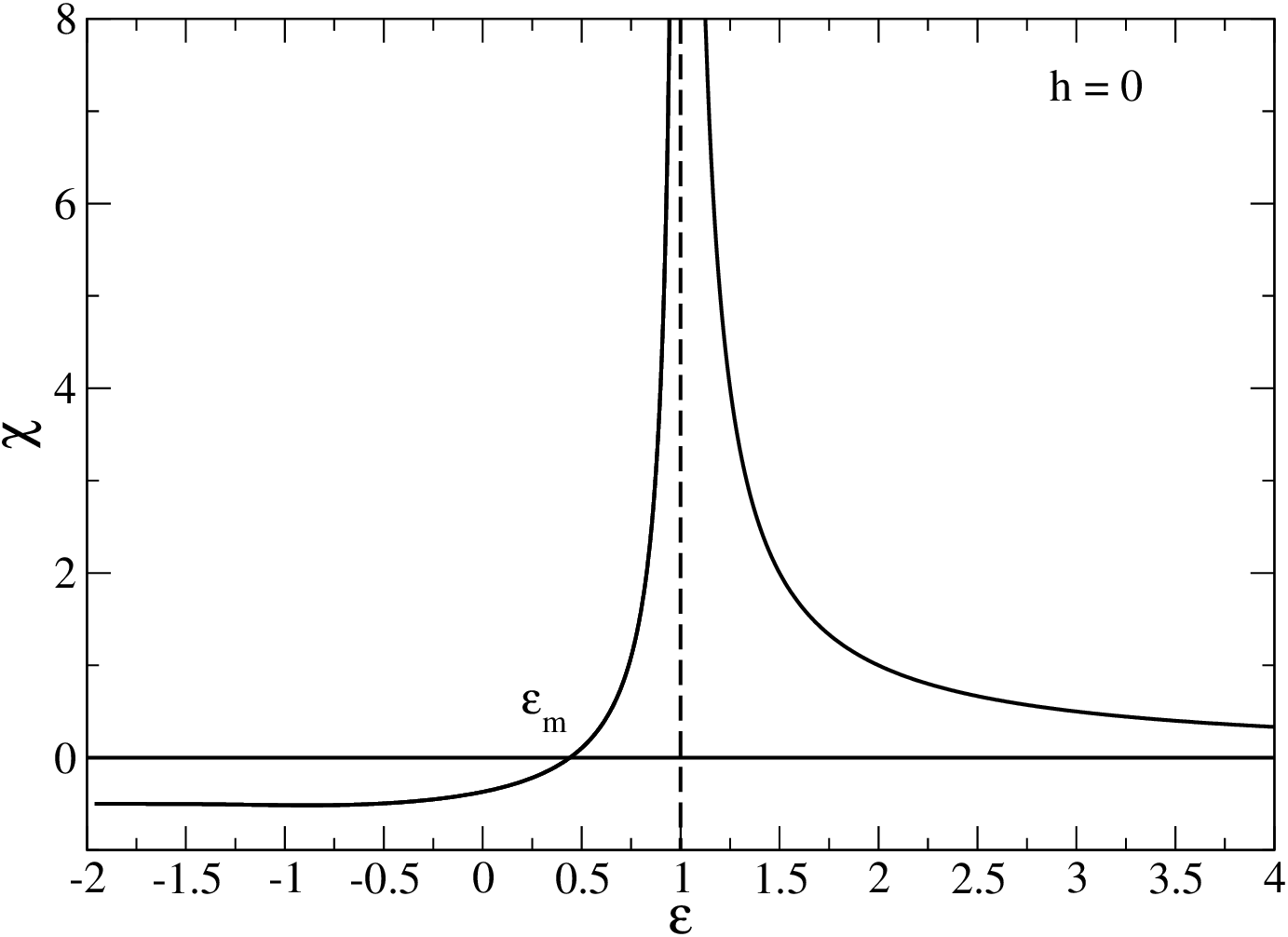}
\caption{Magnetic susceptibility $\chi$ as a function of the energy $\epsilon$ in the microcanonical ensemble for $h\rightarrow 0$. The susceptibility is negative for $\epsilon<\epsilon_m$.}
\label{epsilon-chi-h0}
\end{center}
\end{figure}

Let us now consider the microcanonical ensemble. For any energy $\epsilon>\epsilon_c=1$, we know that $b\rightarrow 0$ when $h\rightarrow 0$ \cite{cvb}. Therefore, we can expand equations (\ref{msm1}) and (\ref{msm2}) in powers of $x\rightarrow 0$ at fixed $\epsilon$. We obtain the equivalents
\begin{eqnarray}
\label{hzero8}
b(x)\sim \frac{x}{2},\qquad h(x)\sim \left (\epsilon-1\right )\frac{x}{2},
\end{eqnarray}
from which we deduce the expression of the magnetic susceptibility
\begin{eqnarray}
\label{hzero9}
\chi=\frac{1}{\epsilon-1}.
\end{eqnarray}
This equation is valid  for any $\epsilon>\epsilon_c=1$. In particular, the magnetic susceptibility diverges at the critical point ($\epsilon\rightarrow \epsilon_c^+=1^+$). For $\epsilon<\epsilon_c=1$ and $h\rightarrow 0$, the magnetic susceptibility $\chi(\epsilon)$ is given by equations (\ref{msm4}), (\ref{t7}) and (\ref{t8}) with $h=0$. These equations can be written
\begin{eqnarray}
\label{hzeor10}
\chi(x)=\frac{1-2xb(x)}{\frac{b(x)}{x b'(x)}-1+2xb(x)},
\end{eqnarray}
\begin{eqnarray}
\label{hzero11}
\epsilon(x)=\frac{2b(x)}{x}-2b(x)^2.
\end{eqnarray}
For $\epsilon\rightarrow \epsilon_c^{-1}=1^{-1}$, we can expand the previous equations in powers of $x\rightarrow 0$. We obtain the equivalents
\begin{eqnarray}
\label{hzero12}
1-\epsilon(x)\sim\frac{5x^2}{8}, \qquad \chi(x)\sim\frac{4}{5x^2},
\end{eqnarray}
so that the magnetic susceptibility diverges at the critical point ($\epsilon\rightarrow \epsilon_c^-=1^-$) like
\begin{eqnarray}
\label{hzero13}
\chi\sim \frac{1}{2(1-\epsilon)}.
\end{eqnarray}
Note that equations (\ref{hzero9}) and (\ref{hzero13}) differ by a factor $2$. The magnetic susceptibility is plotted as a function of $\epsilon$  in Figure \ref{epsilon-chi-h0}.

Finally, we compute the behaviour of the magnetization at the critical point when $h\rightarrow 0$. In the canonical ensemble, taking $\eta=\eta_c=1$ in equations (\ref{msc1}) and (\ref{msc2}), we obtain $b=I_{1}(x)/I_{0}(x)$ and $h=x/2-b(x)$. Considering now the limit $h\rightarrow 0$, corresponding to $x\rightarrow 0$, and using the asymptotic expansions of Appendix \ref{sec_asy}, we obtain $b\sim x/2$ and $h\sim x^3/16$ yielding
\begin{eqnarray}
\label{fin1}
b\sim (2h)^{1/3},
\end{eqnarray}
for $h\rightarrow 0$ at $\eta=\eta_c=1$. This corresponds to the
critical isotherm. In the microcanonical ensemble, taking $\epsilon=\epsilon_c=1$ in equations (\ref{msm1}) and (\ref{msm2}), then considering the limit $h\rightarrow 0$, corresponding to $x\rightarrow 0$, and using the asymptotic expansions of Appendix \ref{sec_asy}, we obtain $b\sim x/2$ and $h\sim 5x^3/16$ yielding
\begin{eqnarray}
\label{fin2}
b\sim \left (\frac{2h}{5}\right )^{1/3},
\end{eqnarray}
for $h\rightarrow 0$ at $\epsilon=\epsilon_c=1$. The exponent $1/3$ is
the same in the two ensembles.

\section{Conclusion}
\label{sec_conclusion}

In this paper, we have studied the thermodynamics of the HMF model
with a magnetic field. In the absence of magnetic field, the
Hamilonian of the HMF model is invariant under the translation
operation $\theta_i\rightarrow \theta_i+\phi$ which is equivalent to
the $U(1)$ rotational symmetry. The $U(1)$ symmetry is spontaneously
broken by the inclusion of an external magnetic field ${\bf H}$. The
fully stable states are aligned with the magnetic field
($BH>0$). Their specific heats are always positive (in canonical and
microcanonical ensembles) but they undergo a discontinuity at the
critical point $(\epsilon_c,\eta_c)=(1,1)$ when $H=0$. Their magnetic
susceptibilities are positive in the canonical ensemble but they are
negative in the microcanonical ensemble for $\epsilon_{min}(h)\le
\epsilon<\epsilon_m\simeq 0.44025$ and they diverge at the critical
point $(\epsilon_c,\eta_c)=(1,1)$ when $H=0$. We have also found
metastable states that are anti-aligned with the magnetic field
($BH<0$). These states have negative specific heats in the
microcanonical ensemble for
$\epsilon(\eta_c(h))<\epsilon<\epsilon_c(h)$ leading to ensembles
inequivalence and zeroth order phase transitions similar to the
isothermal collapse and the gravothermal catastrophe in
astrophysics. Their magnetic susceptibilities are positive in the
canonical ensemble but they are negative in the microcanonical
ensemble for $\epsilon'_{min}(h)\le \epsilon<\epsilon_m$ when
$0<h<h_m\simeq 0.2649$ and for $\epsilon'_{min}(h)\le
\epsilon<\epsilon_c(h)$ when $h_m<h<1$. The magnetic curve $B(H)$
displays therefore hysteretic cycles involving positive or negative
magnetic susceptibilities. We have also shown the existence of gaps in
the magnetization leading to ergodicity breaking. As a result, the HMF
model with a magnetic field presents a rich phenomenology. It would be
interesting to compare these theoretical predictions with direct
numerical simulations. Unfortunately, we have shown that the
``metastable'' states are in fact {\it unstable} with respect to
perturbations that change the phase of the magnetization. Since all
types of perturbations should be considered in principle, this throws
doubts on their physical relevance. This is a pity because much of the
richness of the problem disappears. Indeed, if we consider all types
of perturbations, only fully stable states remain and the system does
not display any phase transition. However, if we consider only
perturbations that are symmetric with respect to the axis determined
by the magnetic field ${\bf H}$, we should be able to observe the
phase transitions between aligned (stable) and anti-aligned
(metastable) states described in this paper.

\appendix

\section{Asymptotic expansions}
\label{sec_asy}

In this section, we give the asymptotic expansions of $b(x)$, $\eta(x)$ and $\epsilon(x)$ and $\chi(x)$ for $x\rightarrow 0$, $x\rightarrow +\infty$, $x\rightarrow -\infty$ and $x\rightarrow x_0(h)$.

For $x\rightarrow 0$, we have
\begin{eqnarray}
\label{asy1}
b(x)=\frac{x}{2}-\frac{x^3}{16}+\frac{x^5}{96}+o(x^6),
\end{eqnarray}
\begin{eqnarray}
\label{asy2}
\eta(x)=\frac{x}{2h}-\frac{x^2}{4h^2}+\frac{x^3}{8h^3}+\frac{(h^2-2)}{32 h^4}x^4+o(x^5),
\end{eqnarray}
\begin{eqnarray}
\label{asy3}
\epsilon(x)=\frac{2h}{x}+1-2hx-\frac{5x^2}{8}+\frac{hx^3}{4}+\frac{7x^4}{48}+o(x^5),\nonumber\\
\end{eqnarray}
\begin{eqnarray}
\label{asy4}
\chi_{cano}(x)=\frac{x}{2h}-\frac{3x^3}{16h}-\frac{x^4}{16h^2}+o(x^5),
\end{eqnarray}
\begin{eqnarray}
\label{asy5}
\chi_{micro}(x)=\frac{x}{2h}-\frac{19x^3}{16h}-\frac{5x^4}{16 h^2}+o(x^{5}).
\end{eqnarray}

For $x\rightarrow +\infty$, we have
\begin{eqnarray}
\label{asy6}
b(x)=1-\frac{1}{2x}-\frac{1}{8x^2}+o(x^{-3}),
\end{eqnarray}
\begin{eqnarray}
\label{asy7}
\eta(x)=\frac{x}{2(1+h)}+\frac{1}{4(1+h)^2}+\frac{3+h}{16(1+h)^3x}+o(x^{-2}),\nonumber\\
\end{eqnarray}
\begin{eqnarray}
\label{asy8}
\epsilon(x)=-2-4h+\frac{4(1+h)}{x}-\frac{2-h}{2x^2}+o(x^{-3}),
\end{eqnarray}
\begin{eqnarray}
\label{asy9}
\chi_{cano}(x)=\frac{1}{2(1+h)x}+\frac{3+h}{4(1+h)^2}\frac{1}{x^2}+o(x^{-3}),
\end{eqnarray}
\begin{eqnarray}
\label{asy10}
\chi_{micro}(x)=-\frac{1}{2(1+h)}+\frac{3h}{8(1+h)^2x}+o(x^{-2}).
\end{eqnarray}

For $x\rightarrow -\infty$, we have
\begin{eqnarray}
\label{asy11}
b(x)=-1-\frac{1}{2x}+\frac{1}{8x^2}+o(x^{-3}),
\end{eqnarray}
\begin{eqnarray}
\label{asy12}
\eta(x)=-\frac{x}{2(1-h)}+\frac{1}{4(1-h)^2}+\frac{3-h}{16(1-h)^3x}+o(x^{-2}),\nonumber\\
\end{eqnarray}
\begin{eqnarray}
\label{asy13}
\epsilon(x)=-2+4h-\frac{4(1-h)}{x}-\frac{2-h}{2x^2}+o(x^{-3}),
\end{eqnarray}
\begin{eqnarray}
\label{asy14}
\chi_{cano}(x)=-\frac{1}{2(1-h)x}+\frac{3-h}{4(1-h)^2}\frac{1}{x^2}+o(x^{-3}),
\end{eqnarray}
\begin{eqnarray}
\label{asy15}
\chi_{micro}(x)=-\frac{1}{2(1-h)}+\frac{3h}{8(1-h)^2x}+o(x^{-2}).
\end{eqnarray}

For $x\rightarrow x_0(h)$, we find that
\begin{eqnarray}
\label{asy16}
b(x)=b_0(h)-b'(x_0(h))(x_0(h)-x)+...,
\end{eqnarray}
\begin{eqnarray}
\label{asy17}
\eta(x)\sim -\frac{x_0(h)}{2b'(x_0(h))(x_0(h)-x)},
\end{eqnarray}
\begin{eqnarray}
\label{asy18}
\epsilon(x)=\epsilon_0(h)-\frac{2b'(x_0(h))}{x_0(h)}(x_0(h)-x)+...
\end{eqnarray}

\section{Absence of critical point with $B_y\neq 0$}
\label{sec_abs}

The density profile corresponding to the distribution function (\ref{c5}) is of the form
\begin{eqnarray}
\label{abs1}
\rho(\theta)=A\, e^{\beta\left \lbrack (B_x+H)\cos\theta+B_y\sin\theta\right \rbrack}.
\end{eqnarray}
Let us write $B_x+H=r\cos\phi$ and $B_y=r\sin\phi$ with $r=\left\lbrack (B_x+H)^2+B_y^2\right\rbrack^{1/2}$. Then, equation (\ref{abs1}) becomes
\begin{eqnarray}
\label{abs2}
\rho(\theta)=\frac{M}{2\pi I_{0}(\beta r)} e^{\beta r \cos(\theta-\phi)},
\end{eqnarray}
where we have used equation (\ref{m8}) to determine the amplitude. Substituting equation (\ref{abs2}) in equations (\ref{m6}) and (\ref{m7}), we obtain the self-consistency relations
\begin{eqnarray}
\label{abs3}
\frac{2\pi B_x}{kM}=\frac{I_1(\beta r)}{I_{0}(\beta r)}\cos\phi,
\end{eqnarray}
\begin{eqnarray}
\label{abs4}
\frac{2\pi B_y}{kM}=\frac{I_1(\beta r)}{I_{0}(\beta r)}\sin\phi,
\end{eqnarray}
determining the magnetization components $B_x$ and $B_y$. Assuming $B_y\neq 0$, we obtain
\begin{eqnarray}
\label{abs5}
\frac{B_x+H}{B_y}=\frac{\cos\phi}{\sin\phi}=\frac{B_x}{B_y}.
\end{eqnarray}
This relation cannot be satisfied unless $H=0$. Therefore, there is no critical point of entropy at fixed mass and energy or critical point of free energy at fixed mass with $B_y\neq 0$.

\section{Eigenvalue equation}
\label{sec_inho}

In \cite{cvb}, it is shown that the thermodynamical stability problem can be reduced to the study of an eigenvalue equation. The eigenvalue equation associated with the maximization problem (\ref{m13}) can be written
\begin{eqnarray}
T\frac{d}{d\theta}\left (\frac{1}{\rho}\frac{dq}{d\theta}\right )+\frac{k}{2\pi}\int_0^{2\pi}q(\theta')\cos(\theta-\theta')\, d\theta'\nonumber\\
=\frac{2V}{MT}\frac{d\Phi_{tot}}{d\theta}+2T\lambda q,
\label{inho1}
\end{eqnarray}
with
\begin{eqnarray}
V=\int_{0}^{2\pi}\frac{d\Phi_{tot}}{d\theta}q(\theta)\,d\theta,
\label{v1}
\end{eqnarray}
and $q(0)=q(2\pi)=0$, where
$q(\theta)=\int_0^{\theta}\delta\rho(\theta')d\theta'$ is the
perturbed integrated density. This corresponds to equations (54) and (55) of \cite{cvb} where we have incorporated the effect of the magnetic field in the potential ($\Phi_{tot}=\Phi+\Phi_{ext}=-(B+H)\cos\theta$). A critical point of entropy at fixed mass and energy is a maximum iff all the eigenvalues $\lambda$ are negative and it is a saddle point if at least one of these eigenvalues is positive. Here, we shall determine the point of marginal stability corresponding to $\lambda=0$. We follow a method similar to the one developed in \cite{cd} and \cite{cc}.

Taking $\lambda=0$ in the eigenvalue equation (\ref{inho1}), we obtain
\begin{eqnarray}
T\frac{d}{d\theta}\left (\frac{1}{\rho}\frac{dq}{d\theta}\right )-\delta B_y \cos\theta+\delta B_x\sin\theta
-\frac{2V}{MT}\frac{d\Phi_{tot}}{d\theta}=0,\nonumber\\
\label{inho2}
\end{eqnarray}
with
\begin{eqnarray}
-\delta B_y=\frac{k}{2\pi}\int_{0}^{2\pi}q(\theta)\cos\theta\, d\theta=-\frac{k}{2\pi}\int_{0}^{2\pi}q'(\theta)\sin\theta\, d\theta,\nonumber\\
\label{inho3}
\end{eqnarray}
\begin{eqnarray}
\delta B_x=\frac{k}{2\pi}\int_{0}^{2\pi}q(\theta)\sin\theta\, d\theta=\frac{k}{2\pi}\int_{0}^{2\pi}q'(\theta)\cos\theta\, d\theta,\nonumber\\
\label{inho4}
\end{eqnarray}
where we have used an integration by parts to obtain the second
equalities. Equation (\ref{inho2}) can be integrated once
to yield
\begin{eqnarray}
\delta\rho=\frac{dq}{d\theta}=\frac{\rho}{T}(\delta B_y \sin\theta+ \delta B_x \cos\theta+C)\nonumber\\
-\frac{2V}{MT^2}(B+H)\rho\cos\theta,
\label{inho5}
\end{eqnarray}
where $C$ is a constant of integration. Another integration with the boundary condition $q(0)=0$ yields
\begin{eqnarray}
q(\theta)=\frac{\delta B_y}{T}\int_{0}^{\theta} \rho \sin\theta'\, d\theta'
+\frac{\delta B_x}{T} \int_{0}^{\theta} \rho \cos\theta'\, d\theta' \nonumber\\
+\frac{C}{T} \int_{0}^{\theta} \rho \, d\theta'-\frac{2V}{MT^2}(B+H)\int_{0}^{\theta} \rho\cos\theta'\, d\theta'.
\label{inho6}
\end{eqnarray}
Then, the condition $q(2\pi)=0$ determines the constant
\begin{eqnarray}
C=\frac{-\delta B_x+\frac{2V}{MT}(B+H)}{M} \int_0^{2\pi}\rho \cos\theta\, d\theta,
\label{inho7}
\end{eqnarray}
where we have used the fact  that the density profile $\rho(\theta)$ is
symmetric with respect to $\theta=0$ to simplify some terms. Using an integration by parts, and the expression of $\Phi_{tot}$, the variable $V$ can be written
\begin{eqnarray}
V=(H+B)\int_{0}^{2\pi} q'(\theta)\cos\theta\,d\theta.
\label{v2}
\end{eqnarray}
Substituting $dq/d\theta$ from equation (\ref{inho5}) in equation (\ref{v2}) and solving for $V$, we obtain
\begin{eqnarray}
V=\frac{H+B}{T}\frac{\delta B_x\int_{0}^{2\pi} \rho \cos^2\theta\, d\theta+C\int_{0}^{2\pi} \rho \cos\theta\, d\theta}{1+\frac{2}{MT^2}(B+H)^2\int_{0}^{2\pi} \rho \cos^2\theta\, d\theta}.\nonumber\\
\label{v3}
\end{eqnarray}
Finally, substituting this expression in equation (\ref{inho7}) and solving for $C$, we get
\begin{eqnarray}
C=-\frac{\delta B_x \int_{0}^{2\pi} \rho \cos\theta\, d\theta}{D},
\label{ccc1}
\end{eqnarray}
where
\begin{eqnarray}
D\equiv M+\frac{2}{T^2}(B+H)^2\int_{0}^{2\pi} \rho \cos^2\theta\, d\theta\nonumber\\
-\frac{2}{MT^2}(B+H)^2\left (\int_{0}^{2\pi} \rho \cos\theta\, d\theta\right )^2.
\label{c1bv}
\end{eqnarray}
Substituting equation (\ref{inho5}) into equation (\ref{inho3}),
we find either that $\delta B_y=0$ or that
\begin{eqnarray}
1=\frac{k}{2\pi T}\int_{0}^{2\pi} \rho \sin^2\theta\, d\theta.
\label{inho8}
\end{eqnarray}
Introducing the dimensionless variables defined in Section \ref{sec_t} and using the relation
\begin{eqnarray}
\frac{1}{M}\int_{0}^{2\pi} \rho \sin^2\theta\, d\theta=\frac{b}{x},
\label{qh3}
\end{eqnarray}
that can be derived from the density profile (\ref{c11}),  we find that the condition (\ref{inho8}) can be rewritten
\begin{eqnarray}
1=2\eta\frac{b}{x}.
\label{qh2}
\end{eqnarray}
When $h=0$, comparing equation (\ref{qh2}) with equation (\ref{t7}), we see that this condition is always satisfied. Therefore, if we restrict ourselves to perturbations such that $\delta B_x=0$ and $\delta B_y\neq 0$, we conclude that the system is always marginally stable \cite{cd}. Such perturbations correspond to a variation of the phase of the magnetization vector ${\bf B}$. This is just a mere rotation of the equilibrium profile. Since the HMF model without magnetic field is invariant by rotation, these perturbations do not change the entropy. By contrast, when $h\neq 0$, this $U(1)$ symmetry is broken and the condition (\ref{qh2}) can never be fulfilled. In fact, the aligned phase is always stable with respect to perturbations of the form   $\delta B_x=0$ and $\delta B_y\neq 0$, while the anti-aligned phase is always unstable by such perturbations (see Appendix \ref{sec_lm}). There is therefore no marginal point corresponding to $\lambda=0$.

Substituting equation (\ref{inho5}) into equation (\ref{inho4}), and using the expressions (\ref{v3}) and  (\ref{ccc1}) of $V$ and $C$ we find either that $\delta B_x=0$ or that
\begin{eqnarray}
1=\frac{k}{2\pi T}\int_{0}^{2\pi} \rho{\cos^2\theta}\, d\theta-\frac{k(B+H)^2}{\pi MT^3E}\left (\int_{0}^{2\pi} \rho\cos^2\theta\, d\theta\right )^2\nonumber\\
-\frac{k}{2\pi T D E}\left (\int_{0}^{2\pi} \rho\cos\theta\, d\theta\right )^2,\qquad\qquad
\label{inho9}
\end{eqnarray}
where
\begin{eqnarray}
E\equiv 1+\frac{2}{MT^2}(B+H)^2\int_{0}^{2\pi} \rho \cos^2\theta\, d\theta.
\label{c1bvw}
\end{eqnarray}
Introducing the dimensionless variables defined in Section \ref{sec_t} and using the relations
\begin{eqnarray}
\frac{1}{M}\int_{0}^{2\pi} \rho \cos\theta\, d\theta=b,
\label{qh2b}
\end{eqnarray}
\begin{eqnarray}
I\equiv \frac{1}{M}\int_{0}^{2\pi} \rho \cos^2\theta\, d\theta=1-\frac{b}{x},
\label{qh1}
\end{eqnarray}
that can be derived from the density profile (\ref{c11}),  we find that the condition (\ref{inho9}) can be rewritten
\begin{eqnarray}
1=2(\eta-x^2)\left (1-\frac{b}{x}\right )-\frac{2\eta b^2}{1+2x^2b'(x)},
\label{qhw}
\end{eqnarray}
where we have used equations (\ref{t7}) and (\ref{msc5}) to simplify some terms. We can easily check that $\eta(x)-x^2=0$ is not solution of this equation. Then, after some transformations (involving a division by $\eta-x^2=0$), we find that equation (\ref{qhw}) is equivalent to
\begin{eqnarray}
\left\lbrack 1-2(\eta-x^2) b'(x)\right \rbrack \left \lbrack 1-\frac{b(x)}{x}+\frac{1}{2x^2}\right \rbrack=0.
\label{az}
\end{eqnarray}
We have checked that the last term is strictly positive. Therefore, the criterion (\ref{inho9}) is equivalent to
\begin{eqnarray}
1-2(\eta-x^2) b'(x)=0.
\label{az2}
\end{eqnarray}
For $h\neq 0$, this relation selects the turning point of energy $\epsilon_c(h)$ of the anti-aligned phase (see equation (\ref{ms10})). For $h=0$, the solution of equation (\ref{az2}) is $x=0$ which corresponds to the bifurcation point $\epsilon_c=1$. Therefore, the condition of marginal stability in the microcanonical ensemble ($\lambda=0$) coincides with the turning point of energy (if $h\neq 0$) or with the bifurcation point (if $h=0$) in agreement with the Poincar\'e theorem.

Let us now consider the canonical ensemble. To that purpose, it suffices to take $V=0$ in the foregoing expressions. This yields equation (\ref{inho8}) and
\begin{eqnarray}
1=\frac{k}{2\pi T}\int_{0}^{2\pi} \rho \cos^2\theta\, d\theta
-\frac{k}{2\pi T M}\left (\int_{0}^{2\pi} \rho \cos\theta\, d\theta\right )^2.\nonumber\\
\label{inho10}
\end{eqnarray}
This is a particular case of equations (F.8) and (F.9) in \cite{cd}. Introducing the dimensionless variables defined in Section \ref{sec_t} and using the relations (\ref{qh2}), (\ref{qh1}) and (\ref{msc5}), we obtain
\begin{eqnarray}
1-2\eta b'(x)=0.
\label{az2b}
\end{eqnarray}
For $h\neq 0$, this relation selects the turning point of temperature $\eta_c(h)$ of the anti-aligned phase (see equation (\ref{cs8})). For $h=0$, the solution of equation (\ref{az2b}) is $x=0$ which corresponds to the bifurcation point $\eta_c=1$. Therefore, the condition of marginal stability in the canonical ensemble ($\lambda=0$) coincides with the turning point of temperature (if $h\neq 0$) or with the bifurcation point (if $h=0$) in agreement with the Poincar\'e theorem.

\section{Local equivalence of the variational problems}
\label{sec_lm}

In this Appendix, we study the local equivalence of the variational problems (\ref{f2}) and (\ref{f11}) in the canonical ensemble and of the variational problems (\ref{sb3}) and (\ref{sb14}) in the microcanonical ensemble.

\subsection{Canonical ensemble}
\label{sec_lmc}

A critical point of (\ref{f2}) is determined by the variational principle $\delta F+\alpha T\delta M=0$
where $\alpha$ is a Lagrange multiplier accounting for the conservation of mass. This leads to the distribution (\ref{c11}). The magnetization $B$ is obtained by substituting equation (\ref{c11}) in equation (\ref{m6}) leading to the self-consistency relation (\ref{c9}). Using the results of \cite{isostab,yukawa} concerning the second variations of free energy, and introducing the dimensionless variables of Section \ref{sec_t}, this critical point is a (local) minimum of $F$ at fixed mass iff
\begin{eqnarray}
\delta^2 f=-2((\delta b_x)^2+(\delta b_y)^2)+\frac{1}{\eta M}\int \frac{(\delta\rho)^2}{\rho}\, d\theta > 0,\quad
\label{lmc3b}
\end{eqnarray}
for all perturbations $\delta\rho$ that conserve mass: $\int \delta\rho\, d\theta=0$. The corresponding variations of magnetization are
\begin{eqnarray}
\delta b_x=\frac{1}{M}\int \delta\rho\cos\theta\, d\theta,
\label{lmc4a}
\end{eqnarray}
\begin{eqnarray}
\delta b_y=\frac{1}{M}\int \delta\rho\sin\theta\, d\theta.
\label{lmc4b}
\end{eqnarray}
We can always write the perturbations in the form
\begin{eqnarray}
\delta\rho=\delta\rho_{\|}+\delta\rho_{\perp},
\label{lmc5}
\end{eqnarray}
where $\delta\rho_{\|}=(\mu+\nu_x\cos\theta+\nu_y\sin\theta)\rho$ and  $\delta\rho_{\perp}\equiv \delta\rho-\delta\rho_{\|}$. The second condition ensures that all the perturbations are considered.  We shall now {\it choose}
the constants $\mu$, $\nu_x$ and $\nu_y$ such that
\begin{eqnarray}
\int \delta\rho_{\|}\, d\theta=0,
\label{lmc6}
\end{eqnarray}
\begin{eqnarray}
\delta b_x=\frac{1}{M}\int
\delta\rho_{\|}\cos\theta\, d\theta,
\label{lmc6b}
\end{eqnarray}
\begin{eqnarray}
\delta b_y=\frac{1}{M}\int
\delta\rho_{\|}\sin\theta\, d\theta.
\label{lmc6c}
\end{eqnarray}
This implies that
\begin{eqnarray}
\int \delta\rho_{\perp}\,
d\theta=0,
\label{lmc7}
\end{eqnarray}
\begin{eqnarray}
\int
\delta\rho_{\perp}\cos\theta\, d\theta=0, \qquad \int
\delta\rho_{\perp}\sin\theta\, d\theta=0.
\label{lmc7b}
\end{eqnarray}
The conditions (\ref{lmc6}), (\ref{lmc6b}) and (\ref{lmc6c}) lead to the relations
\begin{eqnarray}
\mu +\nu_x b=0,
\label{lmc8c}
\end{eqnarray}
\begin{eqnarray}
\delta b_x=\mu b+\nu_x I,
\label{lmc9c}
\end{eqnarray}
\begin{eqnarray}
\delta b_y=\nu_y (1-I),
\label{lmc9bc}
\end{eqnarray}
where we have  defined
\begin{eqnarray}
I\equiv \frac{1}{M}\int \rho\cos^2\theta\, d\theta.
\label{i}
\end{eqnarray}
This forms a system of three algebraic equations that determines the three constants $\mu$, $\nu_x$ and $\nu_y$. Using the equilibrium density profile (\ref{c11}), and the identity
\begin{eqnarray}
I_{n-1}(x)-I_{n+1}(x)=\frac{2n}{x}I_{n}(x),
\label{idw}
\end{eqnarray}
we find that
\begin{eqnarray}
I=1-\frac{b}{x}=b'(x)+b^2,
\label{lmc10}
\end{eqnarray}
where we have used the equations  (\ref{t6}) and (\ref{msc5}).
Solving equations (\ref{lmc8c}), (\ref{lmc9c}) and (\ref{lmc9bc}) for $\mu$, $\nu_x$ and $\nu_y$, and using the result (\ref{lmc10}), we find that
\begin{eqnarray}
\nu_x=\frac{\delta b_x}{b'(x)},\qquad \mu=-\frac{b}{b'(x)}\delta b_x,
\label{lmc16a}
\end{eqnarray}
\begin{eqnarray}
\nu_y=\frac{x}{b}\delta b_y.
\label{lmc16b}
\end{eqnarray}
Therefore, the perturbation $\delta\rho_{\|}$ takes the form
\begin{eqnarray}
\delta\rho_{\|}=-\frac{1}{b'(x)}\rho(\theta) (b-\cos\theta)\delta b_x
+\frac{x}{b}\rho(\theta)\sin\theta\delta b_y.\qquad
\label{lmc17}
\end{eqnarray}
By construction,  $\delta\rho_{\|}$ and $\delta\rho_{\perp}$ are orthogonal for the scalar product weighted by $1/\rho$ in the sense that
\begin{eqnarray}
\int \frac{\delta\rho_{\|}\delta\rho_{\perp}}{\rho}\, d\theta=0.
\label{lmc18}
\end{eqnarray}
Indeed, we have
\begin{eqnarray}
\int \frac{\delta\rho_{\|}\delta\rho_{\perp}}{\rho}\, d\theta=\int (\mu+\nu_x\cos\theta+\nu_y\sin\theta)\delta\rho_{\perp}\, d\theta=0,\nonumber\\
\label{lmc19}
\end{eqnarray}
where we have used equations (\ref{lmc7}) and (\ref{lmc7b}) to get the last equality. As a result, we obtain
\begin{eqnarray}
\int \frac{(\delta\rho)^2}{\rho}\, d\theta=\int \frac{(\delta\rho_{\|})^2}{\rho}\, d\theta+\int \frac{(\delta\rho_{\perp})^2}{\rho}\, d\theta.
\label{lmc20}
\end{eqnarray}
Using equations (\ref{lmc17}) and (\ref{lmc10}), we obtain after simplification
\begin{eqnarray}
\frac{1}{M}\int \frac{(\delta\rho_{\|})^2}{\rho}\, d\theta=\frac{1}{b'(x)}(\delta b_x)^2+\frac{x}{b}(\delta b_y)^2.
\label{lmc21}
\end{eqnarray}
Therefore, the second order variations of free energy given by equation (\ref{lmc3b}) can be written
\begin{eqnarray}
\delta^2f=\left\lbrack \frac{1}{\eta b'(x)}-2\right\rbrack (\delta b_x)^2
+\left (\frac{x}{\eta b}-2\right )(\delta b_y)^2\nonumber\\
+\frac{1}{\eta M}\int \frac{(\delta\rho_{\perp})^2}{\rho}\, d\theta.
\label{lmc22}
\end{eqnarray}
Using equations (\ref{t7}) and (\ref{cs6}), we finally obtain
\begin{eqnarray}
\delta^2 f=\frac{1}{2}f''(b)(\delta b_x)^2+\frac{2h}{b}(\delta b_y)^2
+\frac{1}{\eta M}\int \frac{(\delta\rho_{\perp})^2}{\rho}\, d\theta.\nonumber\\
\label{lmc23}
\end{eqnarray}
For the aligned phase ($h b>0$), we conclude from this expression that $\rho$ is a (local) minimum of $F[\rho]$ at fixed mass iff $b$ is a (local) minimum of $f(b)$ (the argument is essentially the same as the one given in \cite{isostab}). On the other hand, when $h\neq 0$, the anti-aligned phase ($h b<0$) is always unstable with respect to (odd) perturbations that change the phase of the magnetization, i.e. $\delta b_y\neq 0$ (it suffices to take $\delta b_x=0$ and $\delta \rho_{\perp}=0$). Note that, for $h=0$, this instability disappears due to the $U(1)$ rotational symmetry of the system. Finally, if we {\it impose} $\delta b_y=0$, i.e. if we restrict ourselves to even perturbations, then we see from equation (\ref{lmc23}) that  $\rho$ is a (local) minimum of $F[\rho]$ at fixed mass iff $b$ is a (local) minimum of $f(b)$. Combining these results with those of Section \ref{sec_cano}, we conclude that the states (S) are always stable, the states (U) are always unstable and the states (M) are metastable for perturbations with  $\delta b_y=0$ while they are unstable for perturbations with $\delta b_y\neq 0$.

We can also obtain these results by a slightly different method inspired by \cite{campastab}. Considering even perturbations, the stability condition (\ref{lmc3b}) becomes
\begin{eqnarray}
\delta^2 f=-2(\delta b_x)^2+\frac{1}{\eta M}\int \frac{(\delta\rho)^2}{\rho}\, d\theta > 0,\quad
\label{lu1}
\end{eqnarray}
for all perturbations $\delta\rho$ that conserve mass: $\int \delta\rho\, d\theta=0$. We see that when $\int\delta\rho\cos\theta\, d\theta=0$, this condition is automatically verified. Therefore, we can restrict ourselves to perturbations such that $\int\delta\rho\cos\theta\, d\theta\neq 0$. On the other hand since the functional $\delta^2f$ is quadratic in $\delta\rho$, we can impose $\int\delta\rho\cos\theta\, d\theta=1$ without loss of generality. We now look for the perturbation $\delta\rho$ that minimizes $\delta^2 f$ with the constraints  $\int \delta\rho\, d\theta=0$ and $\int\delta\rho\cos\theta\, d\theta=1$. Introducing Lagrange multipliers $\mu$ and $\nu_x$, we find that this perturbation is $\delta \rho=(\mu+\nu_x\cos\theta)\rho$ (it really corresponds to the minimum of $\delta^2f$ since $\delta^2(\delta^2f)=1/(\eta M\rho)>0$). Using the constraints, we obtain $\nu_x=1/(I-b^2)$ and $\mu=-b/(I-b^2)$. Finally, substituting this perturbation is equation (\ref{lu1}), we obtain $(\delta^2 f)_{min}=1/\lbrack \eta(I-b^2)\rbrack-2$. Therefore, inequality (\ref{lu1}) is satisfied iff $1/\lbrack \eta(I-b^2)\rbrack-2\ge 0$. We can check that this condition corresponds to criterion (138) of \cite{campastab}. On the other hand, the equality $1/\lbrack \eta(I-b^2)\rbrack-2=0$ (marginal stability) returns equation (\ref{inho10}). Finally, using equations (\ref{lmc10}) and (\ref{cs6}), we find that $(\delta^2 f)_{min}=1/(\eta b'(x))-2=\frac{1}{2}f''(b)$. Therefore, inequality (\ref{lu1}) is satisfied iff $b$ is a minimum of free energy $f(b)$. Considering now odd perturbations, the stability condition (\ref{lmc3b}) becomes
\begin{eqnarray}
\delta^2 f=-2(\delta b_y)^2+\frac{1}{\eta M}\int \frac{(\delta\rho)^2}{\rho}\, d\theta > 0,\quad
\label{lu2}
\end{eqnarray}
for all perturbations $\delta\rho$. We see that when $\int\delta\rho\sin\theta\, d\theta=0$, this condition is automatically verified. Therefore, using the same argument as above, we can restrict ourselves to perturbations such that $\int\delta\rho\sin\theta\, d\theta=1$. We now look for the perturbation $\delta\rho$ that minimizes $\delta^2 f$ with the constraint $\int\delta\rho\sin\theta\, d\theta=1$. Introducing a Lagrange multiplier $\nu_y$, we find that this perturbation is $\delta \rho=\nu_x\sin\theta\rho$ (it really corresponds to the minimum of $\delta^2f$ since $\delta^2(\delta^2f)=1/(\eta M\rho)>0$). Using the constraint, we obtain $\nu_y=1/(1-I)$. Finally, substituting this perturbation is equation (\ref{lu2}), we obtain $(\delta^2 f)_{min}=1/\lbrack \eta(1-I)\rbrack-2$. Therefore, inequality (\ref{lu2}) is satisfied iff $1/\lbrack \eta(1-I)\rbrack-2\ge 0$. We can check that this condition corresponds to criterion (126) of \cite{campastab}. On the other hand, the equality $1/\lbrack \eta(1-I)\rbrack-2=0$ (marginal stability) returns equation (\ref{inho8}). Finally, using equations (\ref{lmc10}) and (\ref{t7}), we find that $(\delta^2 f)_{min}=x/(\eta b)-2=2h/b$. Therefore, inequality (\ref{lu2}) is satisfied iff $bh\ge 0$.

\subsection{Microcanonical ensemble}
\label{sec_lmm}

A critical point of (\ref{sb3}) is determined by the variational principle $\delta S-\alpha \delta M=0$,
where $\alpha$ is a Lagrange multiplier accounting for the conservation of mass. This leads to the distribution (\ref{c11}) where the temperature is determined by the energy according to equation (\ref{sb2}). The magnetization $B$ is obtained by substituting equation (\ref{c11}) in equation (\ref{m6})  leading to the self-consistency relation (\ref{c9}). Using the results of \cite{isostab,yukawa} concerning the second variations of entropy appropriately generalized to take into account the magnetic field (this amounts to replacing $\Phi$ by $\Phi_{tot}$), and introducing the dimensionless variables of Section \ref{sec_t}, this critical point is a (local) maximum of $S$ at fixed mass  iff
\begin{eqnarray}
\delta^2 s=-\frac{1}{M}\int \frac{(\delta\rho)^2}{2\rho}\, d\theta+\eta \left\lbrack (\delta b_x)^2+(\delta b_y)^2\right\rbrack\nonumber\\
-4\eta^2 (b+h)^2(\delta b_x)^2<0,
\label{lmm3}
\end{eqnarray}
for all perturbations $\delta\rho$ that conserve mass: $\int \delta\rho\, d\theta=0$. We note that the second order variations of entropy (\ref{lmm3}) are related to the second order variations of free energy (\ref{lmc3b}) by
\begin{eqnarray}
\delta^2 s=-\frac{1}{2}\eta \delta^2f-4\eta^2(b+h)^2 (\delta b_x)^2.
\label{lmm4}
\end{eqnarray}
Writing the perturbation $\delta\rho$ in the form  (\ref{lmc5}) with equation (\ref{lmc17}) and using equations (\ref{lmc23}) and (\ref{ms11}), we obtain
\begin{eqnarray}
\delta^2 s=\frac{1}{2}s''(b)(\delta b_x)^2-\frac{\eta h}{b} (\delta b_y)^2-\frac{1}{2M}\int \frac{(\delta\rho_{\perp})^2}{\rho}\, d\theta.\nonumber\\
\label{lmm5b}
\end{eqnarray}
From this identity, we arrive at the same type of conclusions as in the canonical ensemble (see the paragraph following equation (\ref{lmc23})).

We can also proceed as explained at the end of Section \ref{sec_lmc}. Considering even perturbations, we find that $(\delta^2 s)_{max}=-\frac{1}{2}\eta (\delta^2f)_{min}-4\eta^2(b+h)^2=\frac{1}{2}s''(b)$ where we have used equation (\ref{ms11}) to get the last equality. Therefore, the system is microcanonically stable with respect to even perturbations iff $b$ is a maximum of entropy $s(b)$. We can check that this condition corresponds to criterion (134) of \cite{campastab}. On the other hand, the equality $s''(b)=0$ (marginal stability) returns equation (\ref{az2}). Considering odd perturbations, we find that $(\delta^2 s)_{max}=-\frac{1}{2}\eta (\delta^2f)_{min}=-\eta h/b$. Therefore, the system is microcanonically stable with respect to odd perturbations iff $bh>0$. The condition of marginal stability  $(\delta^2 s)_{max}=0$ returns equation (\ref{inho8}).

\section{Partition function and density of states}
\label{sec_heur}

In this Appendix, we compute the partition function and the density of states of the HMF model with a magnetic field and make the connection with the entropy  $s(b)$ and the free energy $f(b)$  of Sections \ref{sec_micro} and \ref{sec_cano} (a more general discussion is given in Appendix C of \cite{isostab}).

In the microcanonical ensemble, the accessible configurations (those
having the proper value of energy) are equiprobable. Therefore, the
 probability density of the configuration
$(\theta_1,v_1,...,\theta_N,v_N)$ is
$P_{N}(\theta_{1},v_1,...,\theta_{N},v_N)=\frac{1}{g(E)}\delta(E-H)$
where $g(E)$ is the density of states
\begin{eqnarray}
g(E)=\int \delta(E-H)\, d\theta_1 dv_1...d\theta_N dv_N.
\label{heur1}
\end{eqnarray}
In other words,  $g(E)dE$ gives the number of microstates with energy between $E$ and $E+dE$.
The entropy is defined by $S(E)=\ln
g(E)$.  Integrating over the velocities in equation (\ref{heur1}), a classical calculation leads to
 \begin{eqnarray}
g(E)=\frac{2\pi^{N/2}}{\Gamma\left (\frac{N}{2}\right )}\int \left\lbrack 2(E-U)\right \rbrack^{\frac{N-2}{2}}\, d\theta_1...d\theta_N,
\label{heur4}
\end{eqnarray}
where $U(\theta_1,...,\theta_N)$ is the potential energy (second and third terms in the r.h.s. of equation (\ref{m1})). Introducing the magnetization vector
\begin{eqnarray}
B_x=\frac{k}{2\pi}\sum_{i=1}^{N}\cos\theta_i, \quad B_y=\frac{k}{2\pi}\sum_{i=1}^{N}\sin\theta_i,
\label{heur7}
\end{eqnarray}
The potential energy can be expressed as
\begin{eqnarray}
U=-\frac{\pi B^2}{k}-\frac{2\pi}{k}HB_x+\frac{kN}{4\pi}.
\label{heur8}
\end{eqnarray}
If we impose the constraint $\sum_{i=1}^{N}\sin\theta_i=0$ (see Remark of Section \ref{sec_poincare}) and introduce the dimensionless variables of Section \ref{sec_t}, the density of states (\ref{heur4}) can be rewritten
\begin{eqnarray}
g(\epsilon)= \int  (\epsilon+2b^2+4bh)^{\frac{N-2}{2}} \Omega(b)\, db,
\label{heur10}
\end{eqnarray}
where $\Omega(b)$ denotes the unconditional number of microstates corresponding to the macrostate ${b}$. The calculation of this integral is classical \cite{cdr}. For $N\rightarrow +\infty$, we get
\begin{eqnarray}
\Omega(b)\sim e^{N\left\lbrack -b\lambda+\ln I_0(\lambda)\right \rbrack},
\label{magn15}
\end{eqnarray}
with
\begin{eqnarray}
b=\frac{I_{1}(\lambda)}{I_{0}(\lambda)}.
\label{magn13}
\end{eqnarray}
Inserting  equation (\ref{magn15}) in equation (\ref{heur10}), we obtain
\begin{eqnarray}
g(\epsilon)= \int e^{N\left\lbrack \frac{1}{2}\ln (\epsilon+2b^2+4bh)-b\lambda+\ln I_0(\lambda)\right\rbrack} \, db.
\label{magn17}
\end{eqnarray}
Therefore, the density of states (\ref{heur1}) can finally be written
\begin{eqnarray}
g(\epsilon)= \int  e^{Ns(b)}\, d{b},
\label{heur13}
\end{eqnarray}
where $s(b)$ is given by equation (\ref{ms1}). In the limit $N\rightarrow +\infty$, we can make the approximation $g(\epsilon)\sim e^{Ns(b_*)}$ where $b_*$ is the global maximum of $s(b)$. We finally obtain $\lim_{N\rightarrow +\infty}\frac{1}{N}S(\epsilon)=s(b_*)$.

In the canonical ensemble, the  probability density of the configuration $(\theta_1,v_1,...,\theta_N,v_N)$  is $P_{N}(\theta_{1},v_1,...,\theta_{N},v_N)=\frac{1}{Z(\beta)}e^{-\beta H}$ where $Z(\beta)$ is the partition function
\begin{eqnarray}
Z(\beta)=\int e^{-\beta H} \, d\theta_1 dv_1...d\theta_N dv_N.
\label{heur14}
\end{eqnarray}
The free energy is defined by $F(\beta)=-\frac{1}{\beta}\ln Z(\beta)$.
Integrating over the velocities in equation (\ref{heur14}), we get
\begin{eqnarray}
Z(\beta)=\left (\frac{2\pi}{\beta}\right )^{N/2}\int e^{-\beta U}\, d\theta_1...d\theta_N.
\label{heur15b}
\end{eqnarray}
Using equation (\ref{heur8}), imposing the constraint $\sum_{i=1}^{N}\sin\theta_i=0$ and introducing the dimensionless variables of Section \ref{sec_t} and the unconditional number of microstates corresponding to the macrostate $b$, the partition function can be rewritten
\begin{eqnarray}
Z(\eta)= \int e^{N \eta (b^2+2bh)} \Omega(b)\, db.
\label{heur16b}
\end{eqnarray}
Inserting  equation (\ref{magn15}) in equation (\ref{heur16b}), we get
\begin{eqnarray}
Z(\eta)= \int e^{-\frac{N\eta}{2}\left \lbrack -2b^2-4bh+\frac{2}{\eta}b\lambda-\frac{2}{\eta}\ln I_0(\lambda)\right\rbrack}\, db.
\label{heur16d}
\end{eqnarray}
Therefore, the partition function can finally be written
\begin{eqnarray}
Z(\eta)= \int  e^{-\frac{N\eta}{2}f(b)}\, d{b},
\label{heur13b}
\end{eqnarray}
where $f(b)$ is given by equation (\ref{cs1}). In the limit $N\rightarrow +\infty$, we can make the approximation $Z(\eta)\sim e^{-\frac{N\eta}{2}f(b_*)}$ where $b_*$ is the global minimum of $f(b)$. We finally obtain $\lim_{N\rightarrow +\infty}\frac{8\pi}{kM^2}F(\eta)=f(b_*)$.

\section{Distribution and variance of the magnetization}
\label{sec_magnetization}

In this Appendix, we determine the distribution and the variance of the magnetization
$b$ for the HMF model with a magnetic field and show its connection with the free energy $f(b)$
and the entropy $s(b)$ of Sections \ref{sec_cano} and \ref{sec_micro}.

The  probability density of the magnetization is defined by
\begin{eqnarray}
P(b)= \int\delta\left (Nb-\sum_{i}\cos\theta_i\right ) \delta\left (\sum_{i}\sin\theta_i\right ) \nonumber\\ \times P_N(\theta_1,...,\theta_N)\, d\theta_1...d\theta_N,
\label{magn1}
\end{eqnarray}
where we have imposed the constraint $b_y=\sum_{i=1}^N\sin\theta_i=0$
(see Remark of Section \ref{sec_poincare}) and written $b$ for
$b_x$. In the canonical ensemble, the $N$-body distribution function
is
$P_{N}(\theta_{1},v_1,...,\theta_{N},v_N)=\frac{1}{Z(\beta)}e^{-\beta
H}$. Integrating over the velocities, we obtain
\begin{eqnarray}
P_N(\theta_1,...,\theta_N)=\frac{1}{Z(\beta)}\left (\frac{2\pi}{\beta}\right )^{N/2}e^{-\beta U},
\label{magn3}
\end{eqnarray}
where $U$ is the potential energy. In the microcanonical ensemble, the $N$-body distribution function is $P_{N}(\theta_{1},v_1,...,\theta_{N},v_N)=\frac{1}{g(E)}\delta(E-H)$. Integrating over the velocities, a classical calculation gives
\begin{eqnarray}
P_N(\theta_1,...,\theta_N)=\frac{1}{g(E)}\frac{2\pi^{N/2}}{\Gamma\left (\frac{N}{2}\right )}\left\lbrack 2(E-U)\right \rbrack^{\frac{N-2}{2}}.
\label{magn2}
\end{eqnarray}
Recalling that the potential energy can be expressed in terms of the magnetization according to equation (\ref{heur8}), and introducing the dimensionless variables of Section \ref{sec_t}, the  probability density of the magnetization in the canonical and microcanonical ensembles is given by
\begin{eqnarray}
P_{CE}(b)=\frac{1}{Z(\eta)} e^{N\eta(b^2+2bh)}\Omega(b),
\label{magn5}
\end{eqnarray}
\begin{eqnarray}
P_{MCE}(b)= \frac{1}{g(\epsilon)}  (\epsilon+2b^2+4bh)^{\frac{N-2}{2}}\Omega(b),
\label{magn4}
\end{eqnarray}
where
\begin{eqnarray}
\Omega(b)= \int\delta\left (Nb-\sum_{i}\cos\theta_i\right )\delta\left (\sum_{i}\sin\theta_i\right )\, d\theta_1...d\theta_N,\nonumber\\
\label{magn6}
\end{eqnarray}
is the unconditional number of microstates with magnetization $b$. It is given, for $N\rightarrow +\infty$, by equation (\ref{magn15}). Substituting equation (\ref{magn15}) in equations (\ref{magn5}) and (\ref{magn4}), we obtain
\begin{eqnarray}
P_{CE}(b)=\frac{1}{Z(\eta)} e^{-\frac{N\eta}{2}\left \lbrack -2b^2-4bh+\frac{2}{\eta}b\lambda-\frac{2}{\eta}\ln I_0(\lambda)\right\rbrack},
\label{magn19}
\end{eqnarray}
\begin{eqnarray}
P_{MCE}(b)=\frac{1}{g(\epsilon)} e^{N\left\lbrack \frac{1}{2}\ln (\epsilon+2b^2+4bh)-b\lambda+\ln I_0(\lambda)\right\rbrack}.
\label{magn17b}
\end{eqnarray}
The distribution of the magnetization can therefore be rewritten
\begin{eqnarray}
P_{CE}({b})=\frac{1}{Z(\eta)}e^{-\frac{N\eta}{2} f(b)},
\label{magn20}
\end{eqnarray}
\begin{eqnarray}
P_{MCE}(b)=\frac{1}{g(\epsilon)}e^{Ns(b)},
\label{magn18}
\end{eqnarray}
where $f(b)$ and $s(b)$ are the free energy and the entropy defined in Sections \ref{sec_cano} and \ref{sec_micro}.

\begin{figure}
\begin{center}
\includegraphics[clip,scale=0.3]{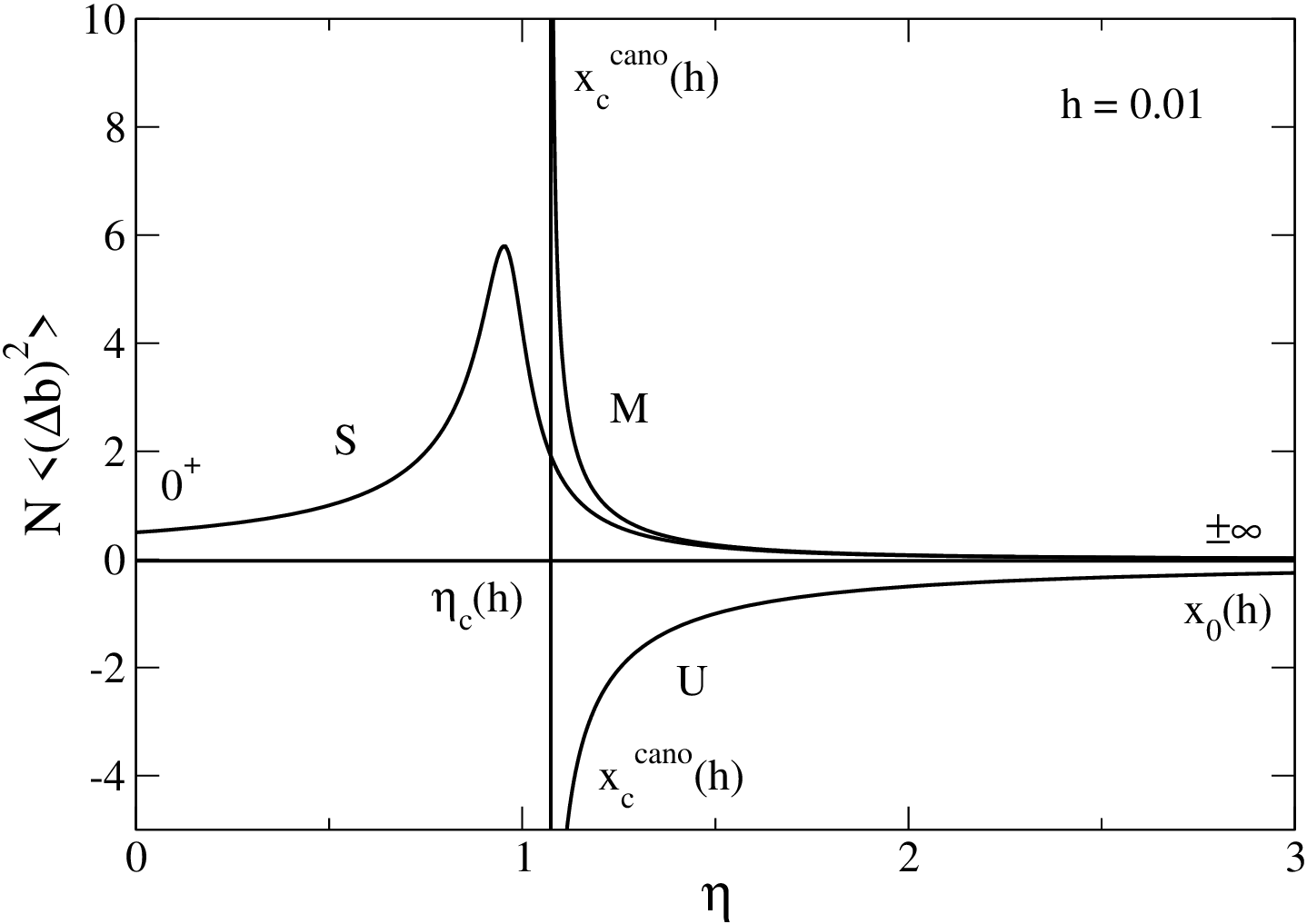}
\caption{Variance of the magnetization as a function of the inverse temperature $\eta$ in the canonical ensemble. Using the results of Appendix \ref{sec_asy}, we find that $N\langle (\Delta b)^2\rangle\rightarrow 1/2$ for $\eta\rightarrow 0$ (corresponding to $x\rightarrow 0$), $N\langle (\Delta b)^2\rangle\sim 1/(8(1+h)^2\eta^2)$ for $\eta\rightarrow +\infty$ (corresponding to $x\rightarrow +\infty$), $N\langle (\Delta b)^2\rangle\sim 1/(8(1-h)^2\eta^2)$ for $\eta\rightarrow +\infty$ (corresponding to $x\rightarrow -\infty$) and $N\langle (\Delta b)^2\rangle\sim -1/(2\eta)$ for $\eta\rightarrow +\infty$ (corresponding to $x\rightarrow x_0(h)$). On the other hand, $N\langle (\Delta b)^2\rangle\propto \pm (\eta-\eta_c(h))^{-1/2}$ when $\eta\rightarrow \eta_c(h)^+$ (corresponding to $x\rightarrow x_c^{cano}(h)$). }
\label{varianceCANOh0.01}
\end{center}
\end{figure}

\begin{figure}
\begin{center}
\includegraphics[clip,scale=0.3]{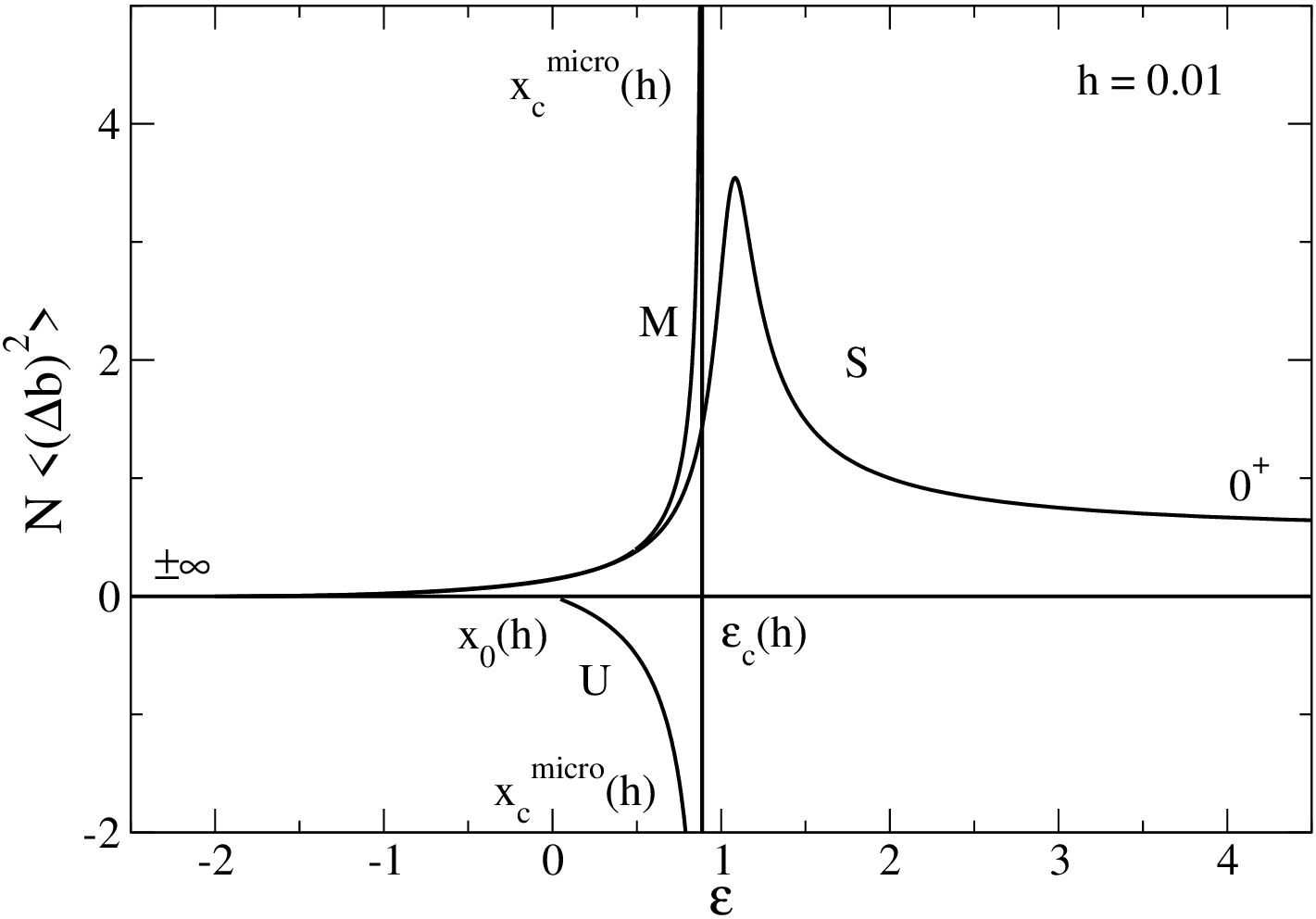}
\caption{Variance of the magnetization as a function of the energy $\epsilon$ in the microcanonical ensemble. Using the results of Appendix \ref{sec_asy}, we find that $N\langle (\Delta b)^2\rangle\rightarrow 1/2$ for $\epsilon\rightarrow +\infty$ (corresponding to $x\rightarrow 0$), $N\langle (\Delta b)^2\rangle\sim (\epsilon-\epsilon_{min}(h))^2/(64(1+h)^2)$ for $\epsilon\rightarrow \epsilon_{min}(h)$ (corresponding to $x\rightarrow +\infty$), $N\langle (\Delta b)^2\rangle\sim (\epsilon-\epsilon'_{min}(h))^2/(64(1-h)^2)$ for $\epsilon\rightarrow \epsilon'_{min}(h)$  (corresponding to $x\rightarrow -\infty$) and $N\langle (\Delta b)^2\rangle\sim -(\epsilon-\epsilon_0(h))/2$ for $\epsilon\rightarrow \epsilon_0(h)$ (corresponding to $x\rightarrow x_0(h)$). On the other hand, $N\langle (\Delta b)^2\rangle\propto \pm (\epsilon_c(h)-\epsilon)^{-1/2}$ when $\epsilon\rightarrow \epsilon_c(h)^-$ (corresponding to $x\rightarrow x_c^{micro}(h)$). }
\label{varianceMICROh0.01}
\end{center}
\end{figure}

For $N\rightarrow +\infty$, $P(b)$ is strongly peaked around the minimum of $f(b)$ or around the maximum of $s(b)$. Therefore, we can make a Gaussian approximation
\begin{eqnarray}
P(b)=\frac{1}{\sqrt{2\pi\langle (\Delta b)^2\rangle}} e^{-\frac{(\Delta b)^2}{2\langle (\Delta b)^2\rangle}},
\label{magn21}
\end{eqnarray}
where the variance is given by
\begin{eqnarray}
N\langle (\Delta b)^2\rangle_{CE}=\frac{2}{\eta f''(b)},
\label{magn22}
\end{eqnarray}
\begin{eqnarray}
N\langle (\Delta b)^2\rangle_{MCE}=-\frac{1}{s''(b)}.
\label{magn23}
\end{eqnarray}
Using equations (\ref{cs6}) and (\ref{ms7}), we obtain
\begin{eqnarray}
N\langle (\Delta b)^2\rangle_{CE}=\frac{1}{\frac{1}{b'(x)}-2\eta},
\label{magn24}
\end{eqnarray}
\begin{eqnarray}
N\langle (\Delta b)^2\rangle_{MCE}=\frac{1}{\frac{1}{b'(x)}-2(\eta-x^2)}.
\label{magn25}
\end{eqnarray}
These quantities are plotted in Figs. \ref{varianceCANOh0.01} and
\ref{varianceMICROh0.01}. In the anti-aligned phase, the variance
diverges at the critical inverse temperature $\eta_c(h)$ in the
canonical ensemble (see equation (\ref{cs8})) and at the critical
energy $\epsilon_c(h)$ in the microcanonical ensemble (see equation
(\ref{ms10})). Of course, the variance of the unstable states is
negative. These results generalize those obtained in \cite{isostab}
for $h=0$.

Comparing equations (\ref{msc4}) and (\ref{magn24}), we find that
$\chi_{cano}=2\eta N \langle (\Delta b)^2\rangle_{CE}$ which
corresponds to the general identity $\chi=\frac{2\pi}{k}\beta \langle
(\Delta B)^2\rangle$ valid in the canonical ensemble. On the other
hand, comparing equations (\ref{msm4}) and (\ref{magn25}) and
evaluating $b\partial b/\partial\epsilon$ from equations (\ref{t7})
and (\ref{t8}), we find that $\chi_{micro}=2\eta N \langle (\Delta
b)^2\rangle_{MCE}+4b\partial b/\partial\epsilon$ which corresponds to
the general identity $\chi=\frac{2\pi}{k}(\beta \langle (\Delta
B)^2\rangle+B(\partial B/\partial E))$ valid in the microcanonical
ensemble for $N\rightarrow +\infty$ \cite{vg}. Note that our approach
provides an alternative derivation of this relation.

\section{Distribution and variance of the energy in the canonical ensemble}
\label{sec_varene}

The distribution of energy in the canonical ensemble is defined by
\begin{eqnarray}
P(E)= \int\delta (E-H(\theta_1,v_1,...,\theta_N,v_N)) \nonumber\\
\times P_N(\theta_1,v_1,...,\theta_N,v_N)\, d\theta_1 dv_1...d\theta_N dv_N,
\label{varene1}
\end{eqnarray}
where $P_N(\theta_1,v_1,...,\theta_N,v_N)=\frac{1}{Z(\beta)}e^{-\beta H}$ is the canonical $N$-body distribution function. Inserting this distribution in equation (\ref{varene1}) and introducing the density of states (\ref{heur1}), we obtain
\begin{eqnarray}
P(E)= \frac{1}{Z(\beta)}g(E) e^{-\beta E}.
\label{varene2}
\end{eqnarray}
Introducing the microcanonical entropy $S(E)=\ln g(E)$ and the dimensionless variables of Section \ref{sec_t}, we get
\begin{eqnarray}
P(\epsilon)= \frac{1}{Z(\eta)} e^{N(s(\epsilon)-\frac{1}{2}\eta \epsilon)}.
\label{varene3}
\end{eqnarray}
Using the results of Section \ref{sec_t}, the microcanonical entropy $s(\epsilon)$ is given, in the limit $N\rightarrow +\infty$, by
\begin{eqnarray}
\label{varene4}
s=\frac{1}{2}\ln \left \lbrack \frac{2(b(x)+h)}{x}\right \rbrack+\ln I_0(x)-x b(x),
\end{eqnarray}
\begin{eqnarray}
\label{varene5}
\epsilon=\frac{2(b(x)+h)}{x}-2b(x)^2-4hb(x),
\end{eqnarray}
\begin{eqnarray}
\label{varene6}
b=\frac{I_1(x)}{I_0(x)}.
\end{eqnarray}
The most probable energy $\epsilon$ at inverse temperature $\eta$ is solution of $s'(\epsilon)=\frac{1}{2}\eta$. Using equations (\ref{varene4})-(\ref{varene6}), we find that it satisfies
the relation
\begin{eqnarray}
\label{varene7}
\eta=\frac{x}{2(b(x)+h)}.
\end{eqnarray}
This returns equation (\ref{t7}). Expanding equation (\ref{varene3}) around this most probable energy, we obtain
\begin{eqnarray}
P(\epsilon)=\frac{1}{\sqrt{2\pi\langle (\Delta\epsilon)^2\rangle}} e^{-\frac{(\Delta \epsilon)^2}{2\langle (\Delta \epsilon)^2\rangle}},
\label{varene8}
\end{eqnarray}
where the variance of energy is given by
\begin{eqnarray}
N\langle (\Delta \epsilon)^2\rangle=-\frac{1}{s''(\epsilon)}.
\label{varene9}
\end{eqnarray}
Using equations (\ref{varene4})-(\ref{varene6}), we can compute $s''(\epsilon)$ at fixed $\eta$. If we evaluate this expression at the most probable energy satisfying equation (\ref{varene7}) and substitute the result in equation (\ref{varene9}), we find that
\begin{eqnarray}
N\langle (\Delta \epsilon)^2\rangle=\frac{2}{\eta^2}\frac{1-2b'(x)(\eta-x^2)}{1-2\eta b'(x)}.
\label{varene10}
\end{eqnarray}
Comparing equation (\ref{varene10}) with equation (\ref{sh4}), we obtain
\begin{eqnarray}
\label{varene11}
N\langle (\Delta \epsilon)^2\rangle=\frac{4}{\eta^2}c,
\end{eqnarray}
which returns the general relation $C=\beta^2\langle (\Delta E)^2\rangle$, where $C=-\beta^2\partial E/\partial\beta$, valid in the canonical ensemble.

\section{Specific heat in canonical and microcanonical ensembles}
\label{sec_shcm}

We consider a general Hamitonian 
\begin{eqnarray}
\label{shcm1}
H=\sum_{i=1}^{N}m\frac{v_i^2}{2}+U({\bf r}_1,...,{\bf r}_{N})=K+U,
\end{eqnarray}
in $d$ dimensions. In the canonical ensemble, the partition function is
\begin{eqnarray}
\label{shcm2}
Z(\beta)=\int e^{-\beta H}\, d{\bf r}_{1}...d{\bf r}_{N}d{\bf v}_{1}...d{\bf v}_{N},
\end{eqnarray}
where $\beta=1/T$ is the inverse temperature (we take the Boltzmann constant equal to unity). The average energy is given by
\begin{eqnarray}
\label{shcm3}
\langle E\rangle=\frac{1}{Z}\int H e^{-\beta H}\, d{\bf r}_{1}...d{\bf r}_{N}d{\bf v}_{1}...d{\bf v}_{N}\nonumber\\
=-\frac{1}{Z}\frac{\partial Z}{\partial \beta}=-\frac{\partial \ln Z}{\partial\beta}=\frac{\partial}{\partial\beta}(\beta F),
\end{eqnarray}
where $F=-(1/\beta)\ln Z$ is the free energy. Similarly, we have
\begin{eqnarray}
\label{shcm4}
\langle E^2\rangle=\frac{1}{Z}\int H^2 e^{-\beta H}\, d{\bf r}_{1}...d{\bf r}_{N}d{\bf v}_{1}...d{\bf v}_{N}=\frac{1}{Z}\frac{\partial^2 Z}{\partial \beta^2}.\nonumber\\
\end{eqnarray}
The specific heat is defined by
\begin{eqnarray}
\label{shcm5}
C=\frac{\partial\langle E\rangle}{\partial T}=-\beta^2\frac{\partial\langle E\rangle}{\partial \beta}=-\beta^2\frac{\partial^2}{\partial\beta^2}(\beta F). 
\end{eqnarray}
Substituting equation (\ref{shcm3}) in equation (\ref{shcm5}) we obtain
\begin{eqnarray}
\label{shcm6}
C=\beta^2\left\lbrack \frac{1}{Z}\frac{\partial^2 Z}{\partial \beta^2}-\frac{1}{Z^2} \left (\frac{\partial Z}{\partial\beta}\right )^2\right\rbrack.
\end{eqnarray}
Using equations (\ref{shcm3}) and (\ref{shcm4}), and introducing the variance of the energy $\langle (\Delta E)^2\rangle=\langle E^2\rangle-\langle E\rangle^2$, we obtain the well-known formula
\begin{eqnarray}
\label{shcm7}
C=\beta^2\langle (\Delta E)^2\rangle,
\end{eqnarray}
which shows in particular that the specific heat is positive in the canonical ensemble. Note that the average value of the kinetic energy and its fluctuations in the canonical ensemble are given by
\begin{eqnarray}
\label{shcm8}
\langle K\rangle =\frac{dN}{2\beta},\qquad \langle (\Delta K)^2\rangle=\frac{dN}{2\beta^2}.
\end{eqnarray}
These relations can easily be obtained from the previous ones by using
the fact that the variables of velocity and position factorize (this
is equivalent to taking $U=0$ in the foregoing formulae).

In the microcanonical ensemble, the density of state is
\begin{eqnarray}
\label{shcm9}
g(E)=\int \delta(E-H)\, d{\bf r}_{1}...d{\bf r}_{N}d{\bf v}_{1}...d{\bf v}_{N}.
\end{eqnarray}
Integration over the velocities yields
\begin{eqnarray}
\label{shcm10}
g(E)=\frac{\pi^{\frac{dN}{2}}}{\Gamma(\frac{dN}{2})}\left (\frac{2}{m}\right )^{\frac{dN}{2}}\int (E-U)^{\frac{dN}{2}-1}\, d{\bf r}_{1}...d{\bf r}_{N}.\nonumber\\
\end{eqnarray}
Introducing the entropy $S(E)=\ln g(E)$, the microcanonical inverse temperature $\beta(E)=1/T(E)$ is given by
\begin{eqnarray}
\label{shcm11}
\beta=\frac{\partial S}{\partial E}=\frac{1}{g(E)}\frac{\partial g}{\partial E}.
\end{eqnarray}
Using equation (\ref{shcm10}), we obtain
\begin{eqnarray}
\label{shcm12}
\beta=\left (\frac{dN}{2}-1\right )\int (E-U)^{-1} P_{N} \, d{\bf r}_{1}...d{\bf r}_{N},
\end{eqnarray}
where $P_N({\bf r}_{1},...,{\bf r}_{N})$ is the microcanonical distribution. Noting that $E-U=K$, this relation can be rewritten
\begin{eqnarray}
\label{shcm13}
\beta=\left (\frac{dN}{2}-1\right )\left\langle \frac{1}{K}\right\rangle,
\end{eqnarray}
and it provides an exact relationship between the inverse microcanonical temperature and the average inverse kinetic energy. For $N\rightarrow +\infty$, we have
\begin{eqnarray}
\label{shcm14}
\left\langle \frac{1}{K}\right\rangle^{-1}\simeq \frac{dN}{2\beta},
\end{eqnarray}
which can be compared with equation (\ref{shcm8}). The specific heat is defined by
\begin{eqnarray}
\label{shcm15}
C=\frac{\partial E}{\partial T}=-\frac{\beta^2}{\frac{\partial\beta}{\partial E}}=-\frac{\beta^2}{S''(E)}. 
\end{eqnarray}
Using equation (\ref{shcm11}), we obtain
\begin{eqnarray}
\label{shcm16}
\frac{1}{C}=-\frac{1}{\beta^{2}}\left\lbrack \frac{1}{g}\frac{\partial^2 g}{\partial E^2}-\frac{1}{g^2} \left (\frac{\partial g}{\partial E}\right )^2\right\rbrack.
\end{eqnarray}
Using equations (\ref{shcm10}) and (\ref{shcm11}), the foregoing relation can be rewritten
\begin{eqnarray}
\label{shcm17}
\frac{1}{C}=-\frac{1}{\beta^{2}}\left\lbrack \left (\frac{dN}{2}-1\right )\left (\frac{dN}{2}-2\right )\left\langle \frac{1}{K^2}\right \rangle -\beta^2\right\rbrack.\nonumber\\
\end{eqnarray}
Introducing the variance of the inverse kinetic energy
\begin{eqnarray}
\label{shcm18}
\left\langle\left (\Delta\frac{1}{K}\right )^2\right\rangle=\left\langle \frac{1}{K^2}\right \rangle-\left\langle \frac{1}{K}\right \rangle^2,
\end{eqnarray}
the specific heat per particle $c=C/N$, and using equation (\ref{shcm13}), we obtain after some rearrangements
the exact relation
\begin{eqnarray}
\label{shcm19}
\frac{1}{c}=\frac{N}{\frac{dN}{2}-1}-\frac{\frac{dN}{2}-2}{\frac{dN}{2}-1}\frac{N \left\langle\left (\Delta\frac{1}{K}\right )^2\right\rangle}{\left\langle \frac{1}{K}\right \rangle^2}. 
\end{eqnarray}
For $N\rightarrow +\infty$, it reduces to
\begin{eqnarray}
\label{shcm20}
\frac{1}{c}\simeq \frac{2}{d}-\frac{N \left\langle\left (\Delta\frac{1}{K}\right )^2\right\rangle}{\left\langle \frac{1}{K}\right \rangle^2}. 
\end{eqnarray}
This returns the relationship obtained by Lebowitz et al \cite{lebowitz}. It clearly shows that the specific heat is not necessarily positive in the microcanonical ensemble. 

\section{Magnetic susceptibility in canonical and microcanonical ensembles}
\label{sec_msce}

Using equation (\ref{heur7}), the Hamitonian (\ref{m1}) can be written 
\begin{eqnarray}
\label{shcm100}
{\cal H}={\cal H}_0-\frac{2\pi}{k}BH,
\end{eqnarray}
where ${\cal H}_0$ is the Hamiltonian of the HMF model without magnetic field.
In the canonical ensemble, the partition function is
\begin{eqnarray}
\label{shcm200}
Z(\beta,H)=\int e^{-\beta ({\cal H}_{0}-\frac{2\pi}{k}BH)}\, d\theta_{1}...d\theta_{N}dv_{1}...dv_{N}.\nonumber\\
\end{eqnarray}
The average magnetization is given by
\begin{eqnarray}
\label{shcm300}
\langle B\rangle=\frac{1}{Z}\int B e^{-\beta ({\cal H}_{0}-\frac{2\pi}{k}BH)} \, d\theta_{1}...d\theta_{N}dv_{1}...dv_{N}\nonumber\\
=\frac{k}{2\pi\beta}\frac{1}{Z}\frac{\partial Z}{\partial H}=\frac{k}{2\pi\beta}\frac{\partial \ln Z}{\partial H}=-\frac{k}{2\pi}\frac{\partial F}{\partial H},\nonumber\\
\end{eqnarray}
where $F=-(1/\beta)\ln Z$ is the free energy. Similarly, we have
\begin{eqnarray}
\label{shcm400}
\langle B^2\rangle=\frac{1}{Z}\int B^2 e^{-\beta ({\cal H}_{0}-\frac{2\pi}{k}BH)}\, d\theta_{1}...d\theta_{N}dv_{1}...dv_{N}\nonumber\\
=\left (\frac{k}{2\pi\beta}\right )^2 \frac{1}{Z}\frac{\partial^2 Z}{\partial H^2}.\qquad
\end{eqnarray}
The magnetic susceptibility is defined by
\begin{eqnarray}
\label{shcm500}
\chi=\frac{\partial\langle B\rangle}{\partial H}=-\frac{k}{2\pi}\frac{\partial^2 F}{\partial H^2}. 
\end{eqnarray}
Substituting equation (\ref{shcm300}) in equation (\ref{shcm500}) we get
\begin{eqnarray}
\label{shcm600}
C=\frac{k}{2\pi\beta}\left\lbrack \frac{1}{Z}\frac{\partial^2 Z}{\partial H^2}-\frac{1}{Z^2} \left (\frac{\partial Z}{\partial H}\right )^2\right\rbrack.
\end{eqnarray}
Using equations (\ref{shcm300}) and (\ref{shcm400}), and introducing the variance of the magnetization $\langle (\Delta B)^2\rangle=\langle B^2\rangle-\langle B\rangle^2$, we obtain the exact relation
\begin{eqnarray}
\label{shcm700}
\chi=\frac{2\pi\beta}{k}\langle (\Delta B)^2\rangle,
\end{eqnarray}
which shows in particular that the magnetic susceptibility is positive in the canonical ensemble.

In the microcanonical ensemble, the density of states is
\begin{eqnarray}
\label{shcm800}
g(E,H)=A\int \left (E-U_0+\frac{2\pi}{k}BH\right )^{\frac{N}{2}-1}\, d\theta_{1}...d\theta_{N},\nonumber\\
\end{eqnarray}
where $U_0$ is the potential energy of the HMF model without magnetic
field and $A=(2\pi)^{\frac{N}{2}}/{\Gamma(\frac{N}{2})}$. The average
magnetization is given by
\begin{eqnarray}
\label{shcm900}
\langle B\rangle=\frac{A}{g(E,H)}\int B \left (E-U_0+\frac{2\pi}{k}BH\right )^{\frac{N}{2}-1}\, d\theta_{1}...d\theta_{N},\nonumber\\
\end{eqnarray}
Taking its derivative with respect to $H$ and $E$, we obtain
\begin{eqnarray}
\label{shcm1000}
\frac{\partial \langle B\rangle}{\partial H}=\left (\frac{N}{2}-1\right )\frac{2\pi}{k}\left\langle \frac{B^2}{K}\right\rangle-\frac{1}{g}\frac{\partial g}{\partial H}\langle B\rangle,
\end{eqnarray}
\begin{eqnarray}
\label{shcm1100}
\frac{\partial \langle B\rangle}{\partial E}=\left (\frac{N}{2}-1\right )\left\langle \frac{B}{K}\right\rangle-\frac{1}{g}\frac{\partial g}{\partial E}\langle B\rangle,
\end{eqnarray}
where $K=E-U_0+\frac{2\pi}{k}BH$ is the kinetic energy. On the other hand, taking the derivative of the density of states (\ref{shcm800}) with respect to $H$ and $E$, we get
\begin{eqnarray}
\label{shcm1200}
\frac{1}{g}\frac{\partial g}{\partial H}=\left (\frac{N}{2}-1\right )\frac{2\pi}{k}\left\langle \frac{B}{K}\right\rangle,
\end{eqnarray}
\begin{eqnarray}
\label{shcm1300}
\frac{1}{g}\frac{\partial g}{\partial E}=\left (\frac{N}{2}-1\right )\left\langle \frac{1}{K}\right\rangle.
\end{eqnarray}
The magnetic susceptibility is defined by
\begin{eqnarray}
\label{shcm1400}
\chi=\frac{\partial\langle B\rangle}{\partial H}.
\end{eqnarray}
Combining the preceding equations, we obtain the exact relation
\begin{eqnarray}
\label{shcm1500}
\frac{k\chi}{2\pi}=\left (\frac{N}{2}-1\right )\left (\left\langle \frac{B^2}{K}\right\rangle-2\left\langle \frac{B}{K}\right\rangle \langle B\rangle+\langle B\rangle^2 \left\langle \frac{1}{K}\right\rangle\right )\nonumber\\
+\langle B\rangle \frac{\partial \langle B\rangle}{\partial E}.\qquad\qquad
\end{eqnarray}
It generalizes the identity 
\begin{eqnarray}
\label{shcm1600}
\frac{k\chi}{2\pi}=\beta \langle (\Delta
B)^2\rangle+\langle B\rangle \frac{\partial \langle B\rangle}{\partial E},
\end{eqnarray}
valid in the microcanonical ensemble for $N\rightarrow +\infty$ (see Appendix
\ref{sec_magnetization}). These relations clearly show that the
magnetic susceptibility is not necessarily positive in the
microcanonical ensemble.


\begin{thebibliography}{}


\bibitem{houches} {\small {\it Dynamics and thermodynamics of systems
with long range interactions}, edited by T. Dauxois {\it et al.},
Lecture Notes in Physics {\bf 602}, (Springer, 2002)}
\bibitem{assisebook} {\small {\it Dynamics and thermodynamics of systems
with long range interactions: Theory and experiments}, edited by
A. Campa {\it et al.}, AIP Conf. Proc. {\bf 970} (AIP, 2008). }
\bibitem{oxford}  {\small  {\it Long-Range Interacting Systems}, edited by T. Dauxois, S. Ruffo and L. Cugliandolo, Les Houches Summer School 2008, (Oxford: Oxford University Press, 2009)}
\bibitem{cdr}  {\small A. Campa, T. Dauxois, S. Ruffo,   Physics Reports {\bf 480}, 57 (2009)}
\bibitem{ms}  {\small J. Messer, H. Spohn, J. Stat. Phys. {\bf 29}, 561 (1982)}
\bibitem{kk}  {\small T. Konishi, K. Kaneko, J. Phys. A {\bf 25}, 6283 (1992)}
\bibitem{ik}  {\small S. Inagaki, T. Konishi, Publ. Astron. Soc. Japan {\bf 45}, 733 (1993)}
\bibitem{inagaki}  {\small S. Inagaki, Prog. Theor. Phys. {\bf 90}, 557 (1993)}
\bibitem{pichon}  {\small  C. Pichon, PhD thesis, Cambridge (1994)}
\bibitem{ar}  {\small  M. Antoni, S.  Ruffo, Phys. Rev. E {\bf  52}, 2361 (1995)  }
\bibitem{cvb}  {\small P.H. Chavanis, J. Vatteville, F. Bouchet, Eur. Phys. J. B {\bf 46}, 61 (2005)}
\bibitem{paddy}  {\small T. Padmanabhan, Phys. Rep. {\bf 188}, 285 (1990)}
\bibitem{katzrev}  {\small J. Katz, Found. Phys. {\bf 33}, 223 (2003)}
\bibitem{ijmpb}  {\small P.H. Chavanis, Int J. Mod. Phys. B {\bf 20}, 3113 (2006)}
\bibitem{lrtbis}  {\small V. Latora, A. Rapisarda, C. Tsallis,  Phys. Rev. E {\bf 64}, 056134 (2001)}
\bibitem{lrt}  {\small V. Latora, A. Rapisarda, C. Tsallis,  Physica A {\bf 305}, 129 (2002)}
\bibitem{choi}  {\small  M. Y. Choi, J. Choi, Phys. Rev. Lett. {\bf 91}, 124101 (2003)}
\bibitem{yamaguchi}  {\small Y.Y. Yamaguchi, J. Barr\'e, F. Bouchet, T. Dauxois, S. Ruffo,  Physica A  {\bf 337}, 36 (2004)}
\bibitem{incomplete}  {\small  P.H. Chavanis, Physica A {\bf 365}, 102 (2006)}
\bibitem{jain}  {\small K. Jain, F. Bouchet, D. Mukamel, J. Stat. Mech., P11008 (2007)}
\bibitem{cd}  {\small P.H. Chavanis, L. Delfini, Eur. Phys. J. B {\bf 69}, 389 (2009)}
\bibitem{campastab}  {\small A. Campa, P.H. Chavanis, J. Stat. Mech., P06001 (2010)}
\bibitem{lb}  {\small  D. Lynden-Bell, Mon. Not. R. astr. Soc.  {\bf 136}, 101 (1967)}
\bibitem{epjb}  {\small P.H. Chavanis, Eur. Phys. J. B {\bf 53}, 487 (2006)}
\bibitem{precommun}  {\small  A. Antoniazzi, D. Fanelli, J. Barr\'e, P.H. Chavanis, T. Dauxois, S. Ruffo, Phys. Rev. E {\bf 75}, 011112 (2007) }
\bibitem{antobis}  {\small A. Antoniazzi, F. Califano, D. Fanelli, S. Ruffo, Phys. Rev. Lett. {\bf 98}, 150602 (2007)}
\bibitem{anto}  {\small A. Antoniazzi, D. Fanelli, S. Ruffo, Y. Yamaguchi, Phys. Rev. Lett. {\bf 99}, 040601 (2007)}
\bibitem{proc}  {\small P.H. Chavanis, G. De Ninno, D. Fanelli, S. Ruffo, {\it Out of equilibrium phase transitions in mean field Hamiltonian dynamics} in {Chaos, Complexity and Transport: Theory and Applications}, edited by C. Chandre, X. Leoncini, G. Zaslavsky  (World Scientific 2008)}
\bibitem{staniscia1}  {\small F. Staniscia, P.H. Chavanis, G. de Ninno, D. Fanelli, Phys. Rev. E {\bf 80}, 021138 (2009)}
\bibitem{bachelard} {\small  R. Bachelard, C. Chandre, D. Fanelli, X. Leoncini, S. Ruffo, Phys. Rev. Lett.  {\bf 101}, 260603 (2008)}
\bibitem{firpo} {\small M.C. Firpo, Europhys. Lett.  {\bf 88}, 30010 (2009)}
\bibitem{leoncini} {\small  X. Leoncini, T.L. Van Den Berg, D. Fanelli, Europhys. Lett.  {\bf 86}, 20002 (2009)}
\bibitem{bachelardbis} {\small R. Bachelard, C. Chandre, A. Ciani,  D. Fanelli, Y.Y. Yamaguchi, Physics Lett. A  {\bf 373}, 4239 (2009)}
\bibitem{prlneg}  {\small F. Staniscia, A. Turchi, D. Fanelli, P.H. Chavanis, G. de Ninno, Phys. Rev. Lett. {\bf 105}, 010601 (2010)}
\bibitem{cc}  {\small P.H. Chavanis, A. Campa, Eur. Phys. J. B {\bf 76}, 581 (2010)}
\bibitem{latoralyap} {\small  V. Latora, A. Rapisarda, S. Ruffo, Phys. Rev. Lett. {\bf 80}, 692 (1998)}
\bibitem{firpolyap} {\small  M.C. Firpo, Phys. Rev. E  {\bf 57}, 6599 (1998)} 
\bibitem{aging}  {\small M. Montemurro, F. Tamarit, C. Anteneodo, Phys. Rev. E {\bf 67}, 031106 (2003)}
\bibitem{glassy}  {\small A. Pluchino, V. Latora, A. Rapisarda, Physica A {\bf 340}, 187 (2004)}
\bibitem{mk}  {\small H. Morita, K. Kaneko, Phys. Rev. Lett. {\bf 96}, 050602 (2006)}
\bibitem{pluch}  {\small A. Pluchino, V. Latora, A. Rapisarda, Europhys. Lett. {\bf 80}, 26002 (2007)}
\bibitem{figue}{\small 	A. Figueiredo, T.M. Rocha Filho, M.A.  Amato,  Europhys. Lett. {\bf 80}, 26002 (2007)}
\bibitem{campadyn}  {\small A. Campa, P.H. Chavanis, A.  Giansanti, G. Morelli,  Phys. Rev. E {\bf 78}, 040102 (2008) }
\bibitem{lrr}  {\small V. Latora, A. Rapisarda, S. Ruffo,  Phys. Rev. Lett. {\bf 83}, 2104 (1999)}
\bibitem{yamaseul}  {\small Y.Y. Yamaguchi,  Phys. Rev. E {\bf 68}, 066210 (2003)}
\bibitem{plr}  {\small A. Pluchino, V. Latora, A. Rapisarda, Physica D {\bf 193}, 315 (2004)}
\bibitem{latora}  {\small A. Pluchino, V. Latora, A. Rapisarda, Physica A {\bf 338}, 60 (2004)}
\bibitem{news}  {\small A. Rapisarda, A. Pluchino, Europhysics News {\bf 36}, 202 (2005)}
\bibitem{bd}  {\small F. Bouchet, T.  Dauxois,  Phys. Rev. E {\bf 72}, 5103 (2005)}
\bibitem{ybd}  {\small Y.Y. Yamaguchi, F. Bouchet, T. Dauxois, 	J. Stat. Mech., P01020 (2007)}
\bibitem{cl}  {\small  P.H. Chavanis, M. Lemou, Eur. Phys. J. B {\bf 59}, 217 (2007)}
\bibitem{cgm}  {\small A. Campa, A. Giansanti, G. Morelli,  Phys. Rev. E {\bf 76}, 041117 (2007)}
\bibitem{bgm}  {\small F. Bouchet, S. Gupta, D. Mukamel,
Physica A, {\bf 389}, 4389 (2010)}
\bibitem{chavkin}  {\small P.H. Chavanis, J. Stat. Mech.,  P05019 (2010)}
\bibitem{bo1}  {\small F. Baldovin, E. Orlandini, Phys. Rev. Lett.
{\bf 96}, 240602 (2006)}
\bibitem{bo2}  {\small F. Baldovin, E. Orlandini, Phys. Rev. Lett.
{\bf 97}, 100601 (2006)}
\bibitem{bco}  {\small F. Baldovin, P.H. Chavanis, E. Orlandini,
Phys. Rev. E {\bf 79}, 011102 (2009)}
\bibitem{cbo}  {\small  P.H. Chavanis, F. Baldovin, E. Orlandini, [arXiv1009.5603]}
\bibitem{gm}  {\small S. Gupta, D. Mukamel, Phys. Rev. Lett. {\bf 105}, 040602 (2010) }
\bibitem{largedev}  {\small J. Barr\'e, F. Bouchet, T. Dauxois, S. Ruffo, J. Stat. Phys. {\bf 119}, 677 (2005)}
\bibitem{isostab}  {\small P.H. Chavanis, [arXiv:1007.4916]}
\bibitem{vg}  {\small L. Velazquez, F. Guzman, [arXiv:0910.2906]}
\bibitem{bh}  {\small W. Braun, K. Hepp, Commun. Math. Phys. {\bf 56}, 101 (1977)}
\bibitem{art}  {\small  M. Antoni, S. Ruffo, A. Torcini, Europhys. Lett., {\bf  66}, 645 (2004)}
\bibitem{metastable}  {\small  P.H. Chavanis, Astron. Astrophys. {\bf  432}, 117 (2005)}
\bibitem{yukawa}  {\small P.H. Chavanis, L. Delfini, Phys. Rev. E {\bf 81}, 051103 (2010)}	
\bibitem{ellis} {\small  R. Ellis, K. Haven, B. Turkington, J. Stat. Phys.  {\bf 101}, 999 (2000)}
\bibitem{mukamel}  {\small D. Mukamel, S. Ruffo, N. Schreiber,  Phys. Rev. Lett. {\bf 95}, 240604 (2005)}
\bibitem{bouchet}  {\small F. Bouchet, T. Dauxois, D. Mukamel, S. Ruffo,  Phys. Rev. E {\bf 77}, 011125 (2008)}
\bibitem{lebowitz}  {\small J.L. Lebowitz, J.K. Percus, L. Verlet,  Phys. Rev. {\bf 153}, 250 (1967)}
\end{thebibliography}
\end{document}